\documentclass[fleqn,usenatbib]{mnras}

\usepackage{newtxtext,newtxmath}

\usepackage[T1]{fontenc}

\DeclareRobustCommand{\VAN}[3]{#2}
\let\VANthebibliography\thebibliography
\def\thebibliography{\DeclareRobustCommand{\VAN}[3]{##3}\VANthebibliography}

\usepackage{graphicx}	
\usepackage{amsmath}	
\usepackage{subcaption}
\usepackage{booktabs}
\usepackage[dvipsnames]{xcolor}

\newcommand*{\kms}{\text{km}\,\text{s}\ensuremath{^{-1}}}

\newcommand*{\mass}{\ensuremath{\rm{M}_{*}}}
\DeclareUnicodeCharacter{2212}{-}

\newcommand*{\http}[1]{\href{http://#1}{#1}}

\defcitealias{licquia2015b}{LNB15}

\title[A UV to IR Portrait of the Milky Way]{Constraining the Milky Way's Ultraviolet to Infrared SED with Gaussian Process Regression}

\author[C. E. Fielder et al.]{
Catherine E. Fielder,$^{1,2}$\thanks{E-mail: cef41@pitt.edu}
Jeffrey A. Newman,$^{1,2}$
Brett H. Andrews,$^{1,2}$
Gail Zasowski,$^{3}$
Nicholas F. Boardman,$^{3}$\newauthor
Tim Licquia,$^{1,4}$
Karen L. Masters,$^{5}$
Samir Salim,$^{6}$
\\
$^{1}$Department of Physics and Astronomy, University of Pittsburgh, Pittsburgh, PA 15260, USA\\
$^{2}$Pittsburgh Particle Physics, Astrophysics, and Cosmology Center (PITT PACC), University of Pittsburgh, Pittsburgh, PA 15260, USA\\
$^{3}$Department of Physics and Astronomy, University of Utah, Salt Lake City, UT, 84112, USA\\
$^{4}$Core R\&D: Information Research, Dow Inc., 633 Washington St, 1776 Building, Midland, MI, 48667, USA\\
$^{5}$Departments of Physics and Astronomy, Haverford College, 370 Lancaster Avenue, Haverford, PA 19041, USA\\
$^{6}$Department of Astronomy, Indiana University, Bloomington, IN, 47405, USA
}

\date{Accepted XXX. Received YYY; in original form ZZZ}

\pubyear{2021}

\begin{document}
\label{firstpage}
\pagerange{\pageref{firstpage}--\pageref{lastpage}}
\maketitle

\begin{abstract}
Improving our knowledge of global Milky Way (MW) properties is critical for connecting the detailed measurements only possible from within our Galaxy to our understanding of the broader galaxy population. We train Gaussian Process Regression (GPR) models on SDSS galaxies to map from galaxy properties (stellar mass, apparent axis ratio, star formation rate, bulge-to-total ratio, disk scale length, and bar vote fraction) to UV (GALEX $FUV/NUV$), optical (SDSS $ugriz$) and IR (2MASS $JHKs$ and WISE $W1/W2/W3/W4$) fluxes and uncertainties. With these models we estimate the photometric properties of the MW, resulting in a full UV-to-IR spectral energy distribution (SED) as it would be measured externally, viewed face-on. We confirm that the Milky Way lies in the green valley in optical diagnostic diagrams, but show for the first time that the MW is in the star-forming region in standard UV and IR diagnostics---characteristic of the population of red spiral galaxies. Although our GPR method  predicts one band at a time, the resulting MW UV--IR SED is consistent with SEDs of local spirals with characteristics broadly similar to the MW, suggesting that these independent predictions can be combined reliably. Our UV--IR SED will be invaluable for reconstructing the MW's star formation history using the same tools employed for external galaxies, allowing comparisons of results from \textit{in situ} measurements to those from the methods used for extra-galactic objects.

\end{abstract}




\section{Introduction}
\label{sec:intro}

Within the Milky Way we have a unique opportunity to study the nuances of galactic properties, allowing us to test galaxy formation and evolution models at an unrivalled level of detail. 
For example, chemical abundances for hundreds of thousands of stars have been obtained from spectroscopic surveys \citep{majewski2017,martell2017}, and
stellar surveys that catalogue distance and dynamical measurements on millions of stars have been performed \citep{gaia2018}. This exhaustive stellar information has helped constrain the Milky Way's evolutionary history.
In turn, high-resolution dynamical simulations have been able to produce galaxies of increased similarity to the Milky Way \citep{guedes2011,swala2016,wetzel2016},  matching  fundamental galaxy properties such as dwarf satellite  populations and reproducing characteristics of our Galaxy's gas, dust, and stellar components. Comparisons between Milky Way stellar data and high-resolution hydrodynamical simulations of Milky Way-like galaxies are an increasingly useful way to improve our understanding of galaxy formation.

However, the marriage between observations and models is delicate: incorrect assumptions on one side can propagate into the other. Simulators must make choices about how to implement crucial parameters that affect the galaxy evolution process, such as the gas density threshold for star formation to occur and the efficiency with which it proceeds; this is sometimes done by attempting to match observed properties of the Milky Way. But without knowing how our Galaxy fits in amongst the broader galaxy population, it is difficult to determine whether simulations match the Milky Way because they have the correct physics or because they have incorrectly tuned parameters that match by coincidence or design. 

This is complicated by the exceptional difficulty of obtaining a global picture of the Milky Way, given our location in the disk and the obscuration caused by interstellar dust. As a result, there are properties that we can easily measure in external galaxies that are impossible to measure directly within our own, making it difficult to determine where we fit within the broader galaxy population.  

Creating an outside-in picture of the Milky Way that spans a multitude of broad-band wavelengths will enable simulators to more accurately tune their physics assumptions, as it will then be possible to test whether quantities that can only be determined from large-scale stellar surveys and those that can be measured directly only for extragalactic objects are reproduced. The most basic, easiest-to-measure intrinsic quantities we can use to study galaxies are their luminosities and colours; once redshift is known, these can be inferred from broad-band photometry. Hence the focus of this paper will be in determining these properties for the Milky Way. This will enable our Galaxy to be placed on standard colour-magnitude and colour-colour diagrams and result in a multi-wavelength spectral energy distribution (SED) for the Milky Way.

Astronomers have found that galaxies in the local Universe predominantly fall into two populations: passively-evolving red galaxies with older stellar populations; or blue galaxies that are still forming stars. In the optical colour-magnitude diagram (CMD), these two galaxy populations are commonly referred to as the ``red sequence'' and the ``blue cloud''. The colour bimodality of galaxies has been observed at both low redshift, \citep[$z\sim0.1$, e.g.,][]{strateva2001,baldry2004} and up to redshift of $z\sim1$ \citep[e.g.,][]{bell2004,weiner2005}. The region of the CMD between these two distinct galaxy populations is often referred to as the ``green valley''. This locus is thought to contain a transitional population of galaxies that are ``passively'' evolving in the sense that no new star formation is occurring \citep[e.g.,][]{bell2004,faber2007}, though they still can contain some younger stars. The increase in the fraction of red galaxies over time has led many astronomers to conclude that a galaxy first lives in the star forming blue cloud and then transitions into the green valley and ultimately into the red sequence in complete quiescence, with the galaxy growing more and more red over time due to the ageing stellar population. Green valley galaxies are presumed to be undergoing some form of quenching of their star formation \citep{salim2007,schawinski2014,smethurst2015} - either late-type galaxies that are gradually running out of their cold gas reservoir or having their star formation suppressed, or early-type galaxies that had their gas reservoirs rapidly destroyed. Objects found within the green valley may also simply be in the tails of the blue cloud or red sequence, rather than being in a transitional state  \citep{taylor2015}. While the precise details of quenching processes and the origin of the galaxy colour bimodality have yet to be determined, a galaxy's location on the CMD remains a very useful tool for determining how a galaxy fits into the broader population. 

The radiation emitted by a galaxy is characterised by its spectral energy distribution (SED), or flux as a function of wavelength. Galaxy SEDs contain the imprints of the physical processes occurring within - the stellar population's ages and abundances (i.e., the star formation history and metallicity of the galaxy), the dust and gas content and the chemistry and physical state of the interstellar medium (ISM). Because different sources dominate the emission at different wavelengths, long-wavelength-baseline SEDs allow one to disentangle the contributing effects. This makes SEDs one of the best direct probes for studying galaxy formation and evolution from both an observational and theoretical modelling perspective.



However, comparing colours and luminosities of the Milky Way to external galaxies is not trivial, regardless of whether we compare to observed galaxies or to mock images from high-resolution hydrodynamical simulations (such as Eris \citealt{guedes2011}, APOSTLE \citealt{swala2016}, and Latte \citealt{wetzel2016}). Much of our view of the Galaxy is obscured by interstellar dust, especially at UV and optical wavelengths \citep[e.g.,][]{cardelli1989,schlegel1998}. Stars outside of the local solar region are reddened as a result of the dust obscuration. Determining the integrated light of stellar populations in the Milky Way is challenging due to the spread of stars over large and varying distances, with correspondingly large and varying dust extinction along lines of sight to the Earth. 
This makes the study of any portions of the Galactic disk beyond the solar neighbourhood exceptionally difficult, and results in a fragmented picture of the Milky Way. Integrated properties that are relatively painless to obtain in external galaxies (though dust obscuration can affect these observations as well, see e.g., \citet{masters2003,masters2010a}), are impossible to obtain directly within our own Galaxy \citep[e.g.,][]{mutch2011}. As a result, simulators often must resort to comparing their simulated Milky Ways to very general galaxy populations (such as sets of Sbc or late type galaxies) that, while superficially resembling the Milky Way, have a wide range of other global properties \citep[e.g.,][]{guedes2011}.

In an effort to circumvent our limited view of the Milky Way, we can study galaxies that mimic the properties of our Galaxy but can be observed from outside, which we label Milky Way Analogues (MWAs). This method hinges on the Copernican assumption that the Milky Way should not be extraordinary among a galaxy population that shares some key properties with it. These comparisons are enabled by working within volume-limited subsets of large surveys, which ensure that the observed objects constitute a representative population. Previous work suggests that galaxies with similar stellar mass and star formation rates are also similar in other properties, as the observed galaxy population is well-matched by models that parameterise galactic star formation histories with a limited collection of curves \citep{behroozi2013,gladders2013,abramson2014,kelson2014}. Even further, \citet{bell2001} showed mass and star formation rate are strongly correlated with the photometric properties of a galaxy. Therefore we can exploit the fact that two galaxies of identical mass and star formation rate should have similar luminosities and colours, with some scatter given the range of galaxy photometric properties at fixed physical parameters. \citet{licquia2015b}, hereafter \citetalias{licquia2015b}, utilised this to constrain the Milky Way's optical colours and magnitudes based on the range of observed properties of MWAs that were matched in stellar mass and star formation rate.
MWAs also allow direct comparison of properties of our Galaxy to its closest peers \citep[e.g.,][]{licquia2016a,licquia2016b,mckelvie2019,boardman2020a,krishnarao2020} and have been a successful tool for improving our understanding of the Milky Way in an extra-galactic context.

The Milky Way, however, has some characteristics that are atypical (at the $<2\sigma$ level) amongst its peers - e.g., the Milky Way has an unusually compact disk (i.e., a small disk scale-length) \citep{bovy2013,bland-hawthorn2016,licquia2016a}, and an unusually quiescent merger history \citep[from observation][]{unavane1996,ruchti2015}, \citep[and simulation, e.g.,][]{fielder2019,carlesi2020}. The deviations of the Milky Way from the average suggest that we should consider parameters beyond just stellar mass and star formation rate in order to identify samples of objects that more closely resemble the Milky Way.

Galaxy morphological characteristics such as disk scale length ($R_{d}$) and bulge-to-total ratio ($B/T$) are tied to a galaxy's evolutionary history and therefore should connect to its photometric properties \citep{cappellari2016,saha2018} as well as to the ways in which the Milky Way is atypical. We would therefore wish to incorporate these properties in addition to stellar mass and star formation rate in defining an MWA. However, as the number of parameters required to match the Milky Way increases, the number of MWAs correspondingly reduces dramatically. For example \citet{mckelvie2019} only found 179 analogues when selecting on stellar mass, bulge-to-total mass ratio, and morphology; \citet{boardman2020b} and \citet{boardman2020a} found no MWAs within $1\sigma$ of the MW when selecting on stellar mass, star formation rate, bulge-to-total ratio, and disk scale length in either the SDSS-IV MaNGA survey \citep{bundy2015} or a larger photometric sample drawn from the GALEX-SDSS-WISE Legacy catalogue (GSWLC; \citet{salim2016}) respectively.

\citetalias{licquia2015b} found that the colour of the Milky Way is consistent the green valley region of the CMD as it has been defined using purely optical passbands. Characterising the UV and IR colours of the Milky Way can provide more sensitive probes of whether it would be classified as in the process of quenching if seen from outside, allowing us to better understand what type of population the Milky Way may belong to. \citetalias[][]{licquia2015b} speculated that the Milky Way might belong to the population of massive ``red spiral'' galaxies, which are characterised by their red optical colours despite ongoing star formation \citep{masters2010,cortese2012}. Galaxies within this population may be  moving into the green valley due to slow quenching (cf. \citet{schawinski2014}). This conjecture can only be fully tested by examining wavelengths outside of the optical range; in $g-r$, the colours of massive spiral galaxies on the star-forming main sequence (a population that should include the Galaxy) overlap with both the red sequence and the blue cloud \citep{cortese2012,salim2014}. However, samples of MWAs that have high-quality photometry over a broader wavelength range will have reduced numbers due to the limited coverage of sufficiently deep photometry in GALEX.

To address the lack of analogues when multi-dimensional parameter spaces are used, and the smaller overall sample size resulting from the increase in wavelength coverage, we introduce a Gaussian process regression (GPR) approach in this work. GPR is an emergent tool in astrophysics.
For example, \citet{bocquet2020} employed GPR to emulate results of cosmological simulations, while \citet{gordon2020} used GPR to detect and classify exoplanets. We can use GPR to leverage information from a wider variety of galaxies, instead of just the closest Milky Way analogues, in order to extract information from large-scale trends between galaxy physical and photometric properties.
Thanks to the probabilistic framework that underlies GPR, we obtain uncertainty estimates for all predicted quantities for free. The primary result from this paper will be an ultraviolet to infrared SED of the Milky Way as viewed face-on, determined via GPR based on  star formation history and structural parameters (i.e., galaxy physical parameters) that have been measured well for both the Milky Way and galaxies from the Sloan Digital Sky Survey \citep{dr82011}.


The paper is organised in the following manner. In \autoref{sec:data} we describe the observational data used, including the external galaxy data in \autoref{subsection:photometry} and \autoref{subsection:strucprops}, and estimated properties of the Milky Way in \autoref{subsection:mwparams}. \autoref{sec:gpr} details our new Gaussian process regression-based methodology.
In \autoref{sec:results} we compare the luminosity and colours of the Milky Way at multiple wavelengths and the Milky Way's predicted SED to properties of other galaxies. Finally we summarise our results, and discuss implications and future work in \autoref{sec:conclusion}. Appendix~\ref{sec:appendix_data} provides a summary of the galaxy parameters and tables that list predicted photometry for the Milky Way. Appendix~\ref{sec:appendix_test_train} describes tests of the accuracy of the GPR procedures used here, and Appendix~\ref{sec:appendix_systematics} describes how we address the systematic corrections needed for $k$-corrections and Eddington bias.

In this paper all magnitudes are reported in the AB system, except for the Johnson-Cousin $UBVRI$ magnitudes which are presented in the Vega system. Absolute magnitudes are derived using a Hubble constant $H_{0} = 100 \;\kms \rm{Mpc}^{-1}$, so they are equivalent to $\rm{M}_{y} - 5\log{h}$ (where $\rm{M}_{y}$ is the $y$-band absolute magnitude and $h = H_{0}/100$) for other values of $h$. 
For other properties in which measurements for the Milky Way are compared to extra-galactic galaxy measurements we assume $H_{0} = 70 \;\kms \rm{Mpc}^{-1}$ ($h=0.7$) in accordance with \citet{licquia2015b}, for a standard flat $\Lambda$CDM cosmology with $\Omega_{m} = 0.3$. Parameters such as log stellar mass and log star formation rate can be modified for different $h$ values by subtracting $2\log{h/0.7}$. We do this to avoid confusion and to allow for potential updates to future $h$ measurements.


\section{Observational Data}
\label{sec:data}

In this section we describe the many galaxy catalogues utilised in this work. We break this up by photometry (\autoref{subsection:photometry}) and inferred galaxy properties (\autoref{subsection:strucprops}), with the Milky Way measurements included in the final subsection (\autoref{subsection:mwparams}).

\subsection{Photometry}
\label{subsection:photometry}

\subsubsection{SDSS Galaxies}
\label{subsubsection:sdss}
The sample of galaxies that we use as a starting point originates from the eighth data release (DR8; \citet{dr82011}) of the Sloan Digital Sky Survey III (SDSS-III; \citet{york2000}). DR8 provides both images and photometry of thousands (almost $10^{6}$) of local galaxies. The optical broadband passbands, $u$, $g$, $r$, $i$, and $z$ were the subjects of previous Milky Way analogue work by \citetalias{licquia2015b} and are used in this study in addition to bands outside of the optical range. 

We make use of both the \texttt{``model''} and \texttt{``cmodel''} magnitudes from SDSS. The former refers to magnitudes derived from the better of either a de Vaucouleurs or an exponential profile fit to the galaxy surface brightness distribution. These types of magnitudes are expected to produce the highest signal-to-noise estimate of galaxy colours; thus when we refer to galaxy colours derived from SDSS we will be using \texttt{model} magnitudes for the calculations. Alternatively, \texttt{cmodel} magnitudes are derived from the best fit to a linear combination of a de Vaucouleurs and exponential profile. These magnitudes provide the best estimate of the total flux of a galaxy in each passband. When we refer to galaxy absolute magnitudes for SDSS bands we will use \texttt{``cmodel''} magnitudes.

$k$-corrections on these passbands to rest-frame $z=0$ were calculated via the \textsc{kcorrect v4.2} software \citep{blanton2007}, as described in \citetalias{licquia2015b}. This provided AB absolute magnitudes for the SDSS $ugriz$ photometry. Additionally, \textsc{kcorrect} was used to convert the SDSS $ugriz$ photometry to restframe Johnsons-Cousins \textit{UVBRI} Vega magnitudes in order to make easy comparisons to literature values. Results are presented with the adoption of the \citet{blanton2007} and \citetalias{licquia2015b} notation, where an absolute magnitude of passband $y$ at redshift $z$ is denoted as $^{z}M_{y}$.

Our main galaxy sample is derived from the volume-limited sample presented in \citetalias{licquia2015b}. A volume-limited sample is required for accurate results from Milky Way analogues in order to alleviate a radial selection effect known as Malmquist bias, i.e., the preferential inclusion of intrinsically bright galaxies. At higher redshifts within the main SDSS sample \citep{strauss2002}, only the most luminous galaxies will be brighter than the sample magnitude limit and followed-up spectroscopically. By using a volume-limited sample we ensure that galaxies within the range of the Milky Way's parameters are included equally at all distances considered. \citetalias{licquia2015b} determined the limits for their volume-limited sample from an initial draw of Milky Way analogues from the full SDSS DR8 parent catalogue without any redshift cuts. Then in $^{0}(g-r)$ vs. $^{0}M_{r}$ (i.e., restframe $g$-$r$ colour derived using $z=0$ passbands versus $r$-band absolute magnitude, again evaluated with the $z=0$ passband) colour-magnitude space a maximum redshift was chosen such that all objects as low in luminosity as the faintest Milky Way analogues would still be included at that $z$. A minimum redshift was also applied to limit the impact of the finite SDSS fiber aperture on measured galaxy properties. The resulting volume-limited sample contains a total of 124,232 target galaxies within the redshift range of $0.03 < z < 0.09$. Some initial cuts on SDSS quality flags were also employed; for further details on the construction of this volume-limited sample refer to \citetalias[][Section 3.1]{licquia2015b}. All cross-matches from SDSS to other catalogues presented here were constructed only using the volume-limited sample. Both the SDSS sample used here and the cross-matched catalogues presented in this paper are available at \href{https://github.com/cfielder/Catalogs}{our catalogue GitHub repository}\footnote{https://github.com/cfielder/Catalogs}.

\subsubsection{GALEX--SDSS--WISE Legacy Catalogue}
\label{subsubsection:gswlc}

Photometry in ultraviolet and infrared wavelengths used in this work comes from the GSWLC \citep{salim2016,salim2018}. We use GSWLC-M2, the medium-deep catalogue of GSWLC-2, which covers $49\%$ of the SDSS DR10 footprint. While this reduces the number of targets for study, the improved signal-to-noise in the UV-imaging over the shallow catalogue enables tighter results. SDSS Photometry between DR7/DR8 and DR10 is the same, so no issue arises between cross matches of the SDSS volume-limited sample and the GSWLC-M2 sample. In order to account for any differences in astrometry we consider matches to be separations within 1.5 arcseconds. 

The ultraviolet sample in the GSWLC catalogue originates from the GALEX survey ($FUV, NUV$), and has been corrected for galactic reddening and calibration errors \citep{galex2005}. UV detections are available for $74\%$ of SDSS targets in the GSWLC-M2 catalogue. The infrared sample in the GSWLC catalogue originated from the 2MASS and WISE surveys. The 2MASS photometry ($JHKs$) is available for $48\%$ of SDSS targets and WISE photometry ($W1, W2, W3, W4$) is available for $41\%$ of SDSS targets.

\subsubsection{DESI Legacy Imaging Surveys}
\label{subsubsection:legacy}

Data release 8 of the DESI (Dark Energy Spectroscopic Instrument) Legacy imaging surveys includes $g$, $r$, $z$, and WISE photometry of  sources detected in DECam or BASS/MzLS imaging \citep{legacy2019}. We use these catalogues for all WISE-band photometry presented here, as the matched-model measurements from the Legacy Surveys {\tt Tractor} catalogues go substantially deeper and have lower errors than other public WISE data products. We also calculate $k$-corrections for the WISE bands using this photometry; we discuss our procedures for this, which allow accurate $k$ corrections in the IR without making any assumptions about SED templates at those wavelengths, briefly in \autoref{subsection:kcorrections}
and more extensively in a forthcoming paper (Fielder, Newman, \& Andrews, in prep.).

DESI Legacy and SDSS contain photometry that comes from differing filter sets and detectors. DESI Legacy North uses BASS and MOSAIC filters while DESI Legacy South used DECam filters \citep{legacy2019}. Our corrections for offsets between the DESI Legacy Survey and SDSS photometry is described in \autoref{subsection:legacy_offset}.


\subsection{SDSS-based Properties}
\label{subsection:strucprops}

Only one of the galaxy properties which we use to create SEDs is recorded in the standard SDSS photometric catalogues and required no further analysis; specifically, the best fit axis ratio ($b/a$) of a galaxy. We use the $b/a$ value determined from an exponential fit to the galaxy's surface brightness density in the $r$-band throughout this work. 

Axis ratio is used here as a proxy for galaxy inclination. Disk galaxies with very small axis ratios are likely heavily tilted from our perspective, in contrast to face-on disks which would typically have axis ratios of $b/a \sim 0.9$ \citep{cho2009,maller2009}. Highly-inclined star forming galaxies suffer greatly from dust obscuration which can effect their colours and magnitudes by making them appear redder and fainter than their intrinsic properties \citep[e.g.,][]{conroy2010,masters2010a,morselli2016,kourkchi2019}. This effect is important to consider when we are making predictions for a disk galaxy like the Milky Way.

\subsubsection{MPA-JHU Masses and Star Formation Rates}
\label{subsubsection:mpajhu}

The MPA-JHU galaxy property catalogue provides total mass and star formation rate estimates for galaxies within SDSS DR7. \citetalias{licquia2015b} developed an updated version of this catalogue using the SDSS DR8 photometry. To summarise, the masses are calculated by fitting stellar population synthesis model to the galaxy's photometry (instead of spectral features, as in \citet{kauffmann2003,gallazzi2005}), similar to \citet{salim2007}. The star formation rates are calculated by emission-line modelling based upon \citet{brinchmann2004} with some updates. In order to account for aperture bias \citetalias{licquia2015b} follows the method of \citet{salim2007} in calculating photometry for the light that falls outside of the fibre and fitting stochastic stellar population synthesis models to it. Thus each galaxy SFR measurement consists of a combination of the SFR measured from inside and outside of the SDSS fibre. For more details on these calculations refer to \citetalias[][Section 2.2.2]{licquia2015b} and \citet{brinchmann2004}. In both cases a Kroupa initial mass function (broken power law) is assumed \citep{kroupa2003}.

Each galaxy in our volume-limited sample is assigned a posterior probability distribution function (PDF) of log stellar mass and log star formation rate, as well as a corresponding cumulative distribution function (CDF) determined by the posterior. Nominal values used for the $\log{\dot{\rm{M}_{*}}}$ and $\log{\rm{SFR}}$ of each object are taken to be the mean value of the posterior. The standard deviation ($\sigma$) on these values are calculated from percentiles ($P$) in the CDF provided in the MPA-JHU measurements; we take $\sigma = (P_{84}-P_{16})/2$, i.e., half the difference between the $16^{\rm{th}}$ and $84^{\rm{th}}$ percentile values. This gives us estimates of the  stellar mass and star formation rates as well as their associated errors for each galaxy in the sample.

The GSWLC-2 catalogues which we use for photometry also provide computations of stellar masses and star formation rates. We chose to use the MPA-JHU results instead in order to avoid any systematic effects that may result from using the same measurements for our galaxy properties that were used to determine these derived quantities. We have tested how our results change if we use GSWLC-2 masses and star formation rates and find minimal differences, as summarised in \autoref{subsection:measurement_impact}.

\subsubsection{Simard et al. Bulge and Disk Decompositions}
\label{subsubsection:simard}

In order to characterise the light-weighted bulge-to-total ratios ($B/T$) and disk scale lengths (R$_{d}$) for SDSS galaxies, we use the catalogue of \citet{simard2011}. This work performed galaxy image decompositions for objects within the Legacy area in SDSS via the \texttt{GIM2D} software package \citep{simard2002}. 

Specifically, we use the catalogue of fits based on composites of a Sersic $n=4$ bulge and a pure exponential disk  (Sersic $n=1$). 
For our analysis we use the $B/T$ computed in $r$ band, as it is expected to be more stable than $g$-based measurements. As a check we have performed our analyses using $B/T_{g}$ instead of $B/T_{r}$ and found minimal differences (cf.  \autoref{subsection:measurement_impact}). It is worth noting that \citet{kruk2018} finds that these bulge+disk decompositions can be somewhat inaccurate when applied to strongly-barred galaxies, which may lead to some biases within our galaxy sample.  However, fits optimised for barred galaxies have been performed for only samples of a few thousand objects, inadequate for our purposes, so we rely on the \cite{simard2011} results here. The \citet{simard2002} bulge and disk decompositions are derived using a $H_{0} = 70$ km$\rm{s}^{-1}\rm{Mpc}^{-1}$.

\subsubsection{Galaxy Zoo 2 Bar Presence}
\label{subsubsection:gz}

The presence of bars have been speculated to be correlated to the star formation history within a galaxy. However, the sense of the effect is unknown; a bar may be related to an increase in star formation \citep[e.g.,][]{hawarden1986,hawarden1996,devereux1987,hummel1990}, no effect \citep{pompea1990,martinet1997,chapelon1999}, or decreased star formation \citep[e.g.,][]{vera2016,diaz-garcia2020,fraser-mckelvie2020}, along with other potential effects on colour and metallicity \citep[see e.g.,][]{masters2011}. Regardless of the source of any correlations, we wish to incorporate any possible differences having a bar may cause when we determine the Milky Way's SED, since our Galaxy does exhibit clear evidence of a bar \citep[e.g.,][]{blitz1991}. The Galaxy Zoo 2 (GZ2) catalogue \citep{willett2013,hart2016} contains identifications of detailed morphological features in disk galaxies, such as spiral arms and bar presence. The galaxies classified in Galaxy Zoo 2 are a sub-sample of the brightest and largest galaxies in the SDSS Main Galaxy Sample. 

The Galaxy Zoo projects are open public projects in which members of the community identify whether they found a variety of galaxy features in the images provided. In the Galaxy Zoo 2 catalogue the number of raw votes for each morphological feature is weighted (to account for user consistency) and adjusted to mitigate the impact of redshift-dependent biases, yielding a corrected fraction of volunteers who identified a given morphological feature for each galaxy. In our case we focus on the debiased fraction of volunteers who identified a galaxy as having a bar, which we denote by $p_{bar}$ (following Galaxy Zoo labelling conventions; however, this is a fraction of votes, not a probability). Above some threshold in $p_{bar}$ (after cuts in related parameters) we can be confident that a given galaxy indeed hosts a bar. \citet{willett2013} developed the initial version of Galaxy Zoo 2 bar thresholds; more conservative thresholds were later defined by \citet{galloway2015}. 

Often, when one uses Galaxy Zoo results it is necessary to consider responses to previous questions that influence whether the question of interest is even presented to the volunteers. For example, as described in \citet{willett2013}, a voter is only asked ``Is there a sign of a bar feature through the centre of the galaxy?'' if they first selected that the object has  ``features or disk'' when asked ``Is the galaxy simply smooth and rounded, with no sign of a disk?'', and then responded in the negative when asked ``Could this be a disk viewed edge-on?''. One would then identify galaxies that have received a large fraction of ``yes'' votes as containing bars. It is worth noting that \citet{willett2013} states galaxies that receive fewer than $10$ net votes for a given question may not have a reliable classification. Therefore to construct a bar sample using the minimum allowances of \citet{willett2013}, using the debiased vote fractions from \citet{hart2016} (which was an improvement to the debiasing of \citet{willett2013}), one would use the cuts in the second and third row of the third column of Table 3 in \citet{willett2013} with $p_{\rm{bar}} \geq 0.3$ as recommended by \citet{galloway2015} and additional voting number thresholds $N_{\rm{not\;edgeon}}\geq10$, and $N_{\rm{bar}}\geq10$.


However, here we do not want to consider only information on the relationship between galaxy star formation history, structural properties, and photometric properties that comes from those galaxies that most definitively host bars. Rather, it is desirable for the training sample for our Gaussian process regression model to include objects spanning a broad range of parameter space: this yielded smaller net prediction errors in Milky Way photometric parameters than when small, restricted samples are used for training. For additionally discussion on this choice refer to \autoref{measurement_impact_gz2}.

\bigskip

To summarise, the primary set of data that we employ in this paper consists of a cross-match between the original SDSS DR8 volume-limited sample reported in \citetalias{licquia2015b}, an updated version of the MPA-JHU catalogue \citep{brinchmann2004} of masses and star formation rates in SDSS, the \citet{simard2011} morphological bulge-disk decomposition catalogue, the GSWLC-2 medium-deep photometry catalogue \citep{salim2016,salim2018}, the Galaxy Zoo 2 catalogue \citep{willett2013,hart2016}, and the DESI Legacy survey DR8 catalogue \citep{legacy2019}. These cross-matches were performed using the $\tt{astropy}$ \texttt{coordinates} package. We required that objects be separated by less than 1.5 arcseconds from a counterpart in the SDSS volume-limited sample to be considered a match, and discard any galaxies that are not included in all catalogues considered. 

Our sample is smaller than the original volume-limited sample as a result. The MPA-JHU catalogue contains all objects from the volume-limited sample, so we begin with the same set of 124,232 galaxies utilised by \citetalias[][]{licquia2015b}. After matching to \citet{simard2011}, 123,167 galaxies remain; some objects are lost due to minor differences between the DR7 and DR8 SDSS catalogues. When we require GSWLC-M2 measurements, we are left with only 60,857 galaxies, as roughly half of the SDSS footprint is covered by medium-deep GALEX, 2MASS, and WISE photometry (see \autoref{subsubsection:gswlc}). After matching to GZ2 29,836 galaxies remain, a consequence of the brighter magnitude limit used to select GZ2 objects compared to the original SDSS Main Galaxy Sample (see \autoref{subsubsection:gz}). We note that due to the brighter magnitude limit of GZ2, our final sample is no longer volume-limited.  This would yield biased results if we were measuring aggregate properties of Milky Way analogues.  However, for our GPR methodology this only modulates the density of our training set within parameter space, causing larger prediction errors due to the sparser sampling, but not leading to a bias.

Finally, matching to DESI Legacy leads to only a minor reduction in galaxy number as it covers a super-set of the SDSS area with deeper photometry (but different reduction algorithms). After these cross matches we then remove objects with photometric values of ``NaN'', infinity, or $-99$, which all indicate missing photometry in one catalogue or another. This is only done when necessary for the evaluation of the GPR. For our WISE $k$-correction calculations (see Appendix~\ref{subsection:kcorrections}) we will exclude objects with large WISE photometry errors in a given band, which we do not propagate into our main galaxy sample.

The final data sample consists of 29,588 galaxies in total from redshift $0.03 < z < 0.09$, which is publicly available at \href{https://github.com/cfielder/Catalogs}{our catalogue GitHub}. The parameters which we will use to predict photometric properties are stellar mass (\mass), star formation rate (SFR),  galaxy axis ratio as a proxy for inclination ($b/a$), bulge-to-total ratio ($B/T$), disk scale length ($R_{d}$), and   corrected bar vote fraction ($p_{\rm bar}$). Covariances amongst these parameters are minimal, as discussed further in Appendix~\ref{subsection:corner}; we show joint distributions for these  parameters in \autoref{fig:corner}.

\subsection{Milky Way Properties}
\label{subsection:mwparams}

A number of the properties of the Milky Way used in this study have been derived using the Bayesian mixture model meta-analysis method first presented in \citet{licquia2015a}. This technique combines information from multiple measurements in order to obtain aggregate constraints on the Milky Way's properties, taking into account the possibility that individual measurements could be incorrect or have their errors miss-estimated. This Bayesian method is combined with Monte Carlo simulations in order to account for uncertainties on the Sun's measured Galactocentric radius, $R_{0}$; the Galactic exponential disk scale length; and uncertainties in the local surface density of stellar mass. The inferred mass of the Milky Way's stellar disk depends upon all three of these parameters. 

\citet{gravitycollab2019} recently obtained a greatly improved geometric measurement of our distance from the centre of the Milky Way, $R_{0} = 8.178 \pm 0.026$ kpc (\citet{licquia2015a} used the value $8.3\pm0.35$ kpc from \citet{gillessen2009}). We have rerun the Bayesian mixture model inference from \citet{licquia2015b} and \citet{licquia2016a} with this updated measurement of the Galactocentric distance to obtain updated measurements of the Milky Way's mass, bulge-to-total mass ratio, and disk scale length; see those papers for all details of the data sets used and the calculations (the star formation rate estimate from \citet{licquia2015a} is not affected by the value of $R_{0}$, so we adopt it unchanged). The resulting Milky Way parameters used in this study are as follows:
\begin{itemize}
    \item Bulge Mass $\mass^{B} = 0.90\pm 0.06 \times 10^{10}\; M_{\sun}$
    \item Disk Mass $\mass^{D} = 4.58^{+1.18}_{-0.94}  \times 10^{10}\; M_{\sun}$
    \item Total Stellar Mass $\mass = 5.48^{+1.18}_{-0.94} \times 10^{10}\; M_{\sun}$
    \item Star Formation Rate SFR$= 1.65 \pm 0.19\; \rm{M}_{\sun}\rm{yr}^{-1}$ \citep{licquia2015a} 
    \item Log Specific Star Formation Rate $\log{\frac{\rm{SFR}}{\mass}} = -10.52 \pm 0.10$
    \item Bulge-to-total Mass Ratio $B/T$ $= 0.16 \pm 0.03$
    \item Disk Scale Length $R_{d} =  2.48^{+0.14}_{-0.15}$ kpc
\end{itemize}

The revised stellar mass estimate for the Milky Way is smaller than the previous estimate, but by an amount that is significantly smaller than the previously-estimated uncertainties.

By design these physical parameters are constructed such that they can be directly compared to the extragalactic catalogues used to predict photometric properties. For example, the stellar mass and star-formation rate estimates for the Milky Way are constructed assuming a Kroupa initial mass function \citep{kroupa2003} and an exponential disk model, which is the same way in which mass and star formation rates were calculated in the MPA-JHU catalogue \citep{brinchmann2004} that we use in this analysis. The only parameter where this is not fully the case is the bulge-to-total ratio, $B/T$. For the Milky Way we can securely estimate only a mass-weighted value for this quantity. In external galaxies, however, the mass-weighted $B/T$ is much more difficult to obtain, and light-weighted measurements tend to be more reliable.  However, our predicted Milky Way properties do not change significantly when we switch from $r$-band-based to $g$-band-based $B/T$ values, even though mass-to-light ratios differ significantly between these bands, suggesting that this is not a major issue.

Our analysis also depends on two additional parameters that are determined independently of any Bayesian mixture model meta-analysis. These quantities have values assigned to them based on their meaning in their respective catalogues and our understanding of the Milky Way:
\begin{itemize}
    \item Axis ratio (inclination proxy) $b/a = 0.9\pm 0.\textbf{}1$
    \item Bar vote fraction $p_{\rm{bar}} = 0.45\pm 0.15$
\end{itemize}

Galaxy inclination has a strong effect on colour and luminosity measurements for disk galaxies. As mentioned in \autoref{subsection:strucprops}, dust alters the observed colours and magnitudes of star-forming galaxies that are highly inclined or edge-on. Our perspective within the Milky Way makes it somewhat equivalent to being edge-on to us (though our position within the Galaxy, rather than outside, does cause some differences). However, photometric properties are most cleanly determined for those objects which are observed face-on. Therefore, we predict the SED that would be observed for the Milky Way for axis ratio values drawn from a uniform distribution spanning from $b/a = 0.8$ to $1.0$, consistent with the intrinsic axis ratios of spiral galaxy disks as described by \citet{maller2009} and in \autoref{subsection:strucprops}. Our results should therefore correspond to the properties of our Galaxy if it were observed face-on. While axis ratio is a good proxy for inclination, it is not a perfect substitute. For example, \citet{vandesande2018} found that disk galaxies that are rounder ($b/a \sim 1$) tend to be older and therefore intrinsically redder. This means that small biases could result from treating the Milky Way as having a face-on axis ratio. However, there is no straightforward way to avoid this, and the effect should be small compared to other sources of error.

The Milky Way exhibits clear evidence that it contains a bar \citep[e.g.,][]{blitz1991,shen2020}. However, very few Galaxy Zoo 2 galaxies have $p_{\rm{bar}} = 1.0$, and those galaxies with the highest bar vote fractions are expected to have \textit{very strong} bars, which may not match our Galaxy. Therefore we assume that in Galaxy Zoo 2 the Milky Way would have a vote fraction above the threshold for defining a bar, but not a value higher than the bulk of barred galaxies. \citet{galloway2015} and \citet{willett2013} find that $p_{\rm{bar}}\geq0.3$ serves as a reliable threshold between bar presence and lack thereof. Galaxies with $0.3 \leq p_{\rm{bar}} < 0.5$ likely have weaker bars while galaxies with $p_{\rm{bar}}>0.5$ likely have stronger bars. Because the bar strength of the Milky Way's bar as it would be determined from outside our Galaxy is not well-constrained, we treat the bar vote fraction for the Milky Way as uniformly distributed between $p_{\rm{bar}} = 0.3$ and $p_{\rm{bar}} = 0.6$. Choosing a larger mean vote fraction has small effect on our results, given the large range of fractions considered (compared to the distribution of $p_{\rm{bar}}$ in GZ2).

\section{Gaussian Process Regression for Predicting Milky Way Photometry}
\label{sec:gpr}

In this subsection we describe how  Gaussian process regression can be used to estimate photometric properties for the Milky Way. First we explain the need to transition from Milky Way analogue-based methods to GPR when we consider higher-dimensionality parameter spaces in \autoref{subsection:analog_limits}. In \autoref{subsection:gpr_defined} we explain the basic concepts behind GPR, and in \autoref{subsection:global_trends} we highlight the fundamental differences between GPR and analogue galaxy methods. \autoref{subsection:kernel} describes the kernel used to set up our GPR, which guides how information is propagated from training objects to predictions. We briefly describe the computational limitations of the GPR implementation we are using in \autoref{subsection:comp_challenges}. Lastly, \autoref{subsection:uncertainty} investigates the contributions of various sources of uncertainty to our GPR predictions.

\subsection{Limitations of Using Analogue Galaxies}
\label{subsection:analog_limits}

Using Milky Way analogues to predict the photometric properties of the Milky Way, as was done in \cite{licquia2015b}, has been a very useful methodology but also has limitations. Of particular concern is the dramatic reduction in MWA sample size that occurs as the number of parameters that must be matched increases; we would like to move from the two parameters considered by \cite{licquia2015b} (\mass and SFR), to a total of six, adding $b/a$, $B/T$, R$_{d}$, and $p_{\rm{bar}}$. Requiring that analogues be Milky Way-like in more ways should reduce the spread in photometric properties of the resulting sample, potentially enabling stronger constraints. There are only limited correlations between these six parameters (cf. \autoref{fig:corner} in Appendix~\ref{subsection:corner}), so degeneracies between them are minimal: they each add new information. For instance, galaxies with stellar mass and star formation rate matching the Milky Way exhibit bulge-to-total ratios ranging from zero to one: structural and star-formation history parameters carry distinct information. However, while matching on additional galaxy parameters produces a population of analogues that must each be closer in properties to the Milky Way, the resulting MWA sample becomes much smaller (with as few as $\sim 5$ analogues in the sample that are within $3\sigma$ of the Milky Way in all of the properties considered, and none within $2\sigma$ for every aspect). 

\begin{figure}
    \centering
    \includegraphics[width=0.9\linewidth]{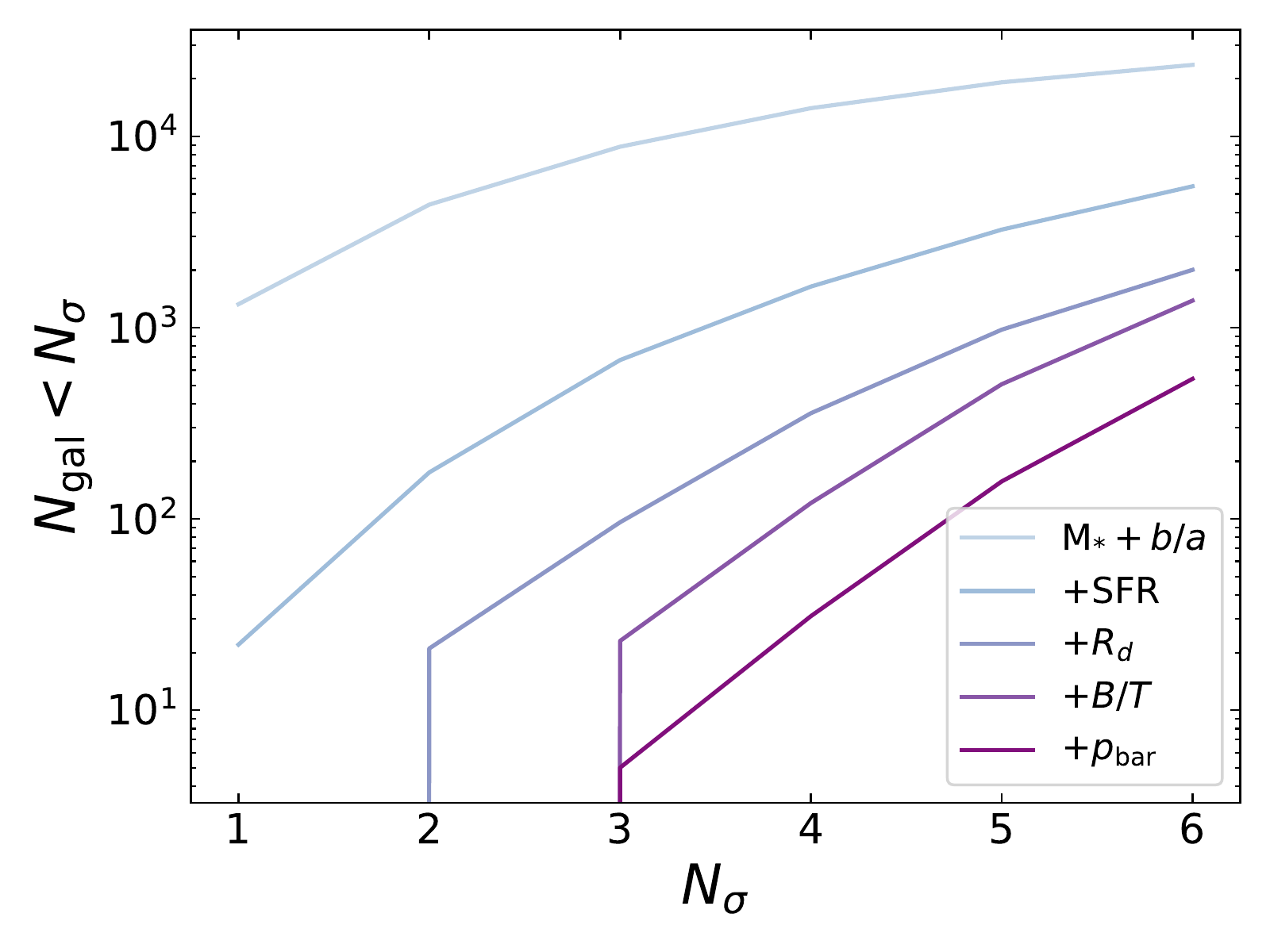}
    \caption{Total number of galaxies within an allowed tolerance ($N_{\rm{gal}}<N_{\sigma}$) as a function of the maximum deviation away from the Galactic values allowed for each parameter, in units of the uncertainty in the Milky Way value of that parameter, $N_{\sigma}$. The lightest shade of blue denotes galaxy counts when analogues are selected using only stellar mass ($\rm{M}_{*}$) and axis ratio ($b/a$). Consecutive parameter additions are cumulative. Thus the darkest purple line incorporates all six parameters of stellar mass, axis ratio, star formation rate (SFR), disk scale length ($R_{d}$), bulge-to-total ratio ($B/T$), and bar vote fraction ($p_{\rm{bar}}$). The $\sigma$ values used for each parameters are defined in \autoref{subsection:mwparams}, except in the case of bar vote fraction, for  which we use a smaller error ($\sigma_{p_{\rm{bar}}} = 0.05$ instead of $0.15$) for illustrative purposes. Note that there are very few galaxies even within large tolerance windows when using several parameters, e.g., only $\sim 10$ MWAs within $3\sigma$ for 5 parameters simultaneously, and no objects that are close to the Milky Way in all aspects, since the MWA method scales poorly with increased dimensionality.}
    \label{fig:mwag_sigma}
\end{figure}

The reduction in the size of analogue galaxy samples as more parameters are considered is illustrated in \autoref{fig:mwag_sigma}. We plot the total number of galaxies that are within a given number of $\sigma$ for every Milky Way parameter considered (where $\sigma$ represents the uncertainty in the Milky Way value for a given property) as a function of the number of $\sigma$ used as a threshold. The lightest shade of blue denotes the number of analogues within a given threshold when only considering stellar mass ($\rm{M}_{*}$) and axis ratio ($b/a$). The consecutive additions indicated in the legend represent the inclusion of the listed parameter in addition to all previous ones; i.e., we consecutively incorporate star formation rate, disk scale length, bulge-to-total ratio, and finally bar vote fraction. Hence, the darkest purple line shows the number of analogues when using all six parameters (which we have used in order of decreasing constraining power on Milky Way colours). The $\sigma$ tolerances used for each parameter are the same Milky Way measurement errors defined in \autoref{subsection:mwparams}, \textit{except} for $p_{\rm{bar}}$. The significant uncertainty we fiducially ascribe to bar vote fraction would cause the 5 parameter and 6 parameter lines to be degenerate with one another; to avoid confusion we use $\sigma_{p_{\rm{bar}}} = 0.05$ instead of $0.15$ when constructing this plot. 

The previous work by \citetalias{licquia2015b} would most closely correspond to the +SFR (3 parameters), medium blue line. If one were to restrict to objects within $\pm 2\sigma$ of the Milky Way value for all parameters employed, there would be a total of zero Milky Way analogue galaxies when using five or more parameters. In the six-dimensional space that we employ below, there are only $~200$ galaxies that are within even $\pm6\sigma$ of the Milky Way value in all six properties; at that extreme, analogue samples would be selecting objects that are not very close to the Milky Way at all. The lack of close analogues in high-dimensional parameter spaces makes constraints on Milky Way properties from the MWA method weak in that limit, with correspondingly large uncertainties.

\subsection{Gaussian Process Regression: A Powerful Method for Interpolation and Prediction}
\label{subsection:gpr_defined}
To address the lack of Milky Way analogue galaxies in our multi-dimensional parameter space, we have developed alternative methods for predicting Milky Way properties based upon Gaussian process regression. In this sub-section we summarise the basic properties of GPR relevant for this work. For in-depth discussion, we refer the reader to \citet{rasmussen2006} and \citet{gortler2019a}.


Gaussian process regression (sometimes called kriging) is effectively a method of interpolation where information from training data is accounted for by a smooth and continuous weighting function, called a ``kernel'' or covariance function. The joint probability distribution of the values of a Gaussian process at any finite set of points in parameter space will be a multi-variate Gaussian (with a number of dimensions set by the number of points in the set); the kernel specifies how the covariance between points depends upon their separation. The kernel should be a smooth and continuous function, with a length scale (which governs how far information propagates from a given point) that is optimised by training on the observed data. We can then \textit{predict} what the value and the uncertainty of the desired quantity would be at any arbitrary point in space by applying this kernel to the training data. This is in contrast to other supervised learning algorithms which typically make single-valued, ``point'' predictions rather than predicting PDFs. It can be shown that GPR yields the minimum variance out of any unbiased interpolation method that depends only linearly on the training data; this makes GPR an \textbf{optimal interpolation algorithm}.

For our application of GPR, the galaxy sample described in \autoref{sec:data} will serve as the training set. The six parameters we defined in \autoref{subsection:strucprops} (stellar mass, star formation rate, etc.) serve as the features we will use for prediction. 
Our goal is to determine an optimised mapping from these physical parameters to a single output photometric parameter in our catalogue, e.g., the $r$-band absolute magnitude, $^{0}M_{r}$. 

Once our training data is selected we then go into the model-selection phase of GPR, during which the mean function and covariance function (or kernel) used for GPR are selected and tuned. 
We detail our selection of the covariance function in \autoref{subsection:kernel}. Effectively, the kernel determines how information from a given training point will be propagated to make predictions at other points in parameter space. Hyper-parameters describing the kernel are tuned at this step to maximise the log-marginal likelihood of the training data.  After this step we consider the model to be ``fit''.

Finally we enter the inference phase of GPR. At any point in our six-dimensional parameter space, we can now determine the posterior probability distribution for the parameter of interest by applying the kernel to the training data.
When evaluated at a single point in parameter space, a Gaussian process corresponds to a 1-D Gaussian; we thus obtain both a predicted mean for the property of interest (e.g., $^{0}M_{r}$) and the standard deviation of the Gaussian describing its uncertainty. In our example we would pass in a set of physical parameters measured for the Milky Way and obtain a predicted value for the $^{0}M_{r}$ of the Milky Way, as well as the uncertainty in that value.

In reality we do not only query the GPR at the mean measured Milky Way properties presented in \autoref{subsection:mwparams}; rather, we perform random draws from the PDFs describing \mass, SFR, $b/a$, $B/T$, $R_{d}$, and $p_{\rm bar}$ in order to incorporate the uncertainties in the Milky Way's measured properties into our analysis. For $\log{\mass}$, $\log{\rm{SFR}}$, $B/T$, and $R_{d}$ we assume a normal distribution. For $b/a$ and $p_{\rm{bar}}$ we draw from uniform distributions, as described in \autoref{subsection:mwparams}. 

We perform these random draws 1000 times (so that we have 1000 full sets of Milky Way parameters). We then evaluate the GPR predictions for Milky Way photometric properties at each of these points in parameter space. This gives us a prediction and error estimate corresponding to each draw from the PDFs of Milky Way characteristics. Thus we end up with 1,000 total predictions for each Milky Way photometric property. Our mean prediction for the Milky Way in a given photometric band corresponds to the arithmetic mean of all these predictions.

In this work all Gaussian Process regression calculations have been done with the \textsc{Python} \texttt{scikit-learn} Gaussian Process module, $\tt{sklearn.gaussian\_process}$ \citep{scikit-learn}.  For details on the implementation of this module, refer to \citet{scikit-learn} and the \href{https://scikit-learn.org/stable/modules/gaussian_process.html#gaussian-process-regression-gpr}{\texttt{scikit-learn} documentation}. 

The following sub-sections provide more details on some aspects of our GPR methods and their advantages.

\subsubsection{Choice of Kernel for GPR}
\label{subsection:kernel}

In this work we use a combination of two kernels for Gaussian Process Regression: a Radial Basis Function kernel and a white noise kernel. The Radial Basis Function (RBF) kernel decreases proportionally to $\exp{-\gamma D^2}$, where $\gamma$ is a free parameter and $D$ is the Euclidean distance between points; this kernel will cause the covariance between the predicted values from GPR at different points in parameter space to decrease as a Gaussian in distance as the separation between those points increases. 

However, there is also scatter in galaxy photometric properties even for objects measured to have the same physical properties. In order to capture that, we also incorporate a white noise kernel, which models the spread in values for the predicted property at a fixed point in parameter space with normally distributed noise \citep{rasmussen2006}.  The net covariance used for the Gaussian process regression is then the sum of the distance-dependent covariance from the RBF kernel and the (diagonal) covariance matrix corresponding to the white noise kernel.

For a given set of training data, there is a nearly endless number of functions that can fit the given data points, each one a realisation of the Gaussian process. The kernel creates a prior on the GP to constrain which functions from that set are most likely to describe the parameter space. The posterior is then determined using the training data values. Due to this probabilistic approach, the Gaussian process provides both predicted values and uncertainties at any points within the parameter space. Uncertainties due to the finite training sample size and its distribution in parameter space and those corresponding to intrinsic variation between training objects that have the same physical parameters are both captured.  In regions of parameter space that are poorly constrained by the training data, the prediction uncertainties are correspondingly larger. 

The kernels we use in this work are available in the $\tt{sklearn.gaussian\_process.kernels}$ base class. For the white noise kernel (\texttt{WhiteKernel} in \texttt{scikit-learn}) we initialise the noise level to be $1$; similarly, we initialise the length scale for the RBF kernel to be $1$ for each parameter. We opt to normalize the output photometric property to have mean zero and variance one across the training set, which helps to ensure that these initial guesses will have the right order of magnitude.  The noise level and length scales are then optimised and the regression model is built via the $\tt{sklearn.gaussian\_process.GaussianProcessRegressor}$ class. In our model we allow the optimiser to restart 10 times in order to find the kernel parameters that maximise the likelihood without being trapped in a false maximum. 

\subsubsection{Optimizing Training Samples}
\label{subsection:comp_challenges}

The computation time and required memory for the $\tt{scikit-learn}$ implementation of GPR scales as the number of data points used to train the model squared and cubed, respectively. As a result, we find that the maximum training sample size we can use without running out of memory on the computers used for this work is $\sim 6,000$; it is infeasible to train from our entire catalogue when using this GPR implementation. 

We have therefore tested the effects of either restricting to objects with physical parameters within some tolerance of the MW fiducial values or  randomly selecting a subset of objects in order to reduce the training set size. We have focused on the root mean squared error (RMSE) of predicted Milky Way photometry for the $NUV$, $r$, and $J$ bands for this optimisation. We use five-fold cross-validation for all the tests; i.e., we always train with 80\% of the data and test with 20\%, but rotate what objects are used for training and testing through the whole dataset, and only retain the values for an object when it was in the test set. This provides unbiased estimates of the RMSE for a training set 80\% as large as the one we actually have. We find that the combination that offered the lowest RMSE across all bands while keeping computational time manageable was to randomly select 2,000 galaxies out of the set of objects that are within $12\sigma$ of the Milky Way for every parameters of interest. We therefore adopted this training strategy for all results below.

\subsubsection{Comparison to Results from Analogue Samples}
\label{subsection:global_trends}

GPR can provide more accurate predictions than many other techniques thanks to its ability to leverage information from both nearby objects in the training set as well as from more distant objects that characterise larger-scale trends. In our application, this allows the GPR to map from the Milky Way's physical properties to its photometric properties much more accurately than if we had only used the few objects that are similar to the Milky Way in all respects (i.e., those which would be classified as MWAs) to inform the mapping.

\begin{figure}
    \centering
    \includegraphics[width=0.9\linewidth]{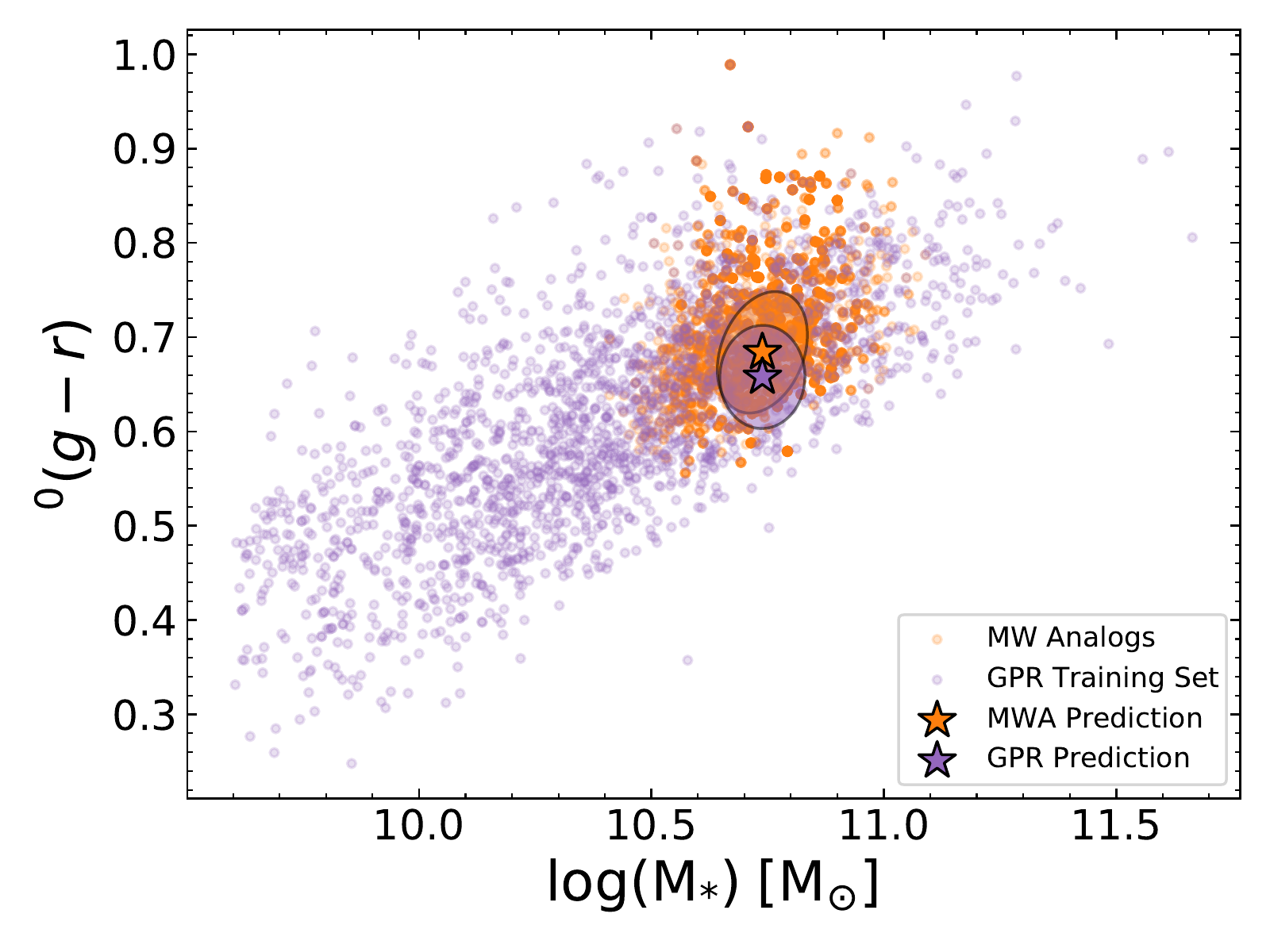}
    \caption{A comparison of galaxy samples and results from the Milky Way analogues method versus Gaussian Process regression. For this analysis analogues were selected based upon stellar mass, star formation rate, and axis ratio, and we use the same parameters to predict colour via GPR. Orange points represent Milky Way analogues selected by methods equivalent to \citet{licquia2015b}, while purple points indicate the galaxies used to train the Gaussian process regression. The predicted mean Milky Way $^{0}(g-r)$ colour is denoted by a star in the corresponding colour, and ellipses depict a $1\sigma$ joint confidence region for colour and stellar mass. Both methods yield a similar predicted $^{0}(g-r)$ colour for the Milky Way when we have large numbers of analogues. In contrast to the small window that the Milky Way analogues lie in, the GPR utilises a wider variety of galaxies to capture larger-scale trends.}
    \label{fig:training_comparison}
\end{figure}

\autoref{fig:training_comparison} illustrates the fundamental difference between how properties are constrained by the Milky Way analogue selection method versus Gaussian process regression. For simplicity's sake we perform this comparison based on only three parameters (stellar mass, star formation rate, and axis ratio, the same ones utilised in \citetalias{licquia2015b}), as the analogue method starts to break down when more parameters are included. We also do not correct for Eddington bias in either measurement (q.v. Appendix~\ref{subsection:eddbias}) for simplicity. Objects included in a set of MWAs based on 5,000 samples from the distribution of possible Milky Way properties via methods equivalent to those from \citet{licquia2015b} are depicted by the orange points. The analogues fall within a narrow range of stellar mass, limiting the set of objects that contribute information. The orange star represents the mean prediction for the Milky Way's $^{0}(g-r)$ colour ($^{0}(g-r) = 0.682$) resulting from this set (derived via the Hodges–Lehmann robust estimator, \citet{hodges1963}). The orange ellipse depicts the $1\sigma$ confidence region for $^{0}(g-r)$ colour and stellar mass. In contrast to the Milky Way analogues, the sample of galaxies used to train the GPR are shown by purple points. These cover a much broader range of parameter space than the MWAs. Similarly, the prediction for the Milky Way's $^{0}(g-r)$ colour using GPR is shown by a purple star (with $^{0}(g-r) = 0.668$), along with the corresponding $1\sigma$ confidence region which is shaded in purple. When we have many analogues, both techniques yield very similar results, but unlike MWAs the GPR technique still provides strong constraints when we consider many parameters at once.

\subsubsection{Characterising Sources of Uncertainty}
\label{subsection:uncertainty}

\begin{figure}
    \centering
    \includegraphics[width=0.99\linewidth]{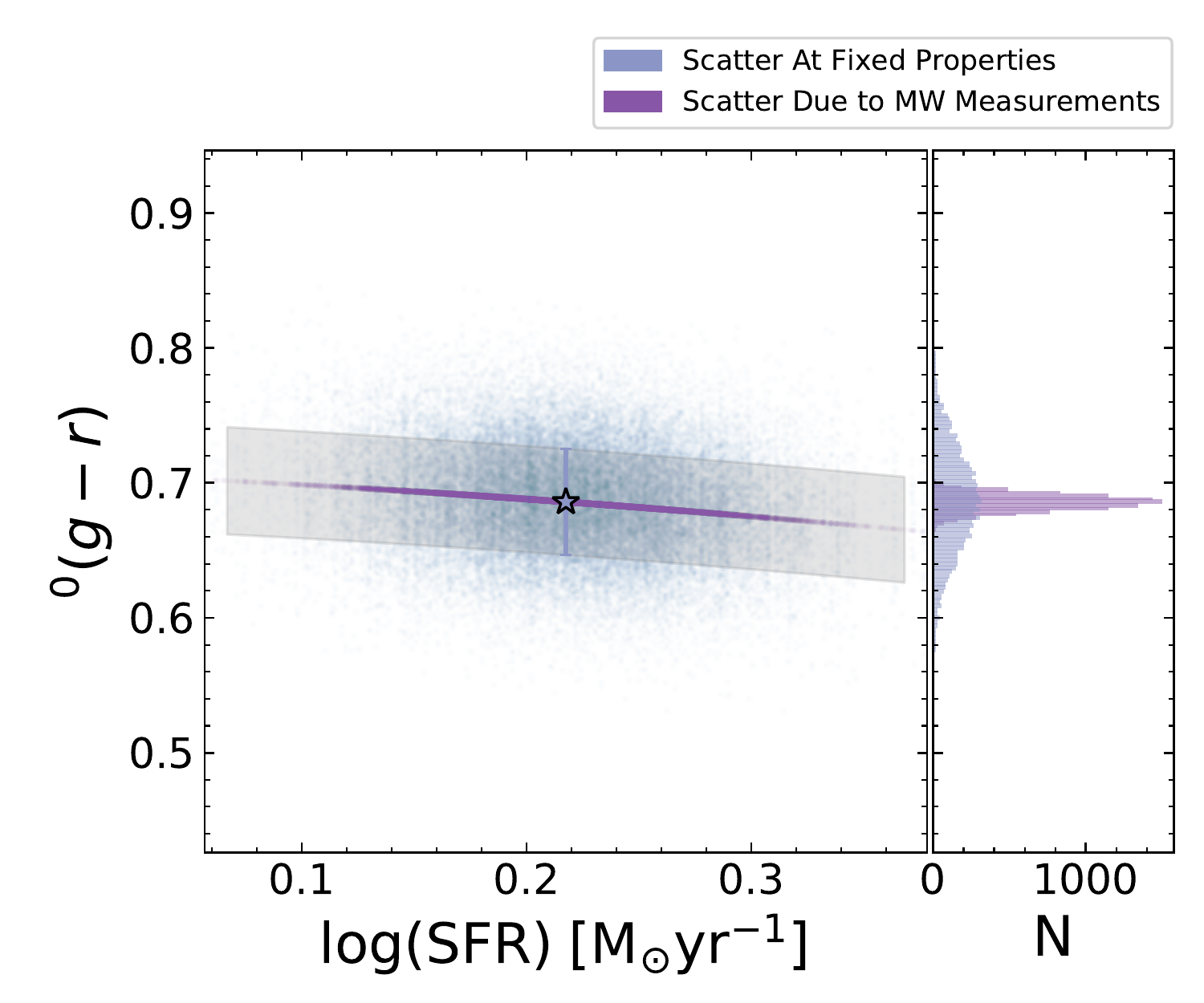}
    \caption{A breakdown of uncertainty due to scatter at fixed properties and scatter due to Milky Way measurement uncertainty. We plot results from a six parameter Gaussian process regression trained to predict $^{0}(g-r)$ colour. For this example we vary only one of the input parameters, the star formation rate (SFR). The light violet colour corresponds to results when we use a fixed training set and evaluate the GPR at fixed MW properties; this isolates the variation attributable to the white noise kernel, corresponding to the scatter in colour at fixed galaxy properties. The dark purple colour corresponds to results when the training set remains fixed but the SFR value is drawn randomly from the fiducial MW PDF, so that the contribution to the colour error attributable to the uncertainty in the  star formation rate of the Milky Way can be evaluated (the other five parameters used to train the GPR are held constant for simplicity for this example). In the left panel the shaded region depicts the $\pm1\sigma$ range predicted by the GPR fit, plotted out to $\pm3\sigma$ of the Milky Way's SFR. The predicted $^{0}(g-r)$ colour and $1\sigma$ error for the Milky Way's fiducial SFR (and hence only including error at fixed properties) is shown by the star-shaped point and error bar in light purple. The predicted $^{0}(g-r)$ colour from each of 5,000 draws from the SFR PDF is shown by the purple points, which accounts for scatter due to the SFR measurement. These predictions perfectly trace the GPR fitand tend to fall within $\sim3\sigma$ of the Milky Way's SFR, by construction. For reference we also show 10 samples from the GPR-predicted PDF for each of the 5,000 random SFR values as faint blue points. The distribution of these points  reflects both the scatter of galaxy colours at fixed properties and the uncertainty in the MW measurements. Histograms of the colours for each sample, whose distributions correspond to the scatter at fixed properties and the scatter due to Milky Way SFR measurement uncertainties, are shown in the right panel. It is evident that the spread in GPR predictions at fixed properties is much larger than the scatter that results from uncertainties in the Milky Way's SFR.}
    \label{fig:error_source}
\end{figure}

We can quantify the contributions of different sources of uncertainty to our GPR estimates by changing how we perform the regression. We illustrate our methods by evaluating how the prediction from a six-parameter Gaussian process regression fit changes as a single parameter varies. The left panel of \autoref{fig:error_source} shows one of the physical parameters being regressed from, star formation rate, on the x axis and the target value, $^{0}(g-r)$ colour, on the y axis. We chose this pair as galaxy star formation rate is expected to correlate with galaxy colour well at fixed stellar mass. 

First we isolate the scatter in colour at fixed properties; this corresponds to the contribution of the white noise kernel to the covariance function of the GPR. To determine the magnitude of this scatter we query the GPR at the Milky Way's fiducial physical properties to obtain a predicted PDF of $^{0}(g-r)$ at this point in parameter space from which we can draw samples. The standard deviation of the colour of these samples corresponds to the scatter encoded in the white noise kernel.

In the panel at the right of \autoref{fig:error_source} we plot a histogram of 10,000 possible $^{0}(g-r)$ colours drawn from the GP's predicted PDF, evaluating it at the fiducial value of the Milky Way's SFR. In the left panel the lavender star denotes the mean predicted $^{0}(g-r)$ for the Milky Way with this model, and the error bar corresponds to the standard deviation of the sample values. This error bar therefore corresponds to the $1\sigma$ scatter in colour at fixed properties. The grey shaded area corresponds to the $\pm1\sigma$ band for the GPR prediction of colour for a range of $\log{(\rm{SFR})}$ values. By construction the half-width of the band must match the standard deviation of samples from the PDF at fixed properties. 

To instead isolate the contribution to errors resulting from the uncertainty in Milky Way properties, we determine the distribution of mean predicted colours evaluated at varying values of SFR drawn from the PDF for the Milky Way. We perform 1000 draws from the fiducial MW $\log{\rm{SFR}}$ PDF in total and evaluate the GPR mean predicted $^{0}(g-r)$ for each. In \autoref{fig:error_source} the dark purple points in the left panel shows the resulting predictions, which all fall on a continuous curve by construction. The panel to the right shows the histogram of this set of predictions in purple. We quantify the scatter in colour attributable to uncertainties in the Milky Way's measurements via the standard deviation of the $^{0}(g-r)$ values at these 1,000 points. 

To illustrate the full range of values obtained via GPR we plot ten samples from the distribution of predictions for each of the thousand Milky Way SFR draws as faint blue points in \autoref{fig:error_source}. These samples vary in colour due to \textit{both} the scatter in colour at fixed properties and due to the uncertainty in Milky Way properties. 

\begin{figure}
    \centering
    \includegraphics[width=0.9\linewidth]{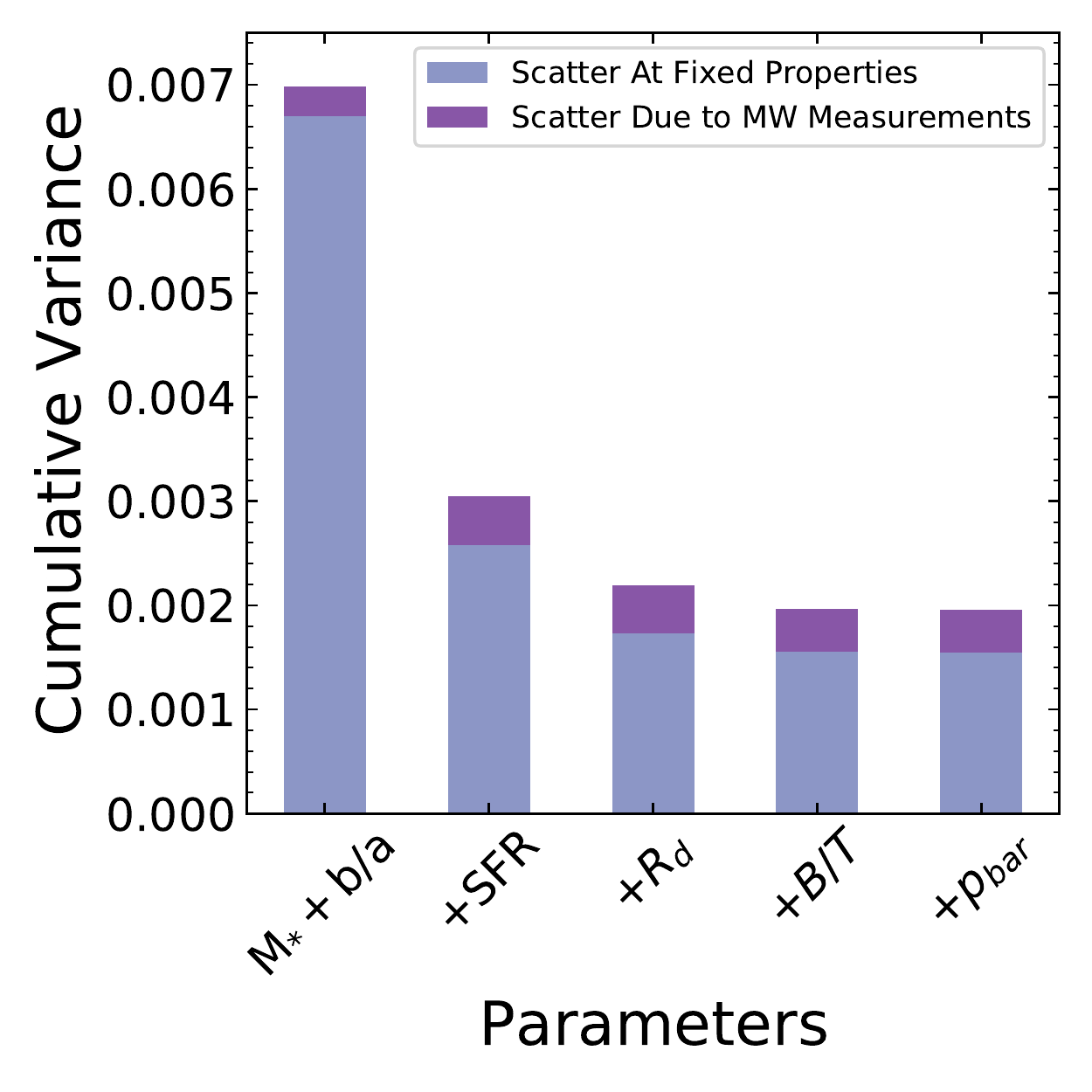}
    \caption{Contributions to the variance in restframe $^{0}(g-r)$ colour for Gaussian process regression employing varying sets of galaxy physical parameters. The x-axis shows the set of parameters used to predict colour; they increase cumulatively as we go from left to right, in order from the most constraining to the least constraining parameter. The variance decreases monotonically as the number of parameters increases. In every case the scatter in colour at fixed properties dominates errors; uncertainties in the physical properties of the Milky Way are sub-dominant. Contributions to uncertainties due to having a finite training set are small enough to be considered negligible, enough so that they would not be visible on this plot if we included them.}
    \label{fig:error_comparison}
\end{figure}

We can extend these same ideas in 1-D to evaluate the relative contributions of uncertainties in Milky Way properties and of the scatter in properties at fixed colour to the error in Milky Way $^{0}(g-r)$ colour, for any GPR model of interest. The key difference from \autoref{fig:error_source} is that, in order to quantify the full scatter due to uncertainties in Milky Way properties, we allow all the parameters to vary, not only SFR. We present the results of this analysis for GPR models based on two to six physical parameters in \autoref{fig:error_comparison}. We use a stacked bar plot to display the contribution to the variance from each error source, where the x-axis is labelled according to the physical parameters used to train the GPR (where additions are all cumulative, so entries further to the right incorporate more parameters). Since independent errors will add in quadrature, the contribution to the net variance from each factor is proportional to the height of its bar. The variance due to the scatter at fixed properties is shown in a lighter violet shade, while that resulting from the uncertainty in Milky Way properties is shown in dark purple. The scatter at fixed properties contributes to the majority of the error in the Gaussian process regression for $^{0}(g-r)$ colour.

We have also evaluated the contribution to uncertainties resulting from the finite size of the training sets used. This scatter is isolated by varying the randomly-selected training sample 100 times. For each training sample we evaluate the mean predicted value from GPR at the Milky Way's fiducial physical parameters. The contribution to uncertainties from the finite size of training samples is obtained by calculating the variance of the GPR predictions across all of the training sets. These values are minute: the variances are an order of magnitude smaller than error attributed to the scatter due to Milky Way measurements. Thus any contribution to errors resulting from the finite training sample size is negligible.

We have performed the same error budget test for colours in the UV, near-IR, and mid-IR. The results mirror those presented in \autoref{fig:error_comparison}: the errors are dominated by scatter at fixed properties, followed by scatter from the Milky Way measurements. In all cases errors attributable to finite training set size are negligible compared to other sources. While the cumulative variance decreases for every parameter added when we predict colours, this is not the case for absolute magnitudes. In that case, the cumulative variance decreases as we add parameters until the sixth parameter, $p_{\rm{bar}}$, is incorporated. At that point, the variance increases and the scatter due to finite training becomes  more important. For this reason all absolute magnitude predictions within this paper are performed using only 5 parameters, excluding $p_{\rm{bar}}$.

While uncertainties in Milky Way characteristics will contribute to the random errors in the derived photometric properties of our Galaxy, uncertainties in the physical parameters of the \textit{training} galaxies can cause \textit{systematic} errors. If the density of objects in parameter space varies quickly (with non-negligible second or higher derivatives), objects will more often scatter from well-populated regions of parameter space into sparser regions than vice versa.  The resulting systematic shift in the measured distribution of parameters compared to the underlying distribution with no scatter is known as Eddington bias. 

In the context of this work, Eddington bias will lead to shifts in the colour and luminosity predicted for the Milky Way. We derive corrections for Eddington bias using methods similar to those of  \citetalias{licquia2015b}; we detail our procedures in Appendix~\ref{subsection:eddbias}. The estimated Eddington bias is subtracted off from the GPR-predicted colours and luminosities for the Milky Way to produce our final estimates for the Galaxy's photometric properties and likewise the uncertainty on the Eddington bias calculations is propagated into our final error estimates. In general Eddington bias has small but nonzero effects on our results ($<1\sigma$ for almost all parameters, as listed in Appendix~\ref{sec:appendix_data}).

In the following sections, the errors on GPR results presented include the contributions from scatter at fixed properties, uncertainties in Milky Way properties, and uncertainty from Eddington bias.

\subsubsection{Summary of the GPR Algorithm for Determining Milky Way Photometric Properties}
\label{subsection:final_algorithm}

Here we summarise the steps taken to predict Milky Way photometric properties via GPR. Our method proceeds as follows:
\begin{enumerate}
    \item Construct the training sample by restricting to objects within $12\sigma$ of the Milky Way in all physical parameters considered and then randomly down-sampling to 2,000 objects.
    \item Adopt the combination of a Radial Basis Function (RBF) kernel and a white noise kernel as the covariance function to be used for GPR.
    \item Train the GPR using a single photometric property (normalised to have mean zero and variance one) as the output or ``y'' value and the physical galaxy parameters as the ``x'' values. This training will tune the hyperparameters of the kernel.
    \item Perform 1,000 random draws from the PDFs that describe the fiducial Milky Way's properties. This will allow us to incorporate uncertainties in the Milky Way measurements into our results.
    \item Use GPR to apply the optimised kernel to the training set and predict the photometric property of interest. For each randomly-drawn set of physical properties for the Milky Way, we obtain the mean prediction, predicted variance, and a set of 1,000 values drawn from the GPR-predicted PDF corresponding to that position in physical parameter space (which we refer to as a set of samples).
    \item The mean photometric prediction for the Milky Way is then calculated as the mean of the set of GPR output means at the position of each  MW draw. The error on the prediction is calculated as the standard deviation of the values from the complete set of samples generated, allowing us to incorporate both uncertainties associated with the scatter at fixed properties \textit{and} errors resulting from the uncertainties in MW properties.
\end{enumerate}

The code used to construct the GPR is provided on our GP GitHub page for public use \href{https://github.com/cfielder/GPR-for-Photometry}{here}\footnote{https://github.com/cfielder/GPR-for-Photometry}. At this site we provide sample code for determining photometry estimates, addressing systematics, and constructing an SED.


\section{Results}
\label{sec:results}

Via GPR predictions for Milky Way photometric properties across the spectrum, we can produce a comprehensive outside-in portrait of the Milky Way SED, allowing comparisons to the colors and luminosiities of other galaxies. In this section we apply a variety of diagnostics from the literature, such as colour-luminosity, colour-mass, and colour-colour diagrams, in order to assess how the Milky Way compares to the broader population. We also construct a multi-wavelength SED for the Milky Way and compare our results to templates from the literature.

\subsection{The Milky Way Compared to the Broader Galaxy Population}
\label{subsection:mw_comparisons}



As discussed in \autoref{sec:gpr}, we have predicted the Milky Way colours and luminosities based upon the six parameters of stellar mass ($\mass$), star formation rate (SFR), axis ratio ($b/a$), bulge-to-total ratio ($B/T$), disk scale length ($R_{\rm{d}}$), and bar vote fraction ($p_{\rm bar}$). In the following colour diagrams, all magnitudes and colours are presented as rest-frame AB magnitudes (evaluating all passbands at redshift zero). 

Our quantitative results are summarised in \autoref{Tab:color_sed}-\autoref{Tab:absmag}. The values provided correspond to the mean rest-frame predictions based upon the Gaussian process regression derived via the methods presented in \autoref{sec:gpr}, and have been corrected for Eddington bias as described in Appendix~\ref{subsection:eddbias}. Colours and magnitudes are all calculated independently of one another. For example, we use GPR to predict $^{0}(g-r)$ galaxy colour directly, as opposed to deriving this value by subtracting the predicted $^{0}M_{r}$ from the predicted $^{0}M_{g}$. For SDSS photometry, our derived colours are based upon \texttt{model} magnitudes, as these yield the most accurate colour estimates for SDSS galaxies; however, the absolute magnitudes provided are based upon \texttt{cmodel} magnitudes, as those most accurately represent the total brightness of an object.

Log-spaced density contours corresponding to the cross-matched galaxy sample of $29,836$ galaxies described in \autoref{subsection:photometry} and \autoref{subsection:strucprops} are plotted in grey-scale on all of the following colour-based diagrams. We also overlay red and blue ellipses which denote the rough locus of the red sequence and blue cloud, respectively, in each plot.  These shadings are intended to guide the eye and should not be interpreted in a quantitative manner. In a corner of each plot we provide error bars that correspond to the mean uncertainties in each galaxy property being plotted for the training set.

In each diagram we also show the locations of the 36 red spiral galaxies selected in \citet{masters2010} that overlap with our cross-matched sample (out of 294 in the original catalogue). This sample of objects was selected based upon their colour, presence of spiral features, and shape/structural parameters from SDSS. They are required to have colour $^{0}(g−r) > 0.63 − 0.02(^{0}M_{r} + 20)$, overlapping the blue edge of the red sequence. They are also selected to have a spiral likelihood $p_{\rm{spiral}}\geq 0.8$ in the prescription of \citet{bamford2009}, and are required to have visible arms in Galaxy Zoo 1 $p_{\rm{CW}} > 0.8$ or $p_{\rm{ACW}} > 0.8$ \citep{lintott2011} in order to ensure they have spiral morphology. These objects are also selected to be approximately face-on (equivalent to an axis ratio requirement $b/a > 0.63$), as dust reddening is expected to have a substantial impact on the apparent colours of spirals \citep{masters2010a}. However, in that paper the axis ratio values wre calculated via $r$--band isophotal measurements, while ours are determined from an exponential profile fit. Therefore we apply a profile-fit-based cut of $b/a > 0.6$ to this sample to enable a more direct comparison to our face-on results for the Milky Way. Finally, \citet{masters2010} requires that the red spiral sample contain galaxies with an SDSS $f_{\rm{deV}}\leq 0.5$, where $f_{\rm{deV}}$ is defined as the weight of the de Vaucouleurs profile in the best-fit linear combination with the exponential profile matched to the object’s image. This ensures that S0 galaxies do not contaminate the sample, although they are already only a small percentage of the GZ1 sample.

The resulting red spiral sample is represented by red points in our plot. We overlay the positions of these objects in each parameter space to help assess the consistency of the inferred properties of the Milky Way with this population. Two objects whose $^{0}(g-r)$ colours in the cross-matched catalogue differed by $>0.1$ mag from the photometry used in \citet{masters2010} due to changes in SDSS pipelines were excluded. 

\subsubsection{Optical Colours}
\label{subsubsection:optical}

\begin{figure}
    \centering
    \includegraphics[width=0.95\linewidth]{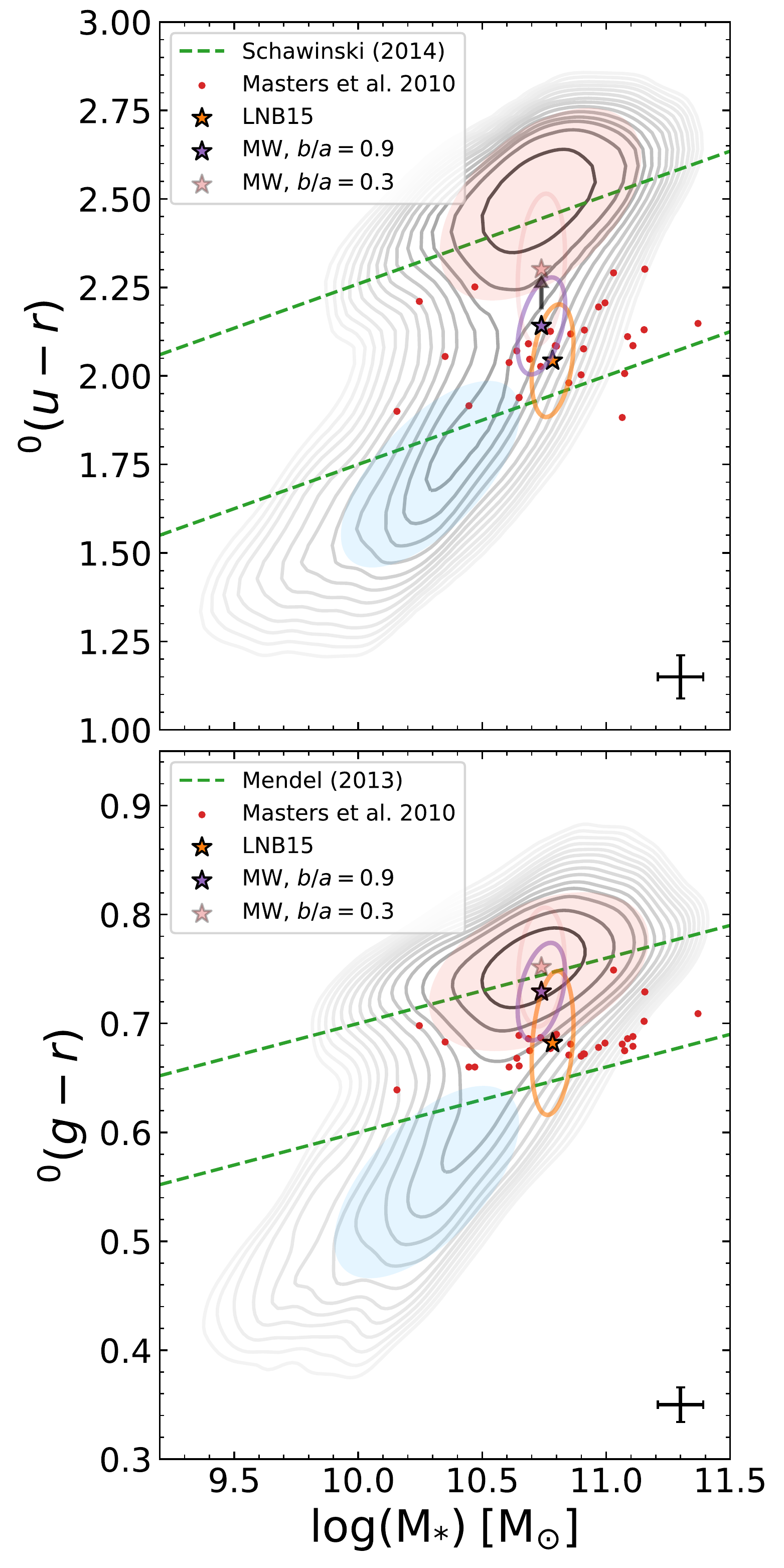}
    \caption{Rest-frame optical colour as a function of stellar mass.  The grey-scale, log-spaced contours depict the density of $0.03 < z < 0.09$ galaxies in our cross-matched sample, without any limiting magnitude applied.  The dashed grey lines correspond to divisions of the galaxy population used in the literature. We expect ``red sequence'' galaxies to be in the upper portion of each plot, with the ``blue cloud'' corresponding to the bluest colours. The region between the two lines corresponds to the ``green valley'' population. For $^{0}(u-r)$ we use the divisions of \citet{schawinski2014} and in $^{0}(g-r)$ we follow \citet{mendel2013} (cf. \autoref{subsubsection:optical}). The results and $1\sigma$ confidence region from \citetalias[][]{licquia2015b} are marked in orange. Our results from applying GPR to all six galaxy physical parameters considered are marked in purple;  the $1\sigma$ region is determined by the covariance between Gaussian process samples. For comparison, in lighter red we show the six-parameter GPR result obtained when setting the axis ratio of the Milky Way to $b/a=0.3\pm0.1$ rather than 0.9. The stellar masses differ between our prediction and \citetalias{licquia2015b} due to the updates to the mass estimate for the Milky Way described in \autoref{subsection:mwparams}. Red points correspond to members of the red spiral galaxy sample of \citet{masters2010}. In the lower right corner of each panel we depict error bars representative of the mean uncertainties for galaxies in the comparison sample.
    Our results are consistent with \citetalias[][]{licquia2015b} and indicate that at optical wavelengths the Milky Way is redder than the typical star-forming spiral galaxy, in addition to being more massive.}
    \label{fig:color_mass}
\end{figure}

We first present results at optical wavelengths, as they allow us to compare directly to previous work done with Milky Way analogues in \citetalias{licquia2015b}. We focus on the SDSS $ugriz$  bands (cf. \autoref{subsubsection:sdss}). \autoref{fig:color_mass} presents predictions for Milky Way optical optical colours as a function of stellar mass (\mass) in solar mass units.

The upper panel shows $^{0}(u-r)$ colour and the lower panel shows $^{0}(g-r)$ colour versus mass. Both panels have overlaid dashed reference lines which can be used to distinguish general regions of the diagrams. The top portion contains galaxies that are on the red sequence, while the middle portion contains green valley galaxies, and the lower portion corresponds to the blue cloud. The dashed lines bracketing the green valley in the upper panel correspond to $^{0}(u-r) = -0.24 + 0.25\mass$ and $^{0}(u-r) = -0.75 + 0.25\mass$ \citep{schawinski2014}. In the lower panel the plotted lines correspond to $^{0}(g-r) = 0.6 + 0.06(\mass - 10)$ and $^{0}(g-r) = (0.6 + 0.06(\mass - 10)) + 0.1$ \citep{mendel2013}.

Previous results from \citetalias{licquia2015b} are plotted in orange. The star represents the mean prediction and the ellipse encompasses the $1\sigma$ confidence region. We remind the reader that these constraints were determined based only on stellar mass and star formation rate, along with a cut on axis ratio. In comparison, the results of the 6-parameter Gaussian process regression are plotted in purple. In red we plot the GPR result evaluated with an axis ratio of $b/a=0.3$ rather than 0.9, to illustrate the impact that the assumed inclination has on the inferred SED. The GPR confidence regions are calculated from the covariance between the the samples drawn from the regression predictions; the distribution of these samples incorporates both uncertainties in Milky Way properties and scatter in colours at fixed properties (cf. \autoref{subsection:uncertainty}). In the lower right corner of each panel we show the mean error in optical colour and log stellar mass amongst the galaxies in our final sample. Per-object uncertainties in the optical colours account for roughly half of the total scatter in our GPR colour prediction for a face-on Milky Way. In contrast, the average error in stellar mass in the training sample does not affect the uncertainty in the stellar mass of the Milky Way (it will, however, contribute to Eddington biases, as discussed in \autoref{sec:appendix_systematics}. Note that \citetalias[][]{licquia2015b} did not use the same stellar mass as we do, reflecting our updated estimate for the mass of the Milky Way (cf. \autoref{subsection:mwparams}). 

In both $^{0}(u-r)$ and $^{0}(g-r)$ our results are consistent with, though marginally redder than, those reported in \citetalias[][]{licquia2015b}. This is no surprise as we do not expect the Milky Way to move far in optical colour space when constraints tighten. Even when we make predictions for a much more inclined Milky Way our results do not change dramatically in the optical, with shifts well within the uncertainties in both our face-on results and those from \citetalias{licquia2015b}, although the colour does become marginally redder as expected. In the optical the Milky Way appears to lie in the ``saddle'' of the the galaxy-colour bimodality, implying that the Milky Way is redder than the average spiral galaxy in the local Universe in the optical bands. That said, if one were to only consider spiral galaxies of similar mass to our Galaxy, the Milky Way is not as unusually red as it would be if compared to lower-mass spiral galaxies.

The green valley (and by extension the galaxy colour-bimodality) has been used as a basic tool to distinguish transitional galaxies from the general galaxy population. Transitional galaxies have lower specific star formations rates (sSFR = ${\rm{SFR}/\mass}$) than a star-forming galaxy of the same mass; specific star formation rate can be used as a proxy for the evolutionary state of a galaxy and its star forming history. \citet{salim2014} defines the transitional region in sSFR space to be below the sSFR of massive Sbc galaxies, as these Sbc's are the earliest galaxy type expected to proceed with regular star formation free of quenching, but above the sSFR at which galaxies appear to no longer be star forming in the UV. As described in that work, this range corresponds to $-11.8 < \log{(\rm{SFR}/\mass)} < -10.8$. Note that here and throughout this paper $\log$ refers to the base 10 logarithm.

\begin{figure}
    \centering
    \includegraphics[width=0.95\linewidth]{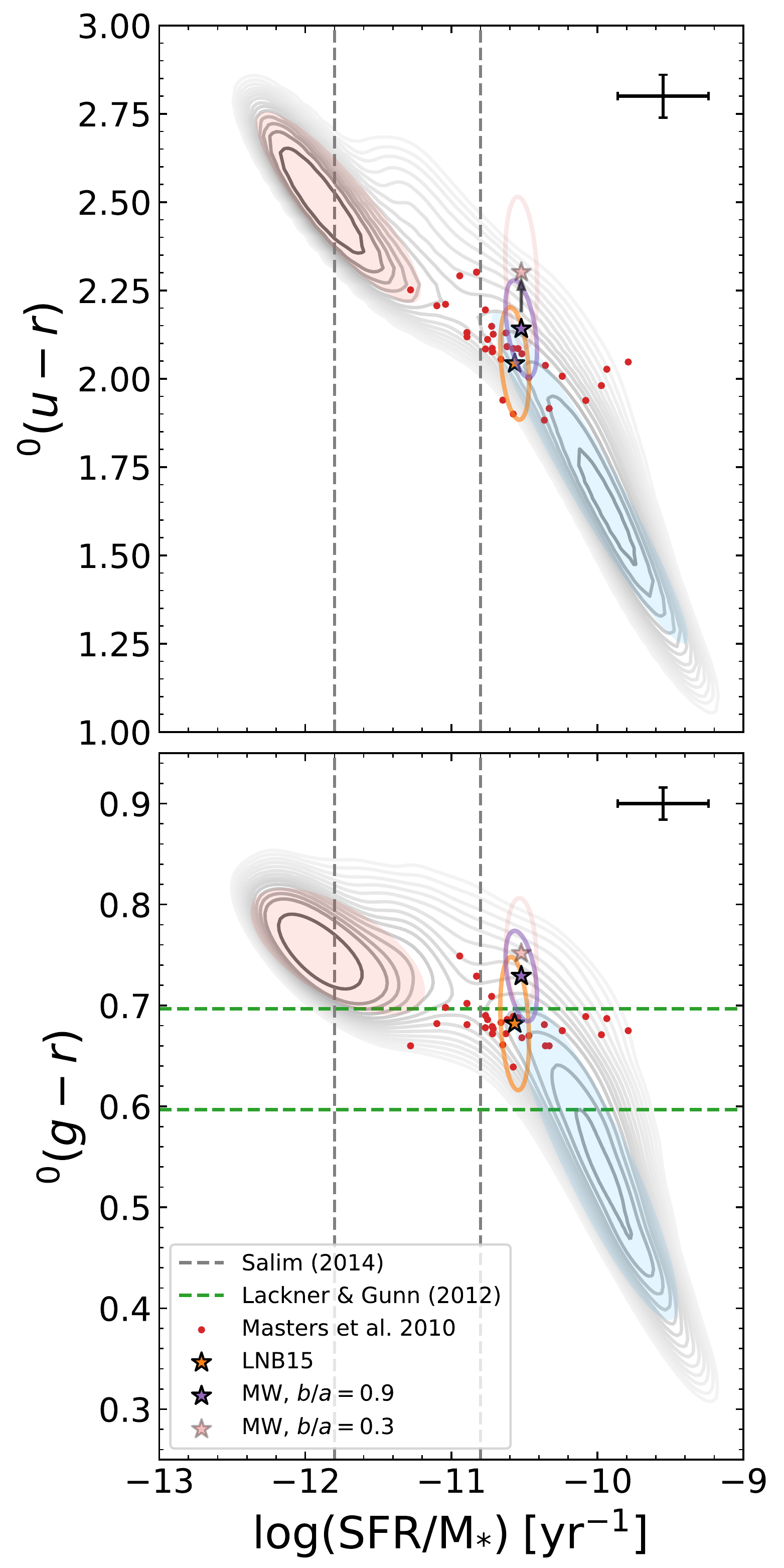}
    \caption{Rest-frame optical colour as a function of log specific star formation rate ($\log{(\rm{SFR}/\mass)}$) in units of $\rm{yr}^{-1}$. As before the grey-scale log contours depict the density of the parent cross-matched sample, red points correspond to the red spiral sample of \citet{masters2010}, the orange star and ellipse correspond to the results of \citet{licquia2015b}, and the purple star and ellipse are the results of our GPR analysis. The vertical dashed grey lines designate divisions of galaxy populations according to their specific star formation rates, following the definitions of \citet{salim2014}: quiescent galaxies are at left, transitional objects are in the middle, and star-forming objects are at right. The Milky Way lies on the star-forming side of these divisions. The green horizontal lines in the bottom panel correspond to the green valley definition of \citet{lackner2012}, evaluated at the $r$-band absolute magnitude of the Milky Way. According to the prescription by \citet{mendez2011} the green valley has a width of 0.1 in $^{0}(g-r)$, leading to the limits shown here here. The Milky Way mean value falls above this ``green valley'' region. 
    While the optical colour of the Milky Way is redder than most star-forming galaxies in the local Universe, based on its specific star formation rate the Milky Way would not be considered a transitional galaxy.
    }
    \label{fig:optical_ssfr}
\end{figure}

In \autoref{fig:optical_ssfr} we plot the same rest-frame colours as in \autoref{fig:color_mass} but as a function of log specific star formation rate. The vertical dashed lines denote the transitional region in $\log{\rm{SFR}/\mass}$, as defined by \citet{salim2014}. The region with log sSFR  above -10.8 corresponds to galaxies that are actively forming stars while objects with log specific star formation rate below -11.8 are quiescent; transitional objects are between them. In the lower panel the green horizontal lines correspond to the green valley definition of \citet{lackner2012} evaluated with the predicted $r$ band absolute magnitude for the Milky Way ($^{0}M_{r} = -20.65 $). Galaxies residing within this range in $^{0}(g-r)$ are expected to reside within the green valley. Galaxies above this designation are expected to lie in or near the red sequence, and galaxies below are expected to lie in or near the blue cloud.

Based on its specific star formation rate, the Milky Way must lie within the star-forming population, rather than in the transitional range. While one might expect an object that meets optical definitions of the green valley to have a transitional sSFR this is not necessarily true, as galaxies of different evolutionary states can share the same optical colour \citep[see e.g.,][]{cortese2012,salim2014}. 

In $^{0}(g-r)$ if we take the green valley to be 0.1 in width, as defined by \citet{mendez2011}, the Galaxy would either fall within or be redder than the green valley, consistent with the results shown in \autoref{fig:color_mass}. Despite ongoing star formation, more massive spiral galaxies tend to be redder in the optical than their lower-mass counterparts \citep[e.g.,][]{masters2010}.
However, our estimated properties for the Milky Way lie in the middle of the distributions of colours, masses, and sSFRs of the red spiral sample from \citet{masters2010}; these objects are redder in the optical than is typical for even the most massive spirals.

As these plots exemplify, differentiating between galaxy populations based only on optical photometry is challenging. For instance, in the lower panel in \autoref{fig:optical_ssfr} we can see that red sequence, transitional, and star-forming objects can all have colours of $^{0}(g-r) \sim 0.7$. In \autoref{fig:color_mass} the blue cloud becomes difficult to distinguish from the red sequence at high masses as the most massive spirals have lower sSFRs and, therefore, redder colours. 
In the following subsections we investigate constraints on the colour of the Milky Way at UV and IR wavelengths where galaxy populations may separate more clearly.

\subsubsection{UV Colours}
\label{subsubsection:uvcolor}

We utilise far-ultraviolet (FUV) and near-ultraviolet (NUV) photometry from GALEX provided in the GSWLC-M2 catalogue \citep{galex2005,salim2016,salim2018}, as discussed in \autoref{subsubsection:gswlc}. Thermal emission from massive stars with lifetimes $<100$ Myr peaks at ultraviolet wavelengths, while lower-mass, longer-lived stars play a larger role at optical wavelengths \citep[][]{salim2014,tuttle2020}. Because UV radiation is produced by short-lived but high-luminosity stars it provides a sensitive indicator of \emph{recent} star formation. As a result, UV photometry can more clearly differentiate star-forming from quiescent galaxies than optical measurements can.

\begin{figure}
    \centering
    \includegraphics[width=0.95\linewidth]{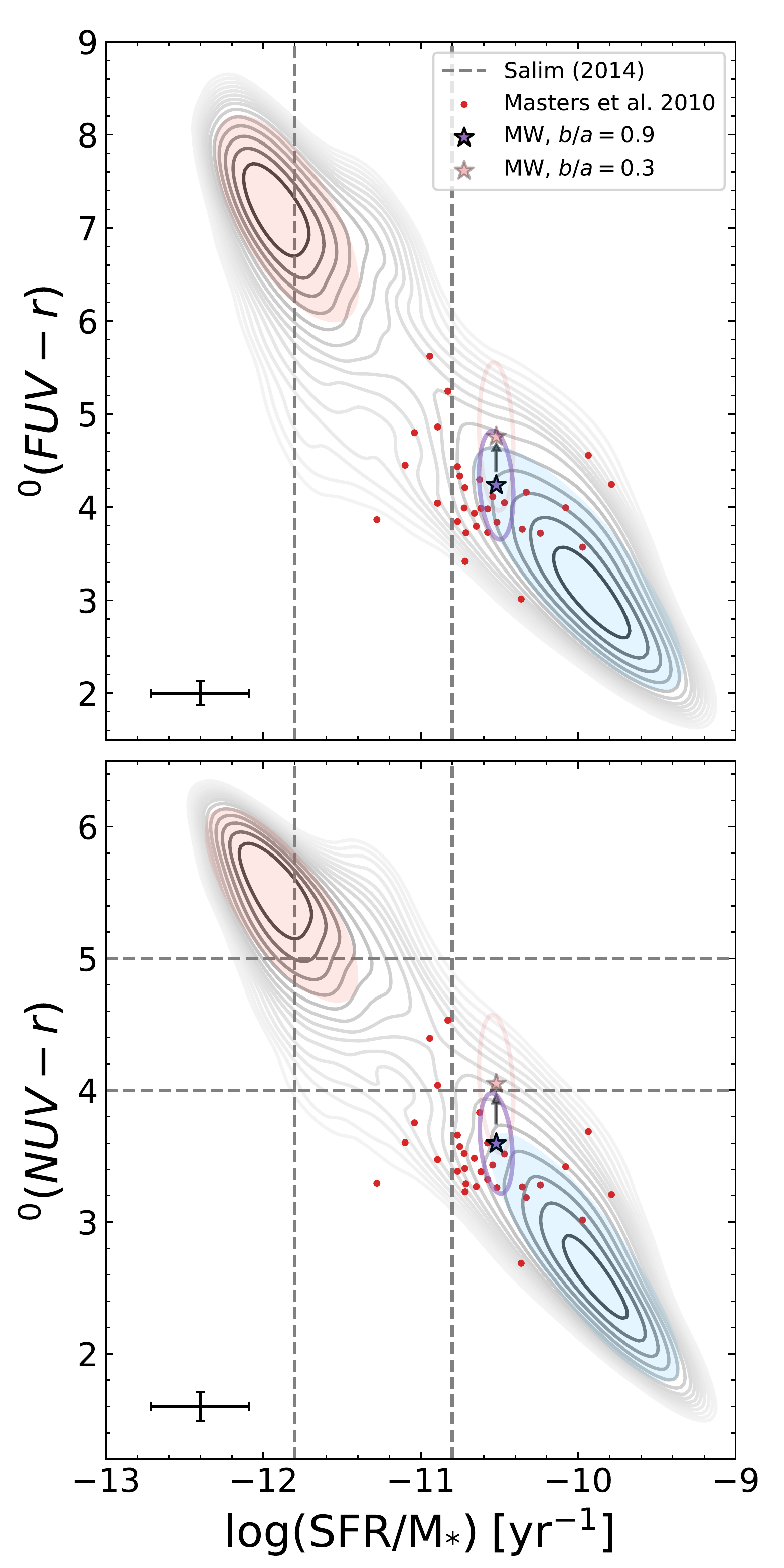}
    \caption{As \autoref{fig:optical_ssfr}, but for UV-optical colours. The vertical reference lines come from \citet{salim2014} where objects with a $\log{\rm{sSFR}} > -10.8$ are actively forming stars, objects with $\log{\rm{sSFR}} < -11.8$ are quiescent, and objects in between are considered transitional. In $^{0}(\rm{NUV}-r)$ we also plot the bounds of the UV-optical ``green valley'' as defined in \citet{salim2014}. Our predictions for the Milky Way show that it lies on the star forming side of the transitional regions, and is most likely on the blue side of the UV-optical green valley border. Our Galaxy lies in a similar region of these colour spaces to the red spiral sample of \citet{masters2010}.}
    \label{fig:uv_ssfr}
\end{figure}

Much as in \autoref{subsubsection:optical}, we can use our GPR results to place the Milky Way on UV-based diagnostic diagrams from the literature.  In \autoref{fig:uv_ssfr} we plot $^{0}(\rm{FUV}-r)$ and $^{0}(\rm{NUV}-r)$ UV-optical colours versus specific star formation rate. The contours and vertical reference lines shown are defined in the same way as in \autoref{fig:optical_ssfr}. For $^{0}(\rm{NUV}-r)$ we show horizontal lines corresponding to the ``green valley'' definition of \citet{salim2014}, bounded at $4 <\; ^{0}(\rm{NUV}-r) < 5$. As before, our face-on MW prediction and the corresponding $1\sigma$ confidence region are plotted in purple and the inclined MW prediction is plotted in red; unlike in the optical, there are no previous estimates of Milky Way properties in this space that we could plot. Much as in the optical, uncertainties in the UV photometry for individual objects account for roughly half of the total scatter ascribed to our Milky Way UV predictions.

Compared to typical star-forming galaxies in the local Universe, the Milky Way has redder than average UV colours and lower than average sSFR. The Milky Way appears to lie on the blue side of the $^{0}(NUV-r)$ green valley border, in contrast to its location in the optical (cf. \autoref{fig:color_mass} and \autoref{fig:optical_ssfr}). This reflects the limited discriminating power of optical colour; the green valley is only 0.1 mag wide in $^{0}(g-r)$, allowing objects to easily scatter over its borders due to even small photometric errors or inclination effects, but it spans an entire magnitude in $^{0}(NUV-r)$. A more inclined Milky Way is predicted to be notably more red in the UV than in the optical, so much so that it could be consistent with the UV green valley in colour.

As in the optical, our estimates for the Milky Way in the UV-sSFR plane lie in the middle of the \citet{masters2010} red spiral population. Red spiral galaxies tend to lie outside of the UV green valley as they have star formation rates comparable to typical blue spirals of the same mass \citep{cortese2012}, and UV colour is more sensitive to recent star formation rate than the optical is.

\subsubsection{Infrared/WISE Colours}
\label{subsubsection:ircolor}

The infrared data for our galaxy sample originates from the 2MASS and WISE surveys, as included in the GSWLC-M2 \citep{salim2016,salim2018} and DESI Legacy catalogues \citep{legacy2019}, respectively (cf. \autoref{subsection:photometry}). Similar to in the ultraviolet, the infrared brightness of a galaxy is sensitive to recent star formation due to re-emission of UV photons absorbed by dust. The infrared colours of galaxies also exhibit a colour bi-modality, but star-forming galaxies exhibit \textit{redder} IR colours than the passively-evolving population, rather than bluer. Instead of the ``green valley,'' the region between the star-forming and quiescent populations in the IR is commonly referred to as the infrared transition zone (IRTZ), following \citet{alatalo2014}.

\begin{figure}
    \centering
    \begin{subfigure}{0.9\linewidth}
        \includegraphics[width=\linewidth]{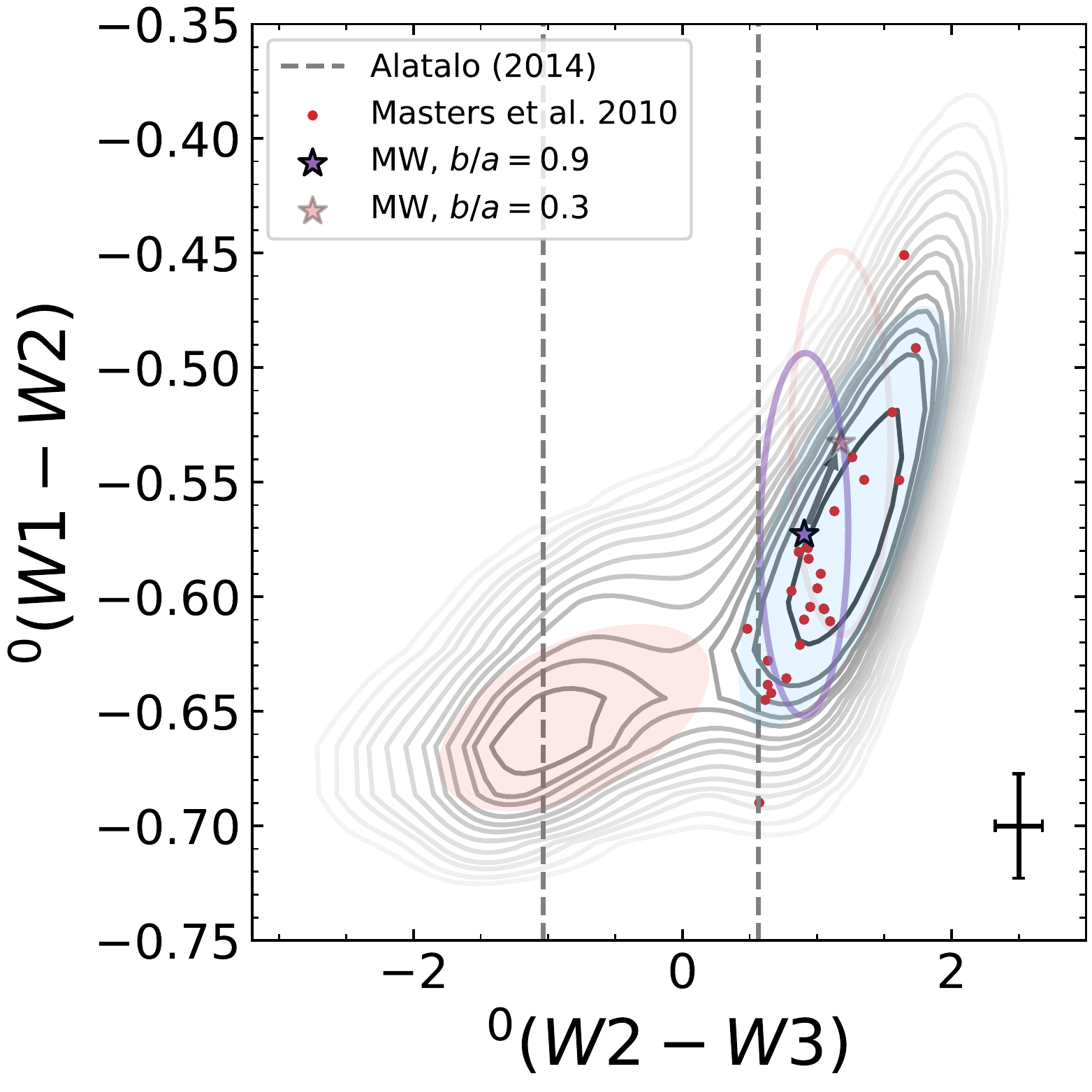}
        \caption{}
        \label{subfig:Wise_color1}
    \end{subfigure}\\%
    \begin{subfigure}{0.9\linewidth}
        \includegraphics[width=\linewidth]{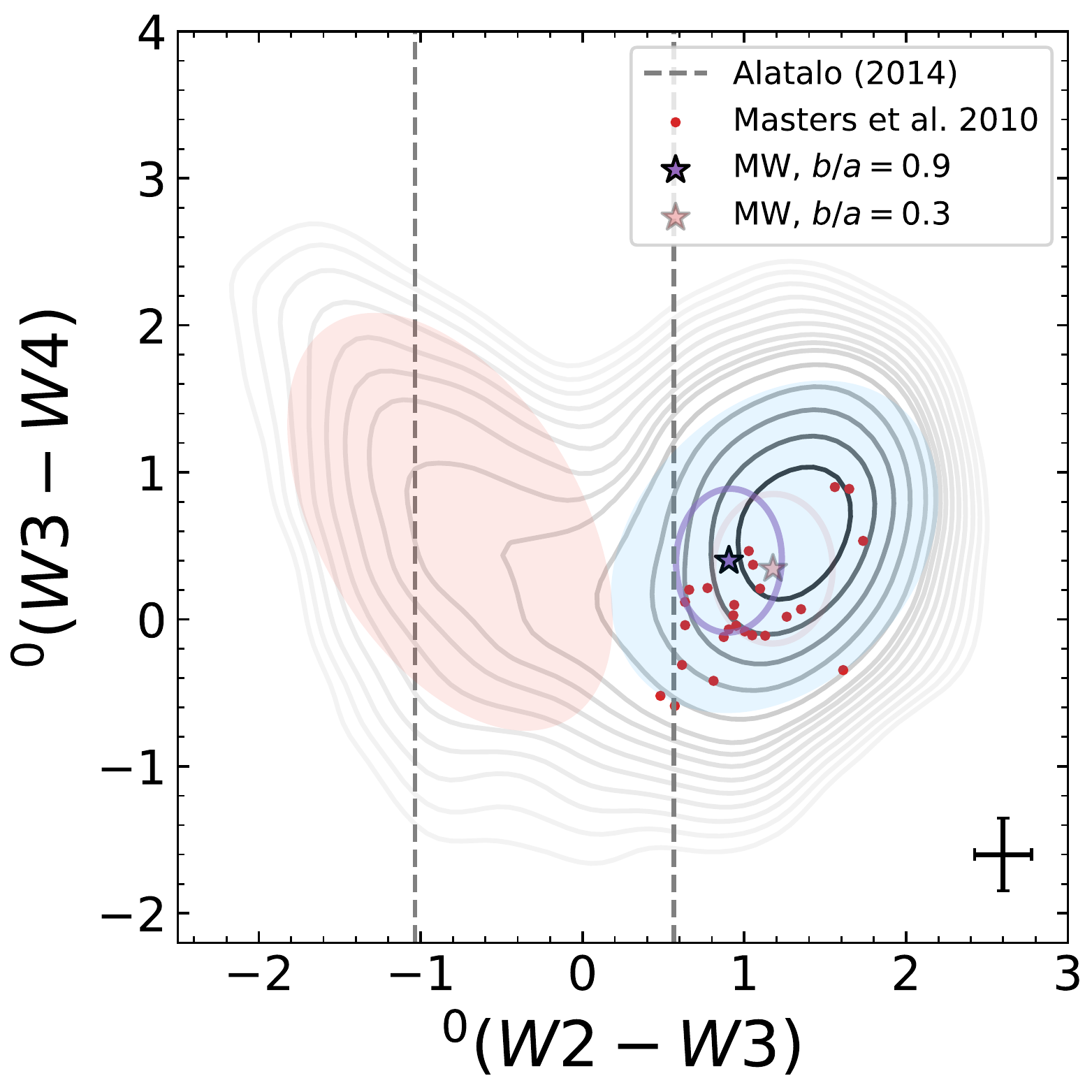}
        \caption{}
        \label{subfig:Wise_color2}
    \end{subfigure}%
\caption{WISE colour-colour diagrams for both our parent sample (log density contours) and the predicted results for the Milky Way from Gaussian process regression. Reference lines from \citet{alatalo2014} designate the infrared transition zone ($1.035 < ^{0}(W2-W3) < 0.565$; IRTZ). The Milky Way appears to lie on the star forming side of the IRTZ, much closer to the median colours of typical spiral galaxies, in contrast to the UV and IR. Again our Galaxy lies in a similar region of these colour spaces to the red spiral sample of \citet{masters2010}. 
} 
\label{fig:ir_color}
\end{figure}

In \autoref{subfig:Wise_color1} and \autoref{subfig:Wise_color2} we plot WISE colour-colour diagrams for the cross-matched galaxy sample in addition to the GPR prediction for the Milky Way. The Milky Way colour is poorly constrained in some WISE bands due to the lower signal-to-noise of these detections. If we compare our covariance ellipse in the $^{0}(W1-W2)$ direction to the average errors in the photometry (lower right error bar) the photometric errors account for a modest fraction of the total uncertainties in our GPR prediction. However, in $^{0}(W2-W3)$ and $^{0}(W3-W4)$ errors in the photometry for individual objects dominate the estimated uncertainties in the Milky Way GPR predictions. The vertical lines in these plots designate the IRTZ from \citet{alatalo2014};  objects with $^{0}(W2-W3) > 0.565$ in AB magnitudes correspond to late-type galaxies, those with $^{0}(W2-W3) < -1.035$ are early-type galaxies, and in between lies the IRTZ. Magnitudes were converted from Vega to AB magnitudes via the prescription of \citet{jarrett2011}.

As before, we show predictions for the Milky Way if it were approximately face-on or more steeply inclined by evaluating the GPR at different axis ratios. An inclined Milky Way appears the be most notably different in the W3 band, which traces prominent dust emission features, particularly those associated with polycyclic aromatic hydrocarbons \citep{wise}. It appears that an inclined spiral galaxy would be measured to be more IR bright compared to a face-on counterpart matching it in all other ways; this may represent a systematic effect related to data processing, since the dust emission would be expected to be optically thin. We find similar results in \autoref{subsection:parameter_impact}.

In both diagrams the prediction for the Milky Way lies on the star-forming side of the infrared transition zone. If we compare \autoref{subfig:Wise_color1} to the classification scheme in Figure 12 of \citet{wise} (note that this requires converting our AB magnitudes to Vega magnitudes), the Milky Way lies within the  region of colour space they label as typical for spiral galaxies, as would be expected. Similarly, the classification scheme of Figure 11.b of \citet{jarrett2017} would place the Milky Way in the intermediate disk region, consistent with the expectation for a massive spiral galaxy. Intermediate disk objects are thought to be in transition towards being quenched due to star formation rates that are decreasing with time; our estimate for the Milky Way's star formation rate is slightly below average for a spiral galaxy of the same mass, consistent with this picture.

The infrared results mirror what we find from the UV: the Milky Way is still forming enough stars to appear bright in the IR due to re-emission from dust. Galaxies are expected to transition in the optical before they do in the infrared \citep{alatalo2014,tuttle2020}. If we follow the narrative of \citet{alatalo2014} and \citet{smethurst2015}, the Milky Way may be in the early transition phase from star-forming to quiescent. It is brighter and redder in the optical compared to the typical star-forming galaxy. In the UV the Milky Way is on the blue side of the green valley but near it. In the IR the Milky Way's inferred colour is more typical for a star-forming galaxy, though uncertainties are substantial. This would track with the expectation that a galaxy transitions in the optical before it does in the IR.  

As before, the inferred colours of the Milky Way in the mid-IR are consistent with the range of values for the \citet{masters2010} red spiral sample, though some WISE bands are not well constraining due to the low signal-to-noise of the underlying measurements used for prediction. Based on this, it remains plausible that the Milky Way is a part of the red spiral population.  We discuss how the Milky Way's colours compare to other galaxy populations further in \autoref{subsection:discussion}.

\subsection{The Multiwavelength Spectral Energy Distribution of the Milky Way}
\label{subsection:sed}

Thus far, we have focused on predictions for a single Milky Way colour at a time. However, we can assemble colour information across all passbands to construct a spectral energy distribution (SED) for the Galaxy. SEDs, which quantify the the total energy of emitted photons as a function of wavelength or frequency, are valuable tools in the study of galaxies. Many physical characteristics of galaxies can alter their SEDs - the age of their stellar population, stellar abundances, gas and dust content, inter-stellar medium (ISM) chemistry, details of star formation history, and the presence of an AGN can all leave distinct signposts that give observers insight into the formation and evolution of a given galaxy \citep[see e.g.,][]{silva2011}. Because these effects each tend to alter the SED at specific portions of the spectral range, with broad enough wavelength coverage and detailed enough spectral information one can disentangle the dominant processes in a given galaxy.

Detailed modelling of the Milky Way's SED will be the focus of a follow-up paper. In this work, we will present a proof-of-concept for a GPR-constructed SED  for the Milky Way and provide an initial analysis of its properties. In the following sub-section we outline our GPR-based methods for determining the SED of the Galaxy before presenting quantitative results and assessing the effects the galaxy physical parameters used for prediction each has on the SED.

\subsubsection{Algorithm for Calculating the SED for the Milky Way}
\label{subsubsection:sed_algorithm}

We work in frequency ($\nu$) space instead of wavelength ($\lambda$) space when calculating the SED of the Milky Way, as spectral energy distributions are most typically presented in units of energy per unit frequency. Our algorithm for calculating the SED proceeds as follows:
\begin{enumerate}
    \item \textbf{Estimate $^{0}M_{r}$ and colours for the Milky Way} -- \\ Our SED is calculated in reference to the $r$-band. Therefore using the GPR (described in \autoref{sec:gpr}) we predict the $r$-band AB absolute magnitude ($^{0}M_{r}$) and all colours with restframe $r$ as the reference band; i.e., $^{0}(x-r)$ where $x$ spans from FUV to W4 (e.g., $^{0}(FUV-r), ..., ^{0}(W4-r)$).\\
    Eddington bias is subtracted off separately from our predicted colours and $^{0}M_{r}$ before we combine them. Similarly, the uncertainty in the Eddington bias is added in quadrature to the uncertainty in the GP calculations (which incorporates both scatter at fixed properties and errors due to the Milky Way property uncertainties; cf. \autoref{subsection:uncertainty}). For details on the Eddington bias calculations refer to Appendix~\ref{subsection:eddbias}. \\
    \item \textbf{Calculate flux ratios} -- \\We calculate the flux in each band $f_{\nu,x}$ relative to the flux in the $r$-band $f_{\nu,r}$ via the relation
    \begin{equation}
        \log{\left( \frac{f_{\nu,x}}{f_{\nu,r}}\right)} = \frac{^{0}(x-r)}{-2.5},
    \label{eq:fracflux}
    \end{equation}
    where $^{0}(x-r)$ is the Eddington bias-corrected colour in the $x$ band compared to the $r$ band, which we have predicted via GPR.\\
    \item \textbf{Calculate luminosity} -- \\From the $r$-band absolute magnitude combined with the flux ratios it is straightforward to convert to luminosity. We first calculate the $r$ band luminosity as
    \begin{equation}
        \log{(L_{\nu,r})} = \frac{(^{0}M_{r} - 34.04)}{-2.5}\; \left[ \log{\left( \frac{W}{Hz}\right)}\right],
    \label{eq:lum_r}
    \end{equation}
    where $^{0}M_{r}$ is the Eddington bias-corrected $r$-band absolute magnitude for the Milky Way obtained via GPR. This formula is derived via the relation for converting flux to AB magnitudes in combination with the area of a 10 pc radius sphere to convert flux to luminosity. The luminosity in any other band can then be calculated via the relation
    \begin{equation}
        \log{(L_{\nu,x})} = \log{\left( \frac{f_{\nu,x}}{f_{\nu,r}}\right)} + \log{(L_{\nu,r})}.
    \label{eq:lum}
    \end{equation} 
    We can then add log frequency ($\log{\nu}$) to obtain 
    $\log{\nu L_{\nu,x}}$.\\
    \item \textbf{Calculate errors} -- \\To convert uncertainties in magnitudes and colours to uncertainties in $\log{\nu L_{\nu,x}}$ we make use of propagation of errors. First to determine $\sigma_{\log{\nu L_{\nu,r}}}$ we calculate the partial derivative of $\log{\nu L_{\nu,r}}$ (which is \autoref{eq:lum_r} $+\log{\nu}$) with respect to $^{0}M_{r}$. This yields
    \begin{equation}
        \sigma_{\log{\nu L_{\nu,r}}} = 0.4\sigma_{^{0}M_{r}}.
    \end{equation}
    Errors in the other bands are calculated in a similar manner, but they depend on both the error in colour and the error in $^{0}M_{r}$: 
    \begin{equation}
    \sigma_{\log{\nu L_{\nu,x}}} = 0.4\times\sqrt{\sigma_{^{0}(x-r)}^{2} + \sigma_{^{0}M_{r}}^{2}}.
    \end{equation}
    In the plots that follow we do not plot the contribution to errors from $\sigma_{^{0}M_{r}}$ as it is fully covariant across all bands; as a result, when templates are normalised to match the observed SED, any error in $^{0}M_{r}$ would simply change the normalisation. We do provide its value for reference. Thus the error bars presented in the Milky Way SED plots are equivalent to $0.4\sigma_{^{0}(x-r)}$.
    

\end{enumerate}

\bigskip

The colour predictions used to derive the luminosities used for our SED are provided for reference in \autoref{Tab:color_sed} and the value of $^{0}M_{r}$ is provided in the absolute magnitude table \autoref{Tab:absmag}. We present our estimated luminosities and associated uncertainties, incorporating Eddington bias corrections in each case, in \autoref{Tab:luminsoity}.

\begin{table}
\centering
\begin{tabular}{rlll}
\hline
Passband & $\lambda_{eff}\;[\mu m]$ & $\log{\nu L_{\nu}}$ [$\log{\rm{W}}$] & $\sigma_{\log{\nu L_{\nu}}}$ [$\log{\rm{W}}$] \\
\hline
FUV &  0.155 & 35.53 &        0.20\\
NUV &  0.2275 & 35.63 &        0.20\\
u   &  0.354  & 36.01 &        0.06 \\
g   &  0.4750 &  36.44 &        0.02 \\
r   &  0.622 &  36.62 &        0.09 \\
i   &  0.763 &  36.66 &        0.01 \\
z   &  0.905 &  36.70 &        0.02 \\
J   &  1.25 & 36.67 &        0.04 \\
H   &  1.65 & 36.54 &        0.05 \\
Ks  &  2.15 & 36.39 &        0.06 \\
W1  &  3.368 & 35.59 &        0.08 \\
W2  &  4.618 & 35.58 &        0.10 \\
W3  &  12.082 & 35.61 &        0.20 \\
W4  &  22.194 & 35.49 &        0.30 \\
\hline
\end{tabular}
\caption{The passbands and corresponding power and uncertainties for the predicted SED of the Milky Way, as plotted in \autoref{fig:sed}. These values have already had Eddington bias subtracted out.}
\label{Tab:luminsoity}
\end{table}

\subsubsection{Interpreting the SED of the Milky Way}
\label{subsubsection:sed_analysis}

\begin{figure*}
    \centering
    \begin{subfigure}{0.5\linewidth}
        \includegraphics[width=\linewidth]{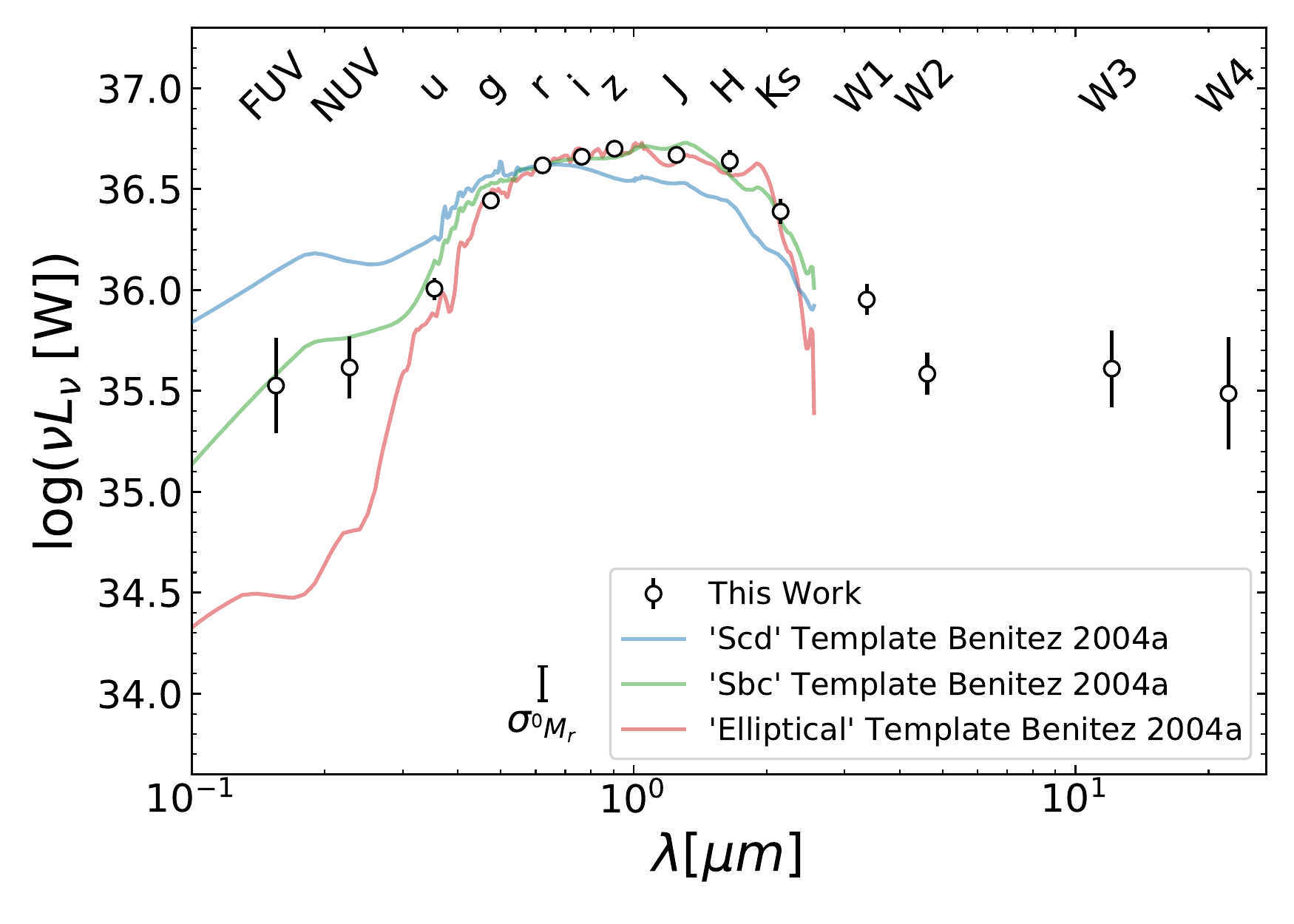}
        \caption{}
        \label{subfig:sed1}
    \end{subfigure}%
    \begin{subfigure}{0.5\linewidth}
        \includegraphics[width=\linewidth]{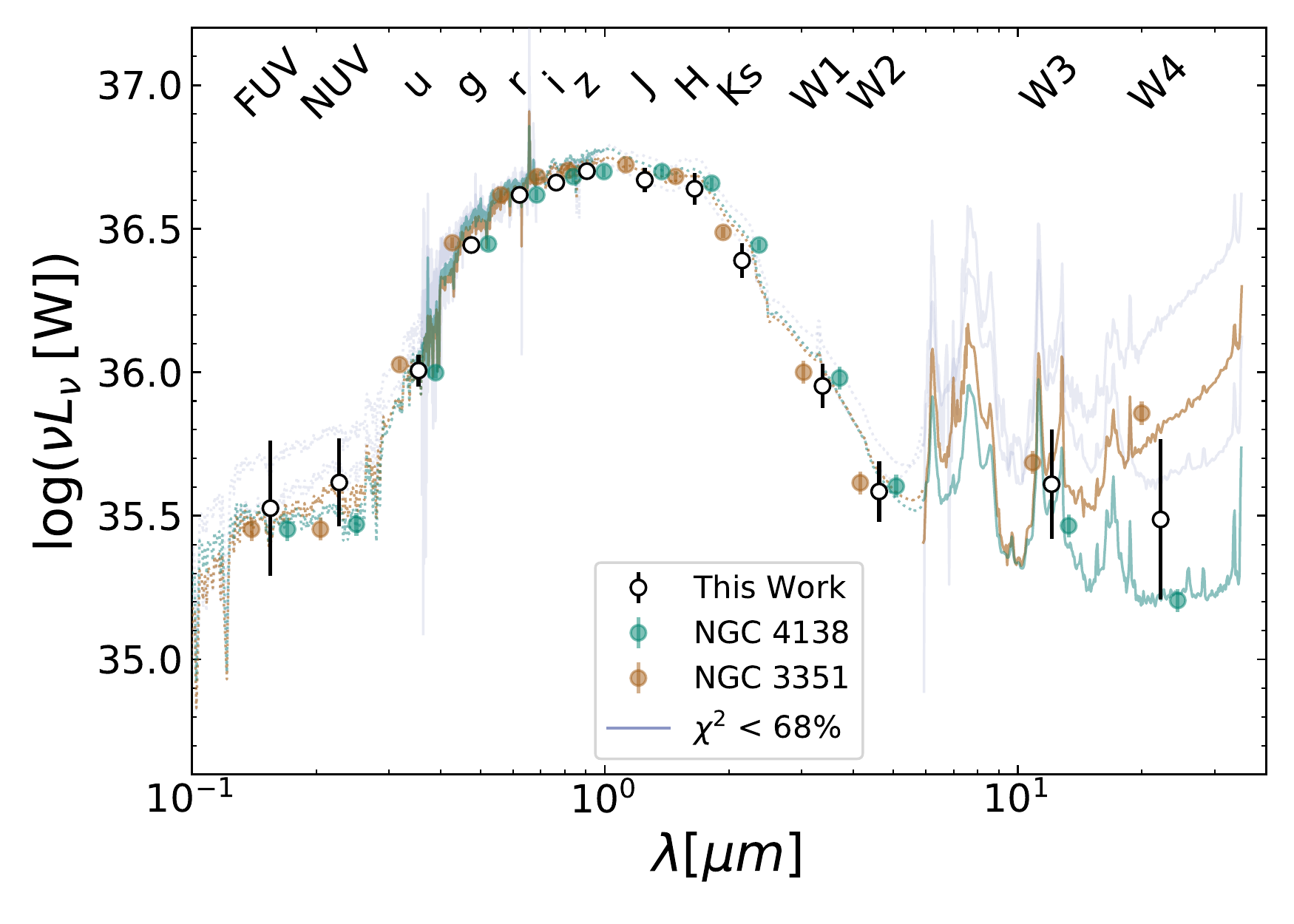}
        \caption{}
        \label{subfig:sed2}
    \end{subfigure}\\%
    \begin{subfigure}{0.5\linewidth}
        \includegraphics[width=\linewidth]{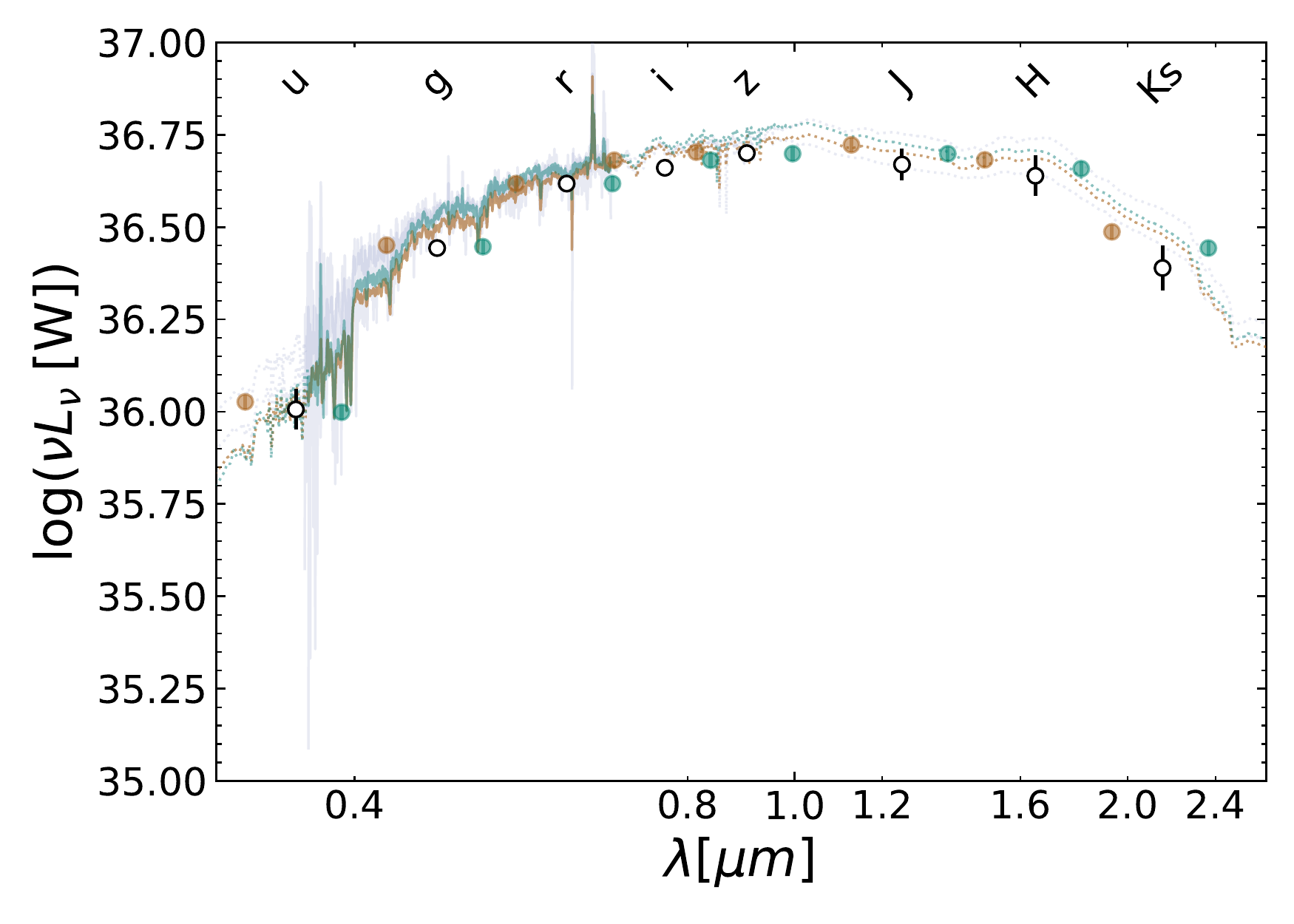}
        \caption{}
        \label{subfig:sed_optical}
    \end{subfigure}%
    \begin{subfigure}{0.5\linewidth}
        \includegraphics[width=\linewidth]{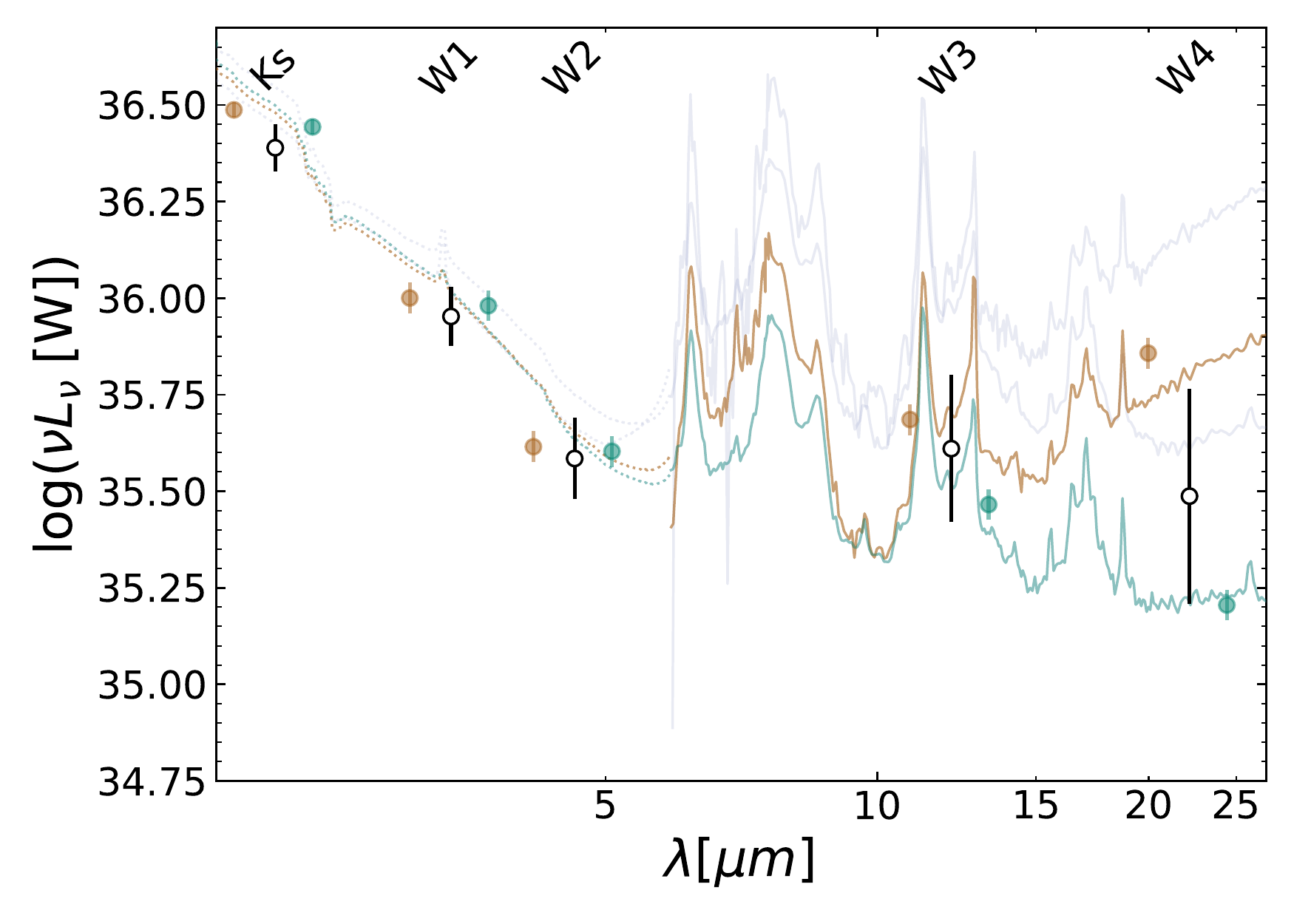}
        \caption{}
        \label{subfig:sed_ir}
    \end{subfigure}%
\caption{The predicted SED for the Milky Way from a six-parameter Gaussian process regression, depicted by black open circles in all panels; the method of calculation is described in \autoref{subsubsection:sed_algorithm}. \textbf{(a)} We compare to empirical galaxy templates from \citet{benitez2004} (re-calibrated from \citet{cww1980}) normalised to match the Milky Way in the $r$-band. The Milky Way SED is consistent with the 'Sbc' template from this set, which is labelled as such because it was originally based on the average of the SEDs of two blue galaxies of morphological type Sbc. 
\textbf{(b)} We compare the SED of the Milky Way to the most closely-matched templates from \citet{brown2014}, which are based on spectra and model fits to bright nearby galaxies. We show all templates whose $\chi^{2}$ values in comparison to the Milky Way SED are below the 95\% upper limit value of $\chi^{2}$ for 13 degrees of freedom; again we normalise in the $r$-band. The two galaxies with the smallest $\chi^{2}$ (which fall below the 68\% upper limit) have their photometry plotted as round points, which are been offset in the  wavelength direction for clarity. Their accompanying spectra from the SED atlas are also plotted, without an offset. 
We also show as fainter curves the SEDs for the remaining two galaxies with potentially matching SEDs, which have $\chi^{2}$ values below the 68\% limit. Images of the four best-fitting galaxies are provided in \autoref{fig:galaxy_tiles}. Portions of the spectra from the SED atlas that are based on models are depicted using dotted lines, while those that are directly based on observations are  shown using solid lines. \textbf{(c)} The optical and near-IR portion of the Milky Way SED, with \citet{brown2014} templates, as in (b). \textbf{(d)} The mid-IR portion of (b). The GPR method produces a Milky Way SED that is consistent in shape with both composite SED templates and individual observed SEDs.%
} 
\label{fig:sed}
\end{figure*}

In \autoref{fig:sed} we present our full predicted spectral energy distribution for the Milky Way, along with a variety of empirical template galaxy spectra from the literature. We plot $\log{\nu L_{\nu}}$ (the power emitted per log interval in frequency) on the vertical axis, in units of log Watts. The horizontal axis corresponds to restframe wavelength in units of $\mu m$; each photometric band used is labelled at its effective wavelength along the top of the plot. The black open circles represent the estimates for the Milky Way's luminosity along with their associated errors, calculated as described in \autoref{subsubsection:sed_algorithm}. We also show the error bar corresponding to the uncertainty in $^{0}M_{r}$ near the bottom of the \autoref{subfig:sed1} for reference. 
The numerical values for the Milky Way SED corresponding to the plotted points are provided in \autoref{Tab:luminsoity}.

\begin{figure}
    \centering
    \includegraphics[width=\linewidth]{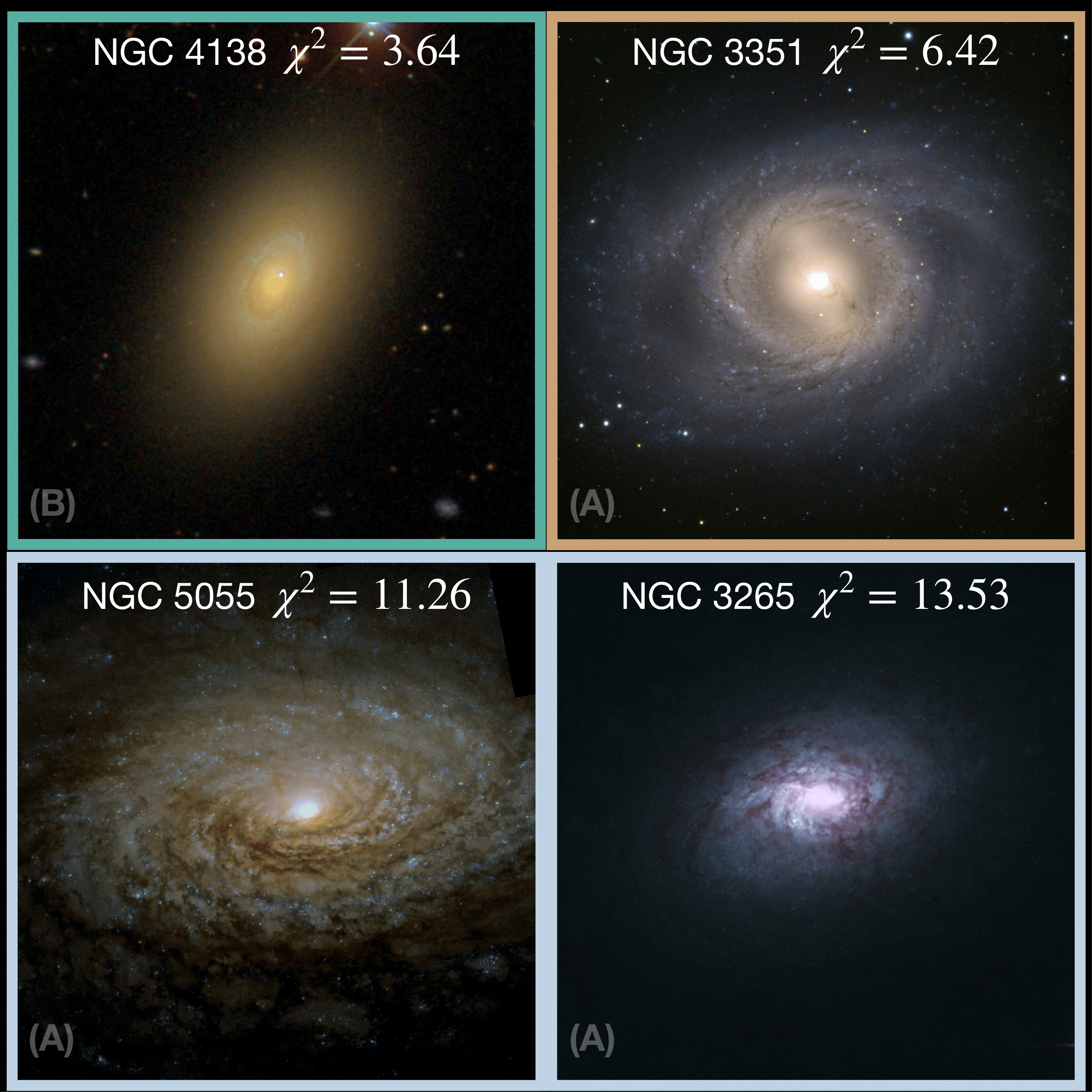}
    \caption{Postage stamp images corresponding to galaxies within the \citet{brown2014} SED atlas with  $\chi^{2}$ values below the 95\% upper limit when compared to the estimated SED of the Milky Way.
    At the top of each image we indicate the galaxy NGC number as well as the $\chi^{2}$ value for the comparison of its SED to the Galactic one; we present them in order from smallest to largest $\chi^{2}$. At the bottom left of each tile is a letter marked (A) or (B) which denote the source of the given image. (A) images are from \citet{ESA_Hubble} and (B) images are from \citet{sdssimg}. 
    The border around each image matches the colour used for its SED in \autoref{subfig:sed2}. 
    It is evident that galaxies that may have similar SEDs to the Milky Way exhibit a broad range of visual morphologies, further emphasising the lack of a unique mapping between morphology and galaxy SED.
    }
    \label{fig:galaxy_tiles}
\end{figure}

While detailed fitting of the Milky Way's SED using physical models lies beyond the scope of this paper, we will compare to observed SEDs of individual galaxies and composite galaxy templates from the literature as a sanity check on the realism of our results. Using photometry for extragalactic samples to constrain the SED of the Milky Way, and then comparing the results to observed galaxy photometry (albeit for different objects) is somewhat circular. However, given that our analysis has treated every band completely independently, there were no guarantees that we should get a sensible SED when combining GPR results across the spectrum.

In \autoref{subfig:sed1} we compare our predicted Milky Way SED to templates from \citet{benitez2004}, which are refinements to the templates from \citet{cww1980} and (for starburst galaxies) \citet{kinney1996}. The \citet{cww1980} templates were based upon averaging the observed SEDs of relatively blue galaxies of a given morphological type. Given the broad range of observed SEDs for objects with similar morphological classification, these should not be considered universally applicable for all galaxies of a given type; however, we use the same labelling as \citet{benitez2004} for consistency with the literature. 

In this and successive plots, we have normalised all templates to match the estimated Milky Way SED in the $r$ band. We normalise in an optical band as those bands have the smallest errors in the predicted SED; which particular optical band we choose has minimal effect on our comparisons. It is apparent that the Milky Way SED is generally consistent with the \citep{benitez2004} ``Sbc'' galaxy template. The galaxies used to construct this template, M51 (NGC 5194) and NGC 2903 are undergoing moderate amounts of star formation, with log specific star formation rates of -10.1 and -10.4 (versus -10.5 for the Milky Way) \citep{munoz-mateos2007}. The Milky Way is generally expected to have an SBbc morphological type (see, e.g., \citet{bland-hawthorn2016} for a recent review on Milky Way structure, as well as \citet{hodge1983,kennicutt2001,efremov2011}), though given how it was constructed, we should not read too much into the agreement of our SED with the \citet{benitez2004} ``Sbc'' template in particular. It is clear, however, that the GPR method yields results that resemble composite SED templates from the literature.

\citet{benitez2004} provides only a sparse set of composite templates that may not match the SEDs of an individual galaxy. Additionally, those templates do not span the full wavelength range of the Milky Way SED we have produced. Thus we also compare to the set of 129 observed galaxy spectral energy distributions from \citet{brown2014}. This SED atlas encompasses a variety of bright galaxies in the very local Universe. Unlike the templates shown in \autoref{subfig:sed1}, these correspond to SEDs of individual galaxies (not averages) with minimal modelling used to interpolate between photometric and spectroscopic coverage.

The data tables from \citet{brown2014} provide extinction-corrected photometry as well as a variety of summary values such as luminosity distance. Because the magnitudes are presented in the AB system we can use the relation:
\begin{equation}
    \log{f_{\nu}} = \frac{m_{AB}-8.9}{-2.5},
\end{equation}
to calculated flux, where $f_{\nu}$ is flux in units of Jansky and $m_{AB}$ is the observed magnitude in each band. We neglect $k$-corrections as these galaxies are very nearby ($3.1 < D_{L} < 249.2 \rm{Mpc}$ at the most extreme), so the corrections are generally negligible. We then can use the flux-luminosity relation $L_{\nu} = 4\pi D_{L}^{2}f_{\nu}$ to calculate the luminosity in each band. Via propagation of errors the uncertainty in $\log{\nu L_{\nu}}$ for these templates is equivalent to $\sigma_{\log{\nu L_{\nu}}} = 0.4\sigma_{m_{AB}}$.

We take a few further steps before comparing the \citet{brown2014} SEDs to our Galaxy's. First, we have obtained the axis ratios (as a proxy for inclination) for 89 out of the 129 galaxies in the \citet{brown2014} sample from the Siena Galaxy Atlas, which have been distributed as part of DESI Legacy imaging surveys Data Release 9. All galaxies with $b/a$ below $0.5$ or unknown axis ratios were excluded from comparisons. Lower axis ratios should correspond to highly-inclined galaxies for which reddening will strongly affect the SED \citep[e.g.,][]{unterborn2008,maller2009}, making them inappropriate comparisons to our face-on SED for the Milky Way. We then normalise the observed SEDs to match the Galactic SED in the $r$-band, as was done for the \citet{benitez2004} templates before. 

Finally, we calculate the $\chi^{2}$ difference between our predicted SED for the Milky Way and each of the galaxy SEDs presented in the \citet{brown2014} atlas. We emphasise that no fitting is performed in this comparison other than matching in the $r$ band. We then calculate $\chi^{2}$ using log quantities, as that is the space in which we perform our predictions and for which errors are (by construction) symmetric:
\begin{equation}
    \chi^{2} = \sum \left( \frac{\log{\nu L_{\nu}^{\rm{atlas}}}-\log{\nu L_{\nu}^{\rm{MW}}}}{\sigma_{\log{\nu L_{\nu}}}} \right)^{2},
\end{equation}
\label{eq:chisq}
where $\sigma_{\log{\nu L_{\nu}}}$ combines in quadrature the total error in the MW SED for a given band, the uncertainties in the \citet{brown2014} photometry, and $\log_{10}$ (1.1), which corresponds to a 10\% error in $\nu L_{\nu}$. This extra error is added to account for systematic uncertainties in the photometry for a given band relative to others; if this were not included, optical bands would dominate the $\chi^{2}$ value due to their small nominal uncertainties. 
We calculate $\chi^{2}$ using the 14 bandpasses in which we have measured the Milky Way's predicted SED, which yields 13 total degrees of freedom (one is lost due to the $r$-band normalisation performed).

\autoref{subfig:sed2} over-plots the \citet{brown2014} SEDs for galaxies whose $\chi^{2}$ values fall within the $68\%$ upper limits for a $\chi^{2}$ distribution with 13 degrees of freedom (corresponding to $\chi^{2}=14.8$). We also examine galaxies that fall below the 95\% limit (with $\chi^{2} < 22.4$), but exclude them from plotting for brevity. Objects below the $68\%$ upper limit (four in total) are clearly consistent with the SED of the Milky Way, while those between the $68$ and $95$ percent limits (comprising three objects) are in some tension with the SED of the Milky Way, but could still be a match. While we calculate $\chi^{2}$ using the broadband photometry for each galaxy, we also plot the full SED from \citet{brown2014} for each of these galaxies for reference. 

In \autoref{subfig:sed2} the two galaxies with the smallest $\chi^{2}$ are labelled and plotted with the highest opacity in teal and gold. For these two galaxies we also plot the observed photometry as points offset in the wavelength direction so they are easier to compare to the Milky Way values. The higher $\chi^2$ objects are plotted with low opacity in pale blue. We also provide more detailed plots of two separate ranges of the SEDs. The optical through near-IR regime is depicted in \autoref{subfig:sed_optical}, while the near- to mid-IR SED is depicted in \autoref{subfig:sed_ir}. Portions of the spectra that are based on observations are plotted with solid lines, while modelled portions are plotted with dotted lines.

We also provide images for the four \citet{brown2014} atlas galaxies that fall below the $68\%$ limit in \autoref{fig:galaxy_tiles}. At the top of each postage stamp image we list the galaxy NGC number, as well as the $\chi^{2}$ value for comparing photometry for that galaxy to our Milky Way SED. The tiles are presented in order from smallest to largest $\chi^{2}$ value. At the bottom left of each tile is a letter marking (A) or (B) which refers to the source for each given image. (A) images are from \citet{ESA_Hubble} and (B) images are from \citet{sdssimg}. 
The borders that surround each image match the colour coding in \autoref{subfig:sed2}. 


It is clear from this comparison that our GPR method produces results whose spectral shape is comparable to observed galaxy SEDs. The small total number of SEDs within the \citet{brown2014} atlas that are consistent with the Milky Way is no surprise; after cutting on inclination we are reduced to only 70 objects, the majority of which are early-type galaxies, leaving a limited number of examples to cover the full range of star-forming galaxies. We emphasise that there remains a need for more careful investigation and modelling of the Milky Way SED we have obtained; this will be the topic of a follow-up paper.

However, we will briefly comment on the galaxies from the \citet{brown2014} atlas whose SEDs are most consistent with the Milky Way. NGC 4138 is an Hubble type SA(r)0 galaxy which contains an AGN and a star-forming ring. Using estimates for mass from \citet{jore1996} ($2.92\times10^{10}\;\rm{M}_{\odot}$) and \citet{kassin2004} ($6.23\times10^{10}\;\rm{M}_{\odot}$) and the star formation rate estimates from \citet{wiegert2014} ($0.14\;\rm{M}_{\odot}\rm{yr}^{-1}$) and \citet{brown2017} ($0.2\;\rm{M}_{\odot}\rm{yr}^{-1}$) yield a log sSFR of $\sim-11.3$ to $-11.5$, significantly smaller than the Milky Way value ($-10.52$). 
NGC 4138 has a $^{0}(g-r)$ colour of $0.73\pm 0.05$ \citet{brown2014}, matching the Milky Way value.

In contrast, NGC 3351/M95 is a galaxy of Hubble type SB(r)b, versus SBb or SBc for the Galaxy; it contains a pseudo-bulge \citep{sandage1994,fisher2009,brown2014}, much as the Milky Way is conjectured to possess \citep[see][and references therein]{bland-hawthorn2016}. According to measurements provided by \citet{leroy2008,george2019} NGC 3351 has a log specific star formation rate of $-10.43$ per year (SFR $=0.940\;\rm{M}_{\odot}\rm{yr}^{-1}$), which is comparable to the log sSFR of the Milky Way of $-10.52$. According to the measurements complied by \citet{brown2014}, NGC 3351 has a $^{0}(g-r)$ colour of $0.74\pm 0.05$, very similar to that of the Milky Way's ($0.73 \pm 0.05$, see \autoref{Tab:color_sed}). It matches the Milky Way SED equally well as NGC 4138 at most wavelengths, with the exception of the longest-wavelength W3 and W4 bands. However, we note that the photometry for NGC 3351 from \citet{dale2017} does not show the strong red slope in the mid-IR seen in the SED atlas spectrum. If the \citet{dale2017} measurements were used in the mid-IR, the agreement with the Milky Way SED would be significantly better.

Two other galaxies in the atlas have SEDs for which the $\chi^2$ value when compared to the Milky Way SED are below the 68\% significance level. 
NGC 5055 is a galaxy of Hubble type SAbc \citep{brown2014} with log sSFR of $-10.53$ as tabulated in \citet{kennicutt2011}. NGC 3265 is of Hubble type SA(rs)0 pec \citep{ann2015} with log sSFR of $-9.12$ \citep{kennicutt2011}. The set of objects whose SEDs are consistent with the Milky Way's span a diverse range of visual morphologies.

\subsubsection{Impact of Physical Parameters on the Estimated SED of the Milky Way}
\label{subsection:parameter_impact}

\begin{figure*}
    \centering
    \includegraphics[width=\linewidth]{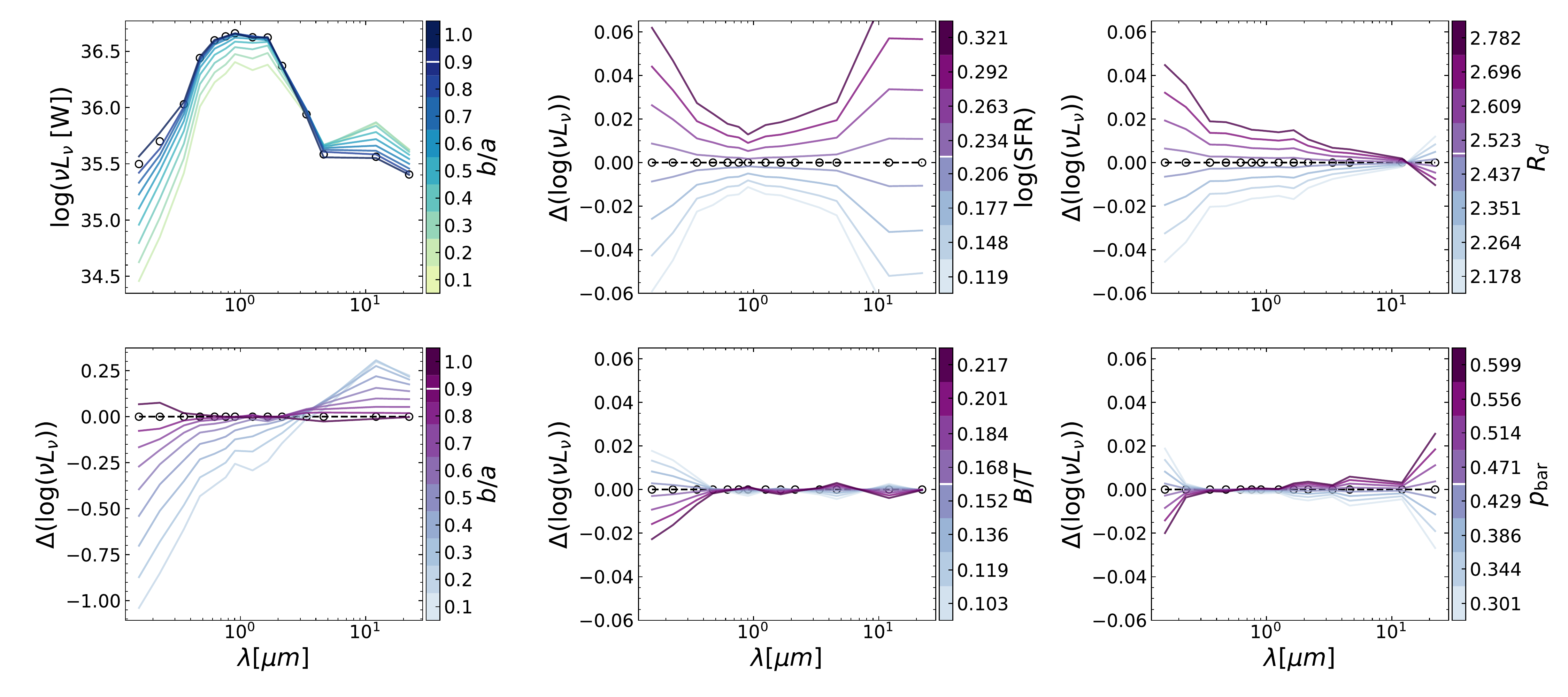}
    \caption{Isolating the contributions of each physical parameter to the SED. Each panels shows the effect of varying one parameter while fixing the other five parameters to the fiducial Milky Way values (see \autoref{subsection:mwparams}). We vary most parameters over a range of $\pm 2\sigma$ from the Milky Way's fiducial values. In the case of axis ratio, we explore a wider range of values to convey better the impact of inclination on the observed SED.
    The upper-left panel depicts the different SEDs that would be inferred for different galaxy axis ratios, with the expected results that an inclined Milky Way would look much fainter than face-on. In the remainder of the panels we depict the log of the predicted SED divided by the SED evaluated for fiducial Milky Way parameters in order to make differences more clear; the plotted quantity is thus $\Delta(\log{(\nu L_{\nu})}) = \log{(\frac{\nu L_{\nu}}{{\nu L_{\nu}^{\rm{MW}}}})}$. The predicted SED for the Milky Way value is plotted with a dashed black line and open circles at the passbands. On the colour bars the Milky Way's value is marked by a horizontal white line. In $b/a$ the GP captures the effects of dust reddening at higher inclinations. If the Milky Way were to have a higher SFR, the SED would be brighter in both the UV and IR. If the Milky Way's disk were more extended we would observe an increased UV brightness and decreased mid-IR brightness. We caution the reader that predictions for disk scale lengths more than $1\sigma$ below the Milky Way value may not be reliable. Relatively few galaxies of the mass of Milky Way have sizes smaller than the Milky Way, causing the GPR
    to be poorly trained for small $R_{d}$ values \citep{licquia2016b}. Changes in the Milky Way's $B/T$ on the SED would be minimal. As we decrease the intensity of the Milky Way's bar it seems to have a minimal effect, with a slightly increased UV brightness. In general these follow our expectations of galaxy evolution which we discuss further in \autoref{subsection:parameter_impact}. %
    }
    \label{fig:varied_sed}
\end{figure*}

Thus far we present results based upon a set of fiducial values (with uncertainties) for the Milky Way's stellar mass, star formation rate, axis ratio, disk scale length, bulge-to-total mass ratio, and bar presence (presented in \autoref{subsection:mwparams}). Because of the generality of the Gaussian process regression fit, we can vary each of these parameters one at a time and test its impact on the inferred SED.

In our previous analyses we randomly drew values from the distributions of Milky Way properties and made predictions based on each of those draws, which were then combined, as described in \autoref{sec:gpr}. This allowed us to incorporate the uncertainties in the Milky Way's physical properties into our results. 
In this subsection, however, we will neglect these uncertainties in order to isolate the effect of changing the central values for each parameter. We have performed a similar analysis to the one presented below by sampling from Milky Way uncertainties as a cross-check; the results are very similar so we do not present them here for brevity.

In this analysis we keep the values for the five Milky Way parameters \textit{not} being studied at their fiducial mean. We then choose a discrete set of values for the sixth parameter at which to evaluate the SED via GPR. In the case of star formation rate, bulge-to-total mass ratio, and bar vote fraction we select 8 values that lie evenly spaced between $\pm2\sigma$ of the fiducial mean value for the Milky Way (inclusive), in addition to evaluating at the nominal value. 
For axis ratio we step through a wide range of possible galaxy axis ratios instead of focusing around $0.9$ in order to capture the full effects of inclination on the Galaxy SED. We have excluded mass from this exercise, as changing mass would radically alter the normalisation of the predicted SED.

For each of the values that we step through for the given test parameter, the GPR is evaluated as before. We apply a $12\sigma$ cutoff on the training sample for all parameters \textit{except} the one being varied. By focusing on one parameter at a time we can explore the impact each has on the predicted SED for the Milky Way and can assess whether the GP is able to capture the expected correlations between galaxy properties.

The results of this analysis are presented in \autoref{fig:varied_sed}.
In the colour bars for each panel the lighter shades correspond to smaller values for the  parameter being varied and darker shades correspond to larger values. In all panels the fiducial value for the Milky Way is marked by a horizontal white line on the colour bar.

The upper left panel shows the GPR-predicted SED for each axis ratio value considered, with axes similar to \autoref{fig:sed}. The SED evaluated with $b/a=0.9$, the fiducial axis ratio value used for the Milky Way, is marked by open points. In this panel one can see the effects of inclination reddening first hand, an effect that has been seen repeatedly in analyses in spiral galaxies \citep[e.g.,][]{shao2007,unterborn2008,maller2009,masters2010}. The cross-section for dust extinction and scattering generally increases with decreasing wavelength, causing reddening effects to be the strongest at the shortest wavelengths. Thus we expect the SEDs of galaxies to appear redder the more inclined they are \citep{xiao2012}. The increased attenuation at higher inclinations for galaxies in the training sample causes the GPR to predict a redder SED as the inclination increases (lower $b/a$). We find a decreased brightness in the UV/optical and an increased brightness in the IR as disks are viewed more edge-on, an effect also observed by SED modellers \citep[see e.g.,][]{noll2009}. 

In the remainder of the panels we plot SEDs \textit{divided by} the SED evaluated at the fiducial Milky Way values, as effects are more subtle and would be difficult to discern in an un-normalised plot. We include a plot of this type based on varying axis ratio in the lower left panel, though we use a larger y-range than the other normalised plots due to the large dynamic range spanned.

An SED which matches the prediction for the fiducial MW values exactly would fall along the horizontal line at $\log{(\nu L_{\nu})}-\log{(\nu L_{\nu}^{\rm{MW}})} = 0$. The photometric predictions for the Milky Way nominal parameters hence correspond to the open circles along this line. 

Star formation rate has the clearest effect on the SED, as seen in the top middle panel. The amount of UV flux is a sensitive indicator of star formation as it is dominated by hot, massive, short-lived stars. These stars contribute to the flux at optical and near-IR wavelengths, but are subdominant there; but in the mid-IR the SED responds strongly to star formation due to light from hot stars that is reprocessed by dust. The GPR predictions reflect all of these phenomena. 

The upper right panel depicts the effect of disk scale length on the predicted SED. We note that the predictions become unstable at shorter scale lengths due to the small number of Milky Way-mass galaxies with radius smaller than our Galaxy in the training samples (cf. \citet{licquia2016b}), so results for disk scale lengths more than $1\sigma$ below the Milky Way value may not be robust. Outside of that regime, it is clear that Milky Way-like galaxies with shorter scale lengths exhibit significantly less flux in the UV than those with longer scale lengths when \mass, SFR, etc. are all held fixed, along with smaller effects at optical-IR wavelengths. 

A smaller disk, with other properties held fixed, could imply that the gas within the disk is denser and dust columns are correspondingly greater. This would in turn cause the SED to look fainter in the UV relative to the IR compared to if the disk were more extended. We have tested the effect of varying $b/a$ and $R_{d}$ simultaneously and find that we can compensate for a change in one with a change in the other almost perfectly. Given the long-standing scenario that inclination effects on the SEDs of disc galaxies are driven by dust \citep[e.g.]{maller2009}, it is reasonable to hypothesize that the effects of varying $R_d$ must relate to varying dust impact as well.



The bottom middle panel shows the effect of varying only the bulge-to-total ratio. Overall, the impact on the SED is small. We see that if the Milky Way were to have a more massive bulge, it would appear slightly fainter in the UV. This could reflect the fact that bulges have older stellar populations than spiral disks; the effect may be subtle here as to first order the effect is captured by variation in specific star formation rate. Bulges are also susceptible to dust reddening \citep{tuffs2004,driver2007}, so a larger bulge may also suffer from greater reddening effects at short wavelengths.

The lower right panel shows the variation in SEDs as we change the bar vote fraction. As this quantity increases we see a decrease in UV brightness and an increase in mid-IR brightness. In general we speculate that most effects caused by a bar may have been captured in the variation with other measured properties (particularly star formation rate). The observed trend with $p_{\rm{bar}}$ could cohere with a narrative of bar-induced star formation suppression \citep[see e.g.,][]{Tubbs1982,haywood2016}, which would affect the UV bands the most and lead to a galaxy looking more red once the bar instability has caused the consumption of cold gas in the disk \citep{masters2010}. However, this would not explain why galaxies with stronger bars are brighter in the mid-IR while simultaneously being fainter in the UV. Alternatively, one could speculate that processes associated with bars could modulate the properties of interstellar dust (e.g., by affecting the ability of gas to cool and form molecular clouds); having more dust (at fixed \mass, SFR, and inclination) should cause a reduction of flux in the UV and an increase in flux in the mid-IR, as observed here.

\subsubsection{Exploration of Other Sources of Physical Parameter Measurements}
\label{subsection:measurement_impact}


As described in \autoref{subsubsection:mpajhu} and \autoref{subsubsection:simard}, there exist multiple options for the values used for the stellar mass, star formation rate, and bulge-to-total ratio for SDSS galaxies. In this subsection we discuss the impact on our results of using different measurements of galaxy parameters or different methods of defining our training samples.

\paragraph{Stellar Mass and Star Formation Rate}

The GSWLC-M2 catalogue \citep{salim2016,salim2018} includes estimates of stellar masses and star formation rates computed based on the photometry within the catalogue. While the \mass\ values in the GSWLC-M2 catalogue closely match those from the MPA-JHU catalogue \citep{brinchmann2004} used for our main results, the SFRs presented in the GSWLC-M2 catalogue are far less bimodal than those presented in the MPA-JHU catalogue; a significant number of galaxies would be classified as star-forming in GSWLC that would be considered quiescent based on the MPA-JHU catalogue. 

We have produced a Milky Way SED using the GSWLC-M2 stellar masses and star formation rates as features in place of the MPA-JHU values, but otherwise following the same methods used to produce the results in \autoref{subsection:mw_comparisons} and \autoref{subsection:sed}. The impact on predictions in the optical and IR is small. The effect is more notable in the UV, where the Milky Way is predicted to be closer to the mean colour of star-forming galaxies with the same sSFR (corresponding to smaller values of $^{0}(FUV-r)$ and $^{0}(NUV-r)$ in \autoref{fig:uv_ssfr}). Even so, the predicted UV colors are still within $0.5\sigma$ of those resulting from using MPA-JHU \mass and SFRs. In addition to being brighter in the NUV and FUV, the predicted SED is also marginally fainter in $W3$ and $W4$ compared to the results presented in \autoref{fig:sed}, with shifts that are again well below $1\sigma$ in each band.

Overall we find that using the stellar masses and star formation rates derived from the GSWLC-M2 catalogue instead of the MPA-JHU catalogue has little effect on our predictions for the Milky Way, and is subdominant to other sources of uncertainty.

\paragraph{Bulge-to-Total Ratio}

There are also multiple options for which band to measure bulge-to-total ratios; the \citet{simard2011} catalogue contains bulge and disk decompositions performed both in the $g$ and $r$ bands. For our main results we use the $r$-band value, $B/T_{r}$, but have also tested the impact of instead using $B/T_{g}$ on our results pesented in \autoref{subsection:mw_comparisons} and \autoref{subsection:sed}. 

Overall we find only very small effects on predicted colours from changing to $B/T_{g}$. All results are well within a few hundredths of a magnitude of the previous values, except in the case of $^{0}(u-r)$. For that value, the predicted value for the Milky Way from the GPR trained on $B/T_{g}$ is larger than when we use $B/T_{r}$ (corresponding to redder color; cf. \autoref{fig:optical_ssfr}), though still within $0.5\sigma$ of our primary result. As a consequence, the predicted SED for the Milky Way using $B/T_{g}$ instead of $B/T_{r}$ is almost identical to before. 


\paragraph{Treatment of Galaxy Zoo 2 Votes}
\label{measurement_impact_gz2}

As discussed in \autoref{subsubsection:gz}, using citizen science votes from Galaxy Zoo 2 requires consideration of responses to previous questions that influence whether the question of interest is even asked. If we wished to select a pure sample of barred galaxies, it would be necessary to ensure not only that a high fraction of people who were asked whether a bar is present voted in the affirmative, but also that a substantial total number of people voted on each of the preceding questions in order to minimise errors in vote fractions (see, e.g., \citealt{willett2013}).
However, with GPR we may gain more information about trends that influence photometric properties by including a broader set of objects, so we do not necessarily wish to exclude non-barred or non-spiral galaxies from the training sample. We explore here how changing the treatment of GZ2 votes influences our GPR predictions.

We have tested what impact restricting the training set to only barred, face-on spirals would have by testing how predicted colours for the Milky Way vary when using training samples with a variety of different constraints: (1) a control sample without any restrictions based on Galaxy Zoo 2 vote results; (2) a sample where if the number of votes on whether or not a galaxy has a bar, $N_{\rm{bar}}$, is less than ten we set $p_{\rm{bar}} = 0$; (3) a sample using the bar selection cuts from \cite{willett2013} Table 3, column 3, rows 2 and 3 in addition to the vote count thresholds mentioned in \autoref{subsubsection:gz}; or (4) a sample using the same cuts as \cite{willett2013} \textit{except} setting $p_{\rm{bar}} = 0$ when $N_{\rm{bar}}<10$, rather than rejecting objects with low $N_{\rm{bar}}$ from the set entirely. 

Applying all of the \cite{willett2013} cuts reduces the size of the training sample by roughly an order of magnitude, degrading the ability of GPR to predict colours and increasing net errors. Furthermore, requiring $N_{\rm{bar}}\geq10$ not only shrinks the size of the sample but also greatly biases the luminosity distribution of the training sample compared to a volume-limited sample, which may result in biases in inferred photometry. 
We have explored how the GPR-predicted Milky Way colours change for each of these four training sample definitions (but otherwise using the  methodologies described in  \autoref{sec:gpr}). The results from (2) compared to our fiducial case, (1), are nearly identical, so the particular values of $p_{bar}$ assigned to objects with poorly-constrained vote counts cannot have had a large systematic impact on our predicted Milky Way photometry. We also find that restricting to training sample (3) or (4) yields much ($\gtrsim 2\times$) larger errors on all predictions. Results from (3) and (4) are still within $1\sigma$ of those from (1) and (2), however. Therefore we conclude that with GPR we get better predictions when we include more objects (including some with noisier vote fractions) than when we instead restrict training to just the best-constrained objects. Therefore we perform \textit{no} GZ2-based cuts on the galaxy sample used for training, and instead include objects spanning the full range of bar, face-on, and features vote fractions in the training set.

\section{Summary and Conclusions}
\label{sec:conclusion}

\subsection{Summary}
\label{subsection:summary}
In this work we have set out to estimate a full SED for the Milky Way, spanning wavelengths from the UV to the IR. Our central motivation is twofold: (1) to improve our understanding of how the Milky Way compares to the general galaxy population and by doing so (2) guide the tuning of parameters in simulations in order to create more realistic galaxies.


The previous work by \citet{licquia2015b} constrained the optical colours and luminosity of the Milky Way using Milky Way analogue galaxies selected based on their stellar mass and star formation rate, obtaining the best constraints on the Milky Way's photometric properties available previous to this work. Here, we have been able to reduce the uncertainties on these constraints further by incorporating information from additional parameters such as disk scale length and bulge-to-total ratio, that also connect to a galaxy's evolutionary history \citep{cappellari2016,saha2018},
and have for the first time developed predictions for Milky Way photometry at wavelengths beyond the optical. 

We have shown that the Milky Way analogue method breaks down when we attempt to match the Galaxy in many physical parameters; the number of Milky Way analogues rapidly approaches zero in higher-dimensional spaces (cf.  \autoref{fig:mwag_sigma}).  
Expanding to a wider wavelength range requires information from datasets that do not cover the full SDSS footprint, making the problem worse. We instead have predicted the photometric properties of the Milky Way using Gaussian process regression, which provides an optimal means of interpolating information from a limited training set. 
We have performed a series of tests throughout this paper that have demonstrated that GPR is able to produce realistic and reliable photometric predictions.


We have compared predictions for the Milky Way to the broader local galaxy population in colour-mass, colour-specific star formation rate, and colour-colour diagrams. As exemplified by \autoref{fig:color_mass}, we obtain similar results in the optical to those reported by \citet{mutch2011} and \citetalias[][]{licquia2015b}, though with reduced errors, further confirming the Milky Way has optical colors consistent with the green valley population.  For the first time we have also predicted UV (\autoref{fig:uv_ssfr}) and IR colours \autoref{fig:ir_color} for the Milky Way, which provide more sensitive diagnostics of the evolutionary status of a galaxy.  We find that in both these regimes the Milky Way appears to lie on the star-forming side of the green valley.

In this work we have determined the luminosity and colours of the Milky Way for GALEX $FUV$ and $NUV$, SDSS $ugriz$, 2MASS $JHKs$, and WISE $W1-W4$ bands in an entirely self-consistent way, giving us unprecedented constraints on its spectral energy distribution. We have constructed the first multi-wavelength SED for the Milky Way. This SED has a shape consistent with both composite galaxy templates (\autoref{subfig:sed1}) and observed SEDs of individual galaxies (\autoref{subfig:sed2}). The GPR method produces a realistic SED with errors and captures previously known galaxy property correlations, such as those between reddening in spiral galaxies and viewing angle or between star-formation rate and UV and IR flux (\autoref{fig:varied_sed}). 
High-resolution hydrodynamical simulators \citep[e.g.,][]{guedes2011,swala2016,wetzel2016} no longer have to compare their mocks of the Milky Way blindly to photometric constraints from broad galaxy populations that span a wide range of properties. Rather, it should now be possible to tune the treatment of star formation efficiency, threshold gas density for star formation, and dust properties to produce galaxies which match the photometric properties of the Milky Way directly, while simultaneously exploiting those properties that we can measure well from inside the Galaxy.

 
\subsection{Discussion: The Milky Way as a Red Spiral} 
\label{subsection:discussion}

As previously suggested in a variety of works (e.g., \citet{salim2014}, \citet{schawinski2014}, and \citet{licquia2015b}), definitions of the green valley that rely only on optical bands may lead to misleading conclusions. The Milky Way has a specific star formation rate that is higher than the canonical values for green valley galaxies, $\log{\rm{sSFR}}$ = $-10.52$ as compared to $\sim -11.8 < \log{\rm{sSFR}} < -10.8$ for transitioning galaxies from \citet{salim2014}, even though it has red optical colours for a star-forming object. However, at UV and IR wavelengths the colours of the Milky Way more clearly place it amongst the star-forming population. This combination of red optical colours when viewed face-on with significant star formation evident at UV and IR wavelengths is characteristic of the previously-identified population of red spiral galaxies.

A population of red spiral galaxies in clusters was first identified by \citet{vandenbergh1976}. Since then, these ``passive spirals'' have been identified at a range of redshifts and in multiple datasets. As noted by to \citet{cortese2012}, for galaxies with a stellar mass above $10^{10}M_{\odot}$ like the Milky Way ($\mass = 5.48^{+1.18}_{-0.94} \times 10^{10}\; M_{\sun}$), the blue cloud and red sequence overlap in their optical colours \citep[this is also consistent with findings by][]{salim2014}. This makes optical photometry a poor choice for constraining the star formation activity for galaxies like our own. However, these massive objects still exhibit a distinct colour bi-modality in the UV, as shown by \citet{wyder2007,salim2014} and is evident from comparing \autoref{fig:optical_ssfr} and \autoref{fig:uv_ssfr}. In comparison to their lower-mass counterparts, massive galaxies produced the great majority of their stars at earlier epochs \citep[e.g.,][]{boselli2001}. This causes the optical colours of massive galaxies to be dominated by relatively old stellar populations as opposed to probing recent star formation activity \citep{wyder2007,chilingarian2012,cortese2012}. \citet{hao2019} and \citet{zhou2021} provide evidence that this is the case for red spirals. Direct or re-radiated light from young stars still dominates the red spiral SEDs at UV and IR wavelengths, however.

Reflecting that, \citet{cortese2012} and \citet{smethurst2015} both find that red spiral galaxies tend to be UV bright; this can be driven by a relatively small amount of total star formation. In order to facilitate comparison of the Milky Way to the red spiral galaxy population, we have over-plotted the red spirals from the \citet{masters2010} catalogue (based on a Galaxy Zoo 2 and optical colour selection) that are also part of our cross-matched galaxy catalogue on all colour-mass, colour-sSFR, and colour-colour diagrams presented here.  In each diagram the Milky Way falls near the middle of the red spiral population.

\citet{cortese2012} notes that $85-90\%$ of objects in their red sample maintain SFRs of $\sim1\rm{M}_{\odot}\;\rm{yr}^{-1}$; \citet{masters2010} found that red spiral galaxies selected from Galaxy Zoo 2 typically had lower rates of ongoing star formation than blue spirals of the same mass, but still non-negligible. 
For comparison the Milky Way has a SFR of  $1.65 \pm 0.19\; \rm{M}_{\sun}\rm{yr}^{-1}$ \citep{licquia2015a}, while the average star formation rate of galaxies of approximately the same mass as the Milky Way ($\pm0.3$ in log stellar mass) with $B/T < 0.75$ (to exclude ellipticals) within our cross-matched galaxy sample is $1.69\; \rm{M}_{\sun}\rm{yr}^{-1}$; the Galaxy is very close to average in this respect. 

\citet{masters2010} also finds that red spiral galaxies have a significantly higher bar fraction compared to blue spirals of the same mass; 70\% versus 27\%. This matches with the clear evidence that the Milky Way possesses a bar \citep[e.g.,][]{blitz1991,bland-hawthorn2016,shen2020}. \citet{masters2010} notes that one possible evolutionary scenario for red spirals is bar-driven gas inflows. This removes gas from the outer disk and funnels it into central star formation, which in turn causes the disk to appear more and more red over time \citep{masters2012,saintonge2012,cheung2013,fraser-mckelvie2020}. 

\citet{masters2011} and \citet{fraser-mckelvie2020} find that barred spirals tend to have redder colours than their unbarred counterparts. It thus may be the case that the bar has played a role in the colours that we observe for the Milky Way in this work. For example, bar quenching may play a role in the development of red spirals, as \citet{bamford2009} finds many high-stellar-mass red spirals in the field and \citet{smethurst2015} notes that field red spirals most likely evolve primarily via secular evolution due to the lack of nearby galaxies. In the case of the Milky Way, work by \citet{haywood2016} and \citet{khoperskov2018} find that the bar may have played a substantial role in the star formation history of the Milky Way (leading to a significant decrease in star formation 9-10 Gyr ago, and thereby causing the observed pattern of chemical abundances in the disk). Although the effect of bar vote fraction in \autoref{fig:varied_sed} is small, it may be that the effects of a bar are primarily captured by other parameters (e.g., SFR).   


We note that \citet{evans2018} studied a population somewhat similar to red spirals, which they labelled ``red misfits''. 
\citet{evans2018} define this population as corresponding to objects with log(sSFR) $> -10.8$ and restframe $g-r > 0.67$ (i.e., specific star formation rate measured to be above the value for the saddle point in the bimodal distribution and colour redder than the saddle point in the colour bimodality). Based on these divisions, the Milky Way almost certainly meets this definition (which is less stringent than most red spiral classifications).   

\subsection{Outlook}

As seen in \autoref{fig:galaxy_tiles}, there is a significant diversity in the set of galaxies that have SEDs consistent with the Milky Way, given the measurement uncertainties in both our results and the Brown SED atlas. The goal of this paper has been to construct the Milky Way's UV-to-IR SED to enable comparisons to samples of external galaxies and to improve the tuning of simulations. However, in a follow-up paper we will fit the estimated SED of the Milky Way using population synthesis models to obtain more detailed constraints on how the star formation history, dust reddening properties, and metallicity of the Galaxy would be interpreted from outside \citep[see e.g.,][]{conroy2013}. This will require proper treatment of covariances between different photometric bands; we will address this by employing multi-output Gaussian process regression in this future work.  
%
%

The longest-wavelength WISE W3 and W4 bands could have substantial discriminating power on what SEDs are consistent with the Milky Way's, if they only had smaller errors, as is evident in \autoref{fig:sed}. However, currently these bands are poorly measured compared to the optical or near-IR; for most objects used in training the Milky Way SED, the signal-to-noise ratio in these bands is below one. Given the low effects of dust extinction in these bands, investigation of the flux ratio (or colour) in these bands across the all-sky WISE imaging, potentially combining modelling of smooth components of the Milky Way with mapping of the contributions from dust, may provide an alternative method to constrain the colour of the Milky Way at the longest wavelengths. If luminosities in the W3 and W4 bands can be measured relative to the luminosity in W2, long-wavelength measurements could be effectively anchored well to the SED presented here; measuring such relative quantities should be affected less by modelling uncertainties than absolute measurements would be.

The SED presented in the paper (or future improved versions) can be used to identify multiwavelength Milky Way analogue galaxies by matching in unresolved photometric properties. If we do not need to require detailed morphological measurements or citizen science inspection of images it would greatly increase the size of the parent catalogues that could be used to identify MWAs, which could be useful for a variety of follow-up studies such as determining gas masses for the Milky Way or studying environments of Milky Way-like galaxies. 

The Milky Way appears to be atypical in its satellite population and mass assembly history. For example, \citet{evans2020} finds that the assembly history of the Milky Way is only reproduced in $0.65\%$ of Milky Way-mass EAGLE galaxies. This assembly history should be closely related to the local environment surrounding our Galaxy, and environment has been found to play a key role in the formation and evolution of galaxies \citep[e.g.,][]{vandenbosch2008,peng2012,bluck2016}.

In future work we plan to explore how incorporating measures of galaxy environment (e.g., measures of the local overdensity of galaxies) within a GPR model affect the predicted SED of the Milky Way.  The noisiness of environment measures \citep{hogg2004} and the impact of SDSS fiber collisions on Local Group-like systems (as typically only one galaxy out of two close neighbours would be observed, causing analogues of a Milky Way-M31 pair to be missed) may limit the information that may be gained from this, however.
While we anticipate the environment to have a small impact on the Galactic SED compared to the dominant effects of stellar mass and star formation rate on galaxy colours \citep[e.g.,][]{grutzbauch2011}, assessing the local environments of the most Milky Way-like galaxies may allow us to explore and to what extent our Galaxy's environment has shaped its exhibited characteristics.

Gaussian process regression can be useful for a variety of studies beyond the photometric estimates for the Milky Way considered here. For this reason the authors have provided their analysis code on \href{https://github.com/cfielder/GPR-for-Photometry}{our project GitHub} for full public access for adaption to any other project, under a CC BY-SA 4.0 license. 


\section*{Acknowledgements}

The authors thank Amelia Fraser-McKelvie for helpful discussion and providing her own WISE $k$-correction code that we used as a starting point before developing our own. We thank Michael Brown and Richard Beare for helpful discussions related to the $k$-correction methods applied in this paper. We also thank Rachel Benzanson for helpful discussions and suggestion throughout the course of this work. Lastly, the authors would like to thank the referee for helpful comments and suggestions.

We gratefully acknowledge support from NASA Astrophysics Data Analysis Program grant number 80NSSC19K0588 which made this research possible.

This work made use of \textsc{Python}, along with many community-developed or maintained software packages, including IPython \citep{ipython}, Jupyter (\http{jupyter.org}), Matplotlib \citep{matplotlib}, NumPy \citep{numpy}, Pandas \citep{pandas}, scikit-learn \citep{scikit-learn}, and SciPy \citep{SciPy2020}.
This research made use of NASA's Astrophysics Data System for bibliographic information.

This paper was based in part on observations made with the NASA Galaxy Evolution Explorer. GALEX is operated for NASA by the California Institute of Technology under NASA contract NAS5-98034.

This publication makes use of data products from SDSS-IV. Funding for the Sloan Digital Sky 
Survey IV has been provided by the 
Alfred P. Sloan Foundation, the U.S. 
Department of Energy Office of 
Science, and the Participating 
Institutions. 

SDSS-IV acknowledges support and 
resources from the Center for High 
Performance Computing  at the 
University of Utah. The SDSS 
website is www.sdss.org.

SDSS-IV is managed by the 
Astrophysical Research Consortium 
for the Participating Institutions 
of the SDSS Collaboration including 
the Brazilian Participation Group, 
the Carnegie Institution for Science, 
Carnegie Mellon University, Center for 
Astrophysics | Harvard \& 
Smithsonian, the Chilean Participation 
Group, the French Participation Group, 
Instituto de Astrof\'isica de 
Canarias, The Johns Hopkins 
University, Kavli Institute for the 
Physics and Mathematics of the 
Universe (IPMU) / University of 
Tokyo, the Korean Participation Group, 
Lawrence Berkeley National Laboratory, 
Leibniz Institut f\"ur Astrophysik 
Potsdam (AIP),  Max-Planck-Institut 
f\"ur Astronomie (MPIA Heidelberg), 
Max-Planck-Institut f\"ur 
Astrophysik (MPA Garching), 
Max-Planck-Institut f\"ur 
Extraterrestrische Physik (MPE), 
National Astronomical Observatories of 
China, New Mexico State University, 
New York University, University of 
Notre Dame, Observat\'ario 
Nacional / MCTI, The Ohio State 
University, Pennsylvania State 
University, Shanghai 
Astronomical Observatory, United 
Kingdom Participation Group, 
Universidad Nacional Aut\'onoma 
de M\'exico, University of Arizona, 
University of Colorado Boulder, 
University of Oxford, University of 
Portsmouth, University of Utah, 
University of Virginia, University 
of Washington, University of 
Wisconsin, Vanderbilt University, 
and Yale University.

This publication makes use of data products from the Two Micron All Sky Survey, which is a joint project of the University of Massachusetts and the Infrared Processing and Analysis Center/California Institute of Technology, funded by the National Aeronautics and Space Administration and the National Science Foundation.

This publication makes use of data products from the Wide-field Infrared Survey Explorer, which is a joint project of the University of California, Los Angeles, and the Jet Propulsion Laboratory/California Institute of Technology, funded by the National Aeronautics and Space Administration.

The Legacy Surveys consist of three individual and complementary projects: the Dark Energy Camera Legacy Survey (DECaLS; Proposal ID \#2014B-0404; PIs: David Schlegel and Arjun Dey), the Beijing-Arizona Sky Survey (BASS; NOAO Prop. ID \#2015A-0801; PIs: Zhou Xu and Xiaohui Fan), and the Mayall z-band Legacy Survey (MzLS; Prop. ID \#2016A-0453; PI: Arjun Dey). DECaLS, BASS and MzLS together include data obtained, respectively, at the Blanco telescope, Cerro Tololo Inter-American Observatory, NSF’s NOIRLab; the Bok telescope, Steward Observatory, University of Arizona; and the Mayall telescope, Kitt Peak National Observatory, NOIRLab. The Legacy Surveys project is honored to be permitted to conduct astronomical research on Iolkam Du’ag (Kitt Peak), a mountain with particular significance to the Tohono O’odham Nation.  NOIRLab is operated by the Association of Universities for Research in Astronomy (AURA) under a cooperative agreement with the National Science Foundation.

BASS is a key project of the Telescope Access Program (TAP), which has been funded by the National Astronomical Observatories of China, the Chinese Academy of Sciences (the Strategic Priority Research Program “The Emergence of Cosmological Structures” Grant \# XDB09000000), and the Special Fund for Astronomy from the Ministry of Finance. The BASS is also supported by the External Cooperation Program of Chinese Academy of Sciences (Grant \# 114A11KYSB20160057), and Chinese National Natural Science Foundation (Grant \# 11433005).

The Legacy Survey team makes use of data products from the Near-Earth Object Wide-field Infrared Survey Explorer (NEOWISE), which is a project of the Jet Propulsion Laboratory/California Institute of Technology. NEOWISE is funded by the National Aeronautics and Space Administration.

The Legacy Surveys imaging of the DESI footprint is supported by the Director, Office of Science, Office of High Energy Physics of the U.S. Department of Energy under Contract No. DE-AC02-05CH1123, by the National Energy Research Scientific Computing Center, a DOE Office of Science User Facility under the same contract; and by the U.S. National Science Foundation, Division of Astronomical Sciences under Contract No. AST-0950945 to NOAO.
This project used data obtained with the Dark Energy Camera (DECam), which was constructed by the Dark Energy Survey (DES) collaboration. Funding for the DES Projects has been provided by the U.S. Department of Energy, the U.S. National Science Foundation, the Ministry of Science and Education of Spain, the Science and Technology Facilities Council of the United Kingdom, the Higher Education Funding Council for England, the National Center for Supercomputing Applications at the University of Illinois at Urbana-Champaign, the Kavli Institute of Cosmological Physics at the University of Chicago, Center for Cosmology and Astro-Particle Physics at the Ohio State University, the Mitchell Institute for Fundamental Physics and Astronomy at Texas A\&M University, Financiadora de Estudos e Projetos, Fundacao Carlos Chagas Filho de Amparo, Financiadora de Estudos e Projetos, Fundacao Carlos Chagas Filho de Amparo a Pesquisa do Estado do Rio de Janeiro, Conselho Nacional de Desenvolvimento Cientifico e Tecnologico and the Ministerio da Ciencia, Tecnologia e Inovacao, the Deutsche Forschungsgemeinschaft and the Collaborating Institutions in the Dark Energy Survey. The Collaborating Institutions are Argonne National Laboratory, the University of California at Santa Cruz, the University of Cambridge, Centro de Investigaciones Energeticas, Medioambientales y Tecnologicas-Madrid, the University of Chicago, University College London, the DES-Brazil Consortium, the University of Edinburgh, the Eidgenossische Technische Hochschule (ETH) Zurich, Fermi National Accelerator Laboratory, the University of Illinois at Urbana-Champaign, the Institut de Ciencies de l’Espai (IEEC/CSIC), the Institut de Fisica d’Altes Energies, Lawrence Berkeley National Laboratory, the Ludwig Maximilians Universitat Munchen and the associated Excellence Cluster Universe, the University of Michigan, NSF’s NOIRLab, the University of Nottingham, the Ohio State University, the University of Pennsylvania, the University of Portsmouth, SLAC National Accelerator Laboratory, Stanford University, the University of Sussex, and Texas A\&M University.

\section*{Data Availability}

The data used in this article is provided publicly on \href{https://github.com/cfielder/Catalogs}{our catalogue GitHub}. We include both the original volume-limited sample of \citetalias{licquia2015b} and the cross-matched final sample used in this work. The ReadMe also provides a detailed description of the columns within these tables, which is also provided by each respective catalogue that our data originates from.



\bibliographystyle{mnras}
\bibliography{refs} 


\pagenumbering{arabic}
\renewcommand*{\thepage}{A\arabic{page}}
\appendix

\section{Data Summary}
\label{sec:appendix_data}

\subsection{Galaxy Properties}
\label{subsection:corner}

In this subsection we summarise the distribution of properties of the galaxy sample used for training the GPR (which is described in \autoref{sec:data}). This sample consists of cross-matches between the SDSS DR8 volume-limited sample reported in \citetalias{licquia2015b}, the MPA-JHU catalogue of galaxy stellar masses and star formation rates \citep{brinchmann2004}, the \citet{simard2011} morphological catalogue, the \citet{salim2016,salim2018} GSWLC-M2 photometric catalogue, the DESI Legacy imaging survey DR8 \citep{legacy2019}, and the Galaxy Zoo 2 catalogue \citep{willett2013,hart2016}.

\begin{figure*}
    \centering
    \includegraphics[width=\linewidth]{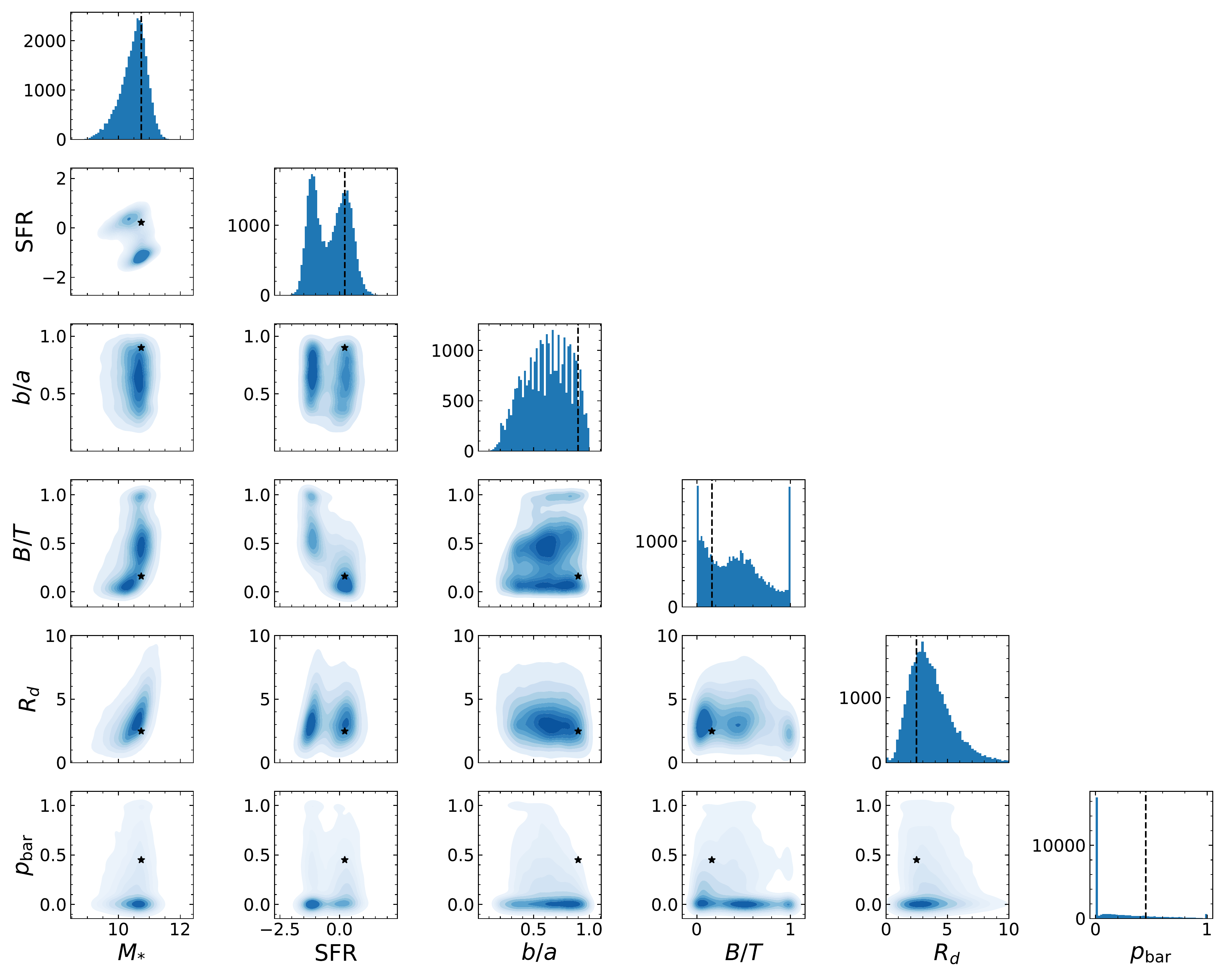}
    \caption{Distribution of the various galaxy physical properties used to predict the Milky Way SED, viewed as both two-dimensional projections and one-dimensional histograms. In the contour plots the fiducial values for the Milky Way are designated by a black star, while in the histograms the fiducial value is indicated by a vertical black dashed line. The covariances are weak in almost all parameter combinations, except for the well-known \mass - $R_{d}$ (mass--size) relation, the bimodal distribution of galaxies in the \mass-SFR plane, and weaker correlations of bulge-to-total ratio and bar vote fraction with star formation rate.}
    \label{fig:corner}
\end{figure*}

In \autoref{fig:corner} we present a corner plot for the distributions of the various galaxy physical properties used to train the GPR: mass (\mass), star formation rate (SFR), axis ratio ($b/a$), bulge-to-total ratio ($B/T$), disk scale length ($R_{d}$), and bar vote fraction ($p_{\rm{bar}}$). In each contour plot the fiducial value for the Milky Way (from \autoref{subsection:mwparams}) is marked by a black star, while in the histograms the fiducial value is marked by a vertical dashed black line. Overall we can conclude that covariances are fairly weak for most parameter pairs, with the exceptions of the well-known \mass - $R_{d}$ (or mass--size) relation, the bimodal distribution of objects in the \mass-SFR plane (corresponding to the red sequence/blue cloud division), and weaker correlations between galaxy bulge-to-total ratio or bar vote fraction and star formation rate.

\subsection{Tabulated Photometric Predictions for the Milky Way}
\label{subsection:tables}

In this section we present the estimates of the colours and luminosities for the Milky Way which we have obtained via Gaussian process regression. \autoref{Tab:color_sed}, \autoref{Tab:color_other}, and \autoref{Tab:absmag} update Table 1 and 3 from in \citet{licquia2015b}, and also incorporate results for the non-SDSS bands used in this work. In addition, \autoref{Tab:color_sed} tabulates the colours used to derive the luminosities presented in \autoref{Tab:luminsoity}, a necessary ingredient for our SED calculation.

In \autoref{Tab:absmag} all absolute magnitudes are calculated assuming a Hubble constant of $H_{0} = 100 h \rm{km}\;\rm{s}^{-1}\;\rm{Mpc}^{-1}$. Results for $FUV,\; NUV,\; ugriz,\; JHKs,$ and $W1-W4$ are all presented in the AB magnitude system, as in the body of the paper. The Johnsons-Cousins $UBVRI$ values are provided in the Vega magnitude system. These were converted from the $ugriz$ measurements via the \textsc{kcorrect v4.2} software on an object-by-object basis, as in \citetalias[][]{licquia2015b}. We remind the reader that absolute magnitudes are computer with $5$ physical parameters instead of 6, due to the increased cumulative variance when $p_{\rm{bar}}$ is incorporated in the GPR for them. Therefore we do $not$ include derivatives with respect to $p_{\rm{bar}}$ for them, as bar vote fraction was not used in the predictions of these quantities.

All tabulated results are presented for rest-frame $z=0$ passbands. We note that each table row is calculated independently from all others; for example, $^{0}(FUV-r)$ is not calculated from subtracting the predicted $^{0}M_{r}$ from the predicted $^{0}M_{FUV}$, but rather from a Gaussian process regression that predicts $^{0}(FUV-r)$ directly. This will in general yield smaller errors for colour calculations, as some of the errors in absolute magnitudes from different bands will be covariant. We derive colour predictions from $\tt{model}$ magnitudes while absolute magnitudes are determined from $\tt{cmodel}$ magnitudes, again to minimise errors. 

In these tables, the ``Corrected Value'' column corresponds to the predicted colour or absolute magnitude from the GPR, derived as the mean predicted value from $1,000$ draws from the fiducial distributions of Milky Way physical properties (as described in \autoref{sec:gpr}), after Eddington bias has been subtracted. The amount subtracted is tabulated for reference in the ``Bias Removed'' column. The errors in the ``Corrected Value'' columns have had the uncertainty in the Eddington bias added in quadrature and represent $1\sigma$ errors.

We also tabulate the derivatives of each photometric property with respect to every galaxy physical property used for the prediction, allowing the photometry presented here to be updated for values of Milky Way parameters that differ from the fiducial values used in this work (or, conversely, to correct for a re-calibration of extragalactic values). The methods applied are directly adapted from those presented in Section 5 of \citetalias[][]{licquia2015b}, but we provide a brief summary here. 

To calculate the derivatives we offset each of the Galaxy's measured properties by $\pm0.1$ multiplied by the fiducial error in that property ($\sigma$). We then evaluate the GPR using these ``offset'' properties, rather than performing random draws from the Milky Way distributions. For example, in order to calculate $\partial(^{0}M_{r})/\partial(\log{\rm{SFR}})$ we would query the GPR for predictions of the Milky Way values at $\log{\rm{SFR}}-\sigma_{\rm{SFR}}/10$, the nominal $\log{\rm{SFR}}$, and $\log{\rm{SFR}}+\sigma_{\rm{SFR}}/10$, while all other parameters (stellar mass, disk scale length, etc.) are held at the fiducial Milky Way values. This yields three total predictions, two with offset SFR values and one at the nominal SFR. We then use these three values to calculate the derivative using the three-point Lagrangian interpolation method. Because these derivatives are small, they are sensitive to the limited training sample size required for the $\tt{scikit-learn}$ implementation of GPR (refer to \autoref{subsection:comp_challenges} for details). In order to mitigate this effect we repeat this process for ten different training sets, re-evaluating the GPR prediction at the offset physical parameter and nominal value each time and calculating the derivative again. The derivatives presented here for each parameter are then computed as the average of the ten derivative values.

\begin{table*}
\centering
\resizebox{\textwidth}{!}{\begin{tabular}{lrrrrrrrr}
\hline
\hline
 Property    &   Corrected Value &  Bias Removed & $\partial/\partial (\log{\mass})$ &   $\partial/\partial(\log{\rm{SFR}})$ &   $\partial/\partial (b/a)$ &   $\partial/\partial R_{d}$ &   $\partial/\partial (B/T)$ &   $\partial/\partial p_{\rm{bar}}$ \\
\hline
$^{0}(FUV-r)$ &              $4.24 \pm 0.59$ &           $0.16 \pm 0.05$ &                          1.70  &                              -1.03 &                   -1.50  &                     -0.15 &                    0.51 &                        0.14 \\
$^{0}(NUV-r)$ &              $3.60  \pm 0.38$ &           $0.16 \pm 0.05$ &                          1.34 &                              -0.85 &                   -1.13 &                     -0.12 &                    0.76 &                        0.09 \\
$^{0}(u-r)$   &              $2.14 \pm 0.14$ &           $0.06 \pm 0.01$ &                          2.02 &                              -0.96 &                   -0.58 &                     -0.29 &                   -0.19 &                       -0.05 \\
$^{0}(g-r)$   &              $0.73 \pm 0.04$ &           $0.05 \pm 0.01$ &                          0.13 &                              -0.13 &                   -0.09 &                     0.00    &                    0.07 &                        0.01 \\
 $^{0}(i-r)$   &             $-0.33 \pm 0.03$ &           $0.03 \pm 0.00$    &                         -0.08 &                               0.05 &                    0.06 &                      0.01 &                   -0.03 &                       0.00    \\
 $^{0}(z-r)$   &             $-0.61 \pm 0.06$ &           $0.04 \pm 0.01$ &                         -0.16 &                               0.09 &                    0.14 &                      0.01 &                   -0.03 &                       -0.02 \\
 $^{0}(J-r)$   &             $-0.89 \pm 0.11$ &           $0.02 \pm 0.01$ &                         -0.24 &                               0.05 &                    0.16 &                      0.03 &                    0.03 &                       -0.01 \\
 $^{0}(H-r)$   &             $-1.11 \pm 0.14$ &           $0.03 \pm 0.01$ &                         -0.28 &                               0.05 &                    0.09 &                      0.03 &                    0.02 &                       -0.03 \\
 $^{0}(Ks-r)$  &             $-0.78 \pm 0.15$ &           $0.02 \pm 0.02$ &                         -0.29 &                               0.05 &                    0.20  &                      0.03 &                    0.03 &                       -0.02 \\
 $^{0}(W1-r)$  &             $-0.17 \pm 0.19$ &           $0.02 \pm 0.01$ &                         -0.30  &                              -0.12 &                    0.44 &                      0.09 &                   -0.09 &                       -0.03 \\
 $^{0}(W2-r)$  &              $0.41 \pm 0.26$ &           $0.00    \pm 0.01$ &                         -0.25 &                              -0.19 &                    0.67 &                      0.11 &                   -0.11 &                       -0.03 \\
 $^{0}(W3-r)$  &             $-0.30  \pm 0.48$ &           $0.42 \pm 0.04$ &                          0.40  &                              -1.18 &                    0.54 &                      0.12 &                    0.06 &                        0.07 \\
 $^{0}(W4-r)$  &             $-1.01 \pm 0.70$  &          $ 0.06 \pm 0.03$ &                          1.08 &                              -1.54 &                    0.84 &                      0.23 &                   -0.19 &                       -0.38 \\
\hline
\end{tabular}}
\caption{$X-r$ colour estimates for each photometric band $X$ used in this work; these values are used to calculate the luminosities presented in \autoref{Tab:luminsoity}, and have been determined by the methods described in \autoref{sec:gpr}. The corrected value column refers to our final estimate after accounting for Eddington bias (cf.  \autoref{subsection:eddbias}). Colours presented here are in the AB magnitude system.}
\label{Tab:color_sed}
\end{table*}

\begin{table*}[h]
\centering
\resizebox{\textwidth}{!}{\begin{tabular}{lrrrrrrrrrr}
\hline
\hline
 Property    &   Corrected Value &  Bias Removed & $\partial/\partial(\log{\mass})$ &   $\partial/\partial(\log{\rm{SFR}})$ &   $\partial/\partial (b/a)$ &   $\partial/\partial R_{d}$ &   $\partial/\partial (B/T)$ &   $\partial/\partial p_{\rm{bar}}$ \\
\hline
$^{0}(u-g)$    &              $1.49 \pm 0.11$ &           $0.12 \pm 0.01$ &                          0.65 &                              -0.47 &                   -0.40  &                     -0.15 &                   -0.35 &                       -0.02 \\
 $^{0}(r-i)$   &              $0.30  \pm 0.03$ &           $0.01 \pm 0.00$    &                          0.07 &                              -0.04 &                   -0.05 &                     -0.01 &                    0.02 &                        0.00    \\
 $^{0}(i-z)$   &              $0.27 \pm 0.03$ &          $0.01 \pm 0.00$    &                          0.06 &                              -0.03 &                   -0.11 &                     0.00    &                    0.01 &                        0.00    \\
 $^{0}(U-V)$   &              $1.11 \pm 0.12$ &           $0.13 \pm 0.01$ &                          0.42 &                              -0.41 &                   -0.24 &                      0.08 &                   -0.25 &                        0.02 \\
 $^{0}(U-B)$   &              $0.30  \pm 0.08$ &           $0.12 \pm 0.02$ &                          0.69 &                              -0.28 &                   -0.39 &                      0.04 &                   -0.68 &                        0.09 \\
 $^{0}(B-V)$   &              $0.80  \pm 0.05$ &           $0.04 \pm 0.01$ &                          0.17 &                              -0.16 &                   -0.08 &                     -0.01 &                    0.08 &                        0.01 \\
 $^{0}(V-R)$   &              $0.54 \pm 0.03$ &           $0.00    \pm 0.00$    &                          0.06 &                              -0.04 &                   -0.06 &                     0.00    &                    0.03 &                        0.01 \\
 $^{0}(R-I)$   &              $0.60  \pm 0.04$ &           $0.01 \pm 0.00$    &                          0.08 &                              -0.06 &                   -0.09 &                     -0.01 &                    0.02 &                        0.01 \\
 $^{0}(W1-W2)$ &             $-0.57 \pm 0.08$ &           $0.00    \pm 0.01$ &                          0.03 &                               0.10  &                   -0.07 &                     -0.03 &                    0.01 &                        0.02 \\
 $^{0}(W2-W3)$ &              $0.91 \pm 0.33$ &           $0.00    \pm 0.02$ &                         -0.33 &                               0.92 &                   -0.08 &                     -0.02 &                   -0.38 &                       -0.02 \\
 $^{0}(W3-W4)$ &              $0.40  \pm 0.49$ &           $0.02 \pm 0.01$ &                         -0.65 &                               0.09 &                    0.25 &                     -0.02 &                    0.21 &                        0.34 \\
\hline
\end{tabular}}
\caption{Additional colour estimates for the Milky Way, updating values from Table 1 and Table 3 of \citetalias[][]{licquia2015b} in addition to values for WISE colours (as plotted in \autoref{fig:ir_color}). The SDSS $ugriz$ and WISE $W1 - W4$ values are presented in AB magnitudes. The Johnsons-Cousins $UBVRI$ bands have been converted from the $ugriz$ measurements via \textsc{kcorrect} on an object-by-object basis, as in \citetalias[][]{licquia2015b}, and are presented in the Vega magnitude system.}
\label{Tab:color_other}
\end{table*}

\begin{table*}[h]
\centering
\resizebox{\textwidth}{!}{\begin{tabular}{lrrrrrrrrr}
\hline
\hline
 Property & Corrected Value & Bias Removed & $\partial/\partial (\log{\mass})$ &   $\partial/\partial(\log{\rm{SFR}})$ &   $\partial/\partial(b/a)$ &   $\partial/\partial R_{d}$ &   $\partial/\partial (B/T)$ \\
\hline
$^{0}M_{FUV}-5logh$ &            $-16.34 \pm 0.66$ &           $0.37 \pm 0.06$ &                          -0.04 &                                -1.31 &                   -1.84 &                     -0.27 &                    0.71   \\
 $^{0}M_{NUV}-5logh$ &            $-17.12 \pm 0.49$ &           $0.25 \pm 0.06$ &                          -0.23 &                                -1.12 &                   -1.52 &                     -0.26 &                    0.48   \\
 $^{0}M_{u}-5logh$   &            $-18.53  \pm 0.33$ &           $0.24 \pm 0.03$ &                          -1.69 &                                -1.14 &                   -1.08 &                     -0.35 &                    0.36  \\
 $^{0}M_{g}-5logh$    &            $-19.91  \pm 0.24$ &           $0.17 \pm 0.02$ &                          -1.24 &                                -0.52 &                   -0.7  &                     -0.15 &                   -0.04 \\
 $^{0}M_{r}-5logh$    &            $-20.68 \pm 0.53$ &           $0.23 \pm 0.03$ &                          -1.29 &                                -0.52 &                   -0.52 &                     -0.14 &                   -0.17 \\
 $^{0}M_{i}-5logh$    &            $-20.95 \pm 0.21$ &           $0.10  \pm 0.02$ &                          -1.34 &                                -0.41 &                   -0.35 &                     -0.16 &                    0.01 \\
 $^{0}M_{z}-5logh$    &            $-21.34 \pm 0.24$ &           $0.14 \pm 0.03$ &                          -1.48 &                                -0.38 &                   -0.33 &                     -0.14 &                    0.13  \\
 $^{0}M_{U}-5logh$    &            $-19.47    \pm 0.39$ &           $0.24 \pm 0.03$ &                          -0.48 &                                -1.02 &                   -1.11 &                     -0.36 &                   -0.59  \\
 $^{0}M_{B}-5logh$    &            $-19.59  \pm 0.25$ &           $0.16 \pm 0.02$ &                          -1.15 &                                -0.65 &                   -0.58 &                     -0.16 &                   0.00     \\
 $^{0}M_{V}-5logh$    &            $-20.48  \pm 0.24$ &           $0.09 \pm 0.02$ &                          -1.24 &                                -0.50  &                   -0.46 &                     -0.18 &                   -0.06  \\
 $^{0}M_{R}-5logh$    &            $-20.91 \pm 0.21$ &           $0.04 \pm 0.02$ &                          -1.34 &                                -0.46 &                   -0.53 &                     -0.18 &                   -0.09  \\
 $^{0}M_{I}-5logh$    &            $-21.65 \pm 0.2$ &           $0.02 \pm 0.02$ &                          -1.28 &                                -0.42 &                   -0.29 &                     -0.13 &                    0.06  \\
 $^{0}M_{J}-5logh$    &            $-21.65 \pm 0.20$ &           $0.01 \pm 0.02$ &                          -1.72 &                                -0.40  &                   -0.35 &                     -0.12 &                   -0.11  \\
 $^{0}M_{H}-5logh$    &            $-21.75 \pm 0.21$ &           $0.06 \pm 0.02$ &                          -1.85 &                                -0.39 &                   -0.18 &                     -0.10  &                   -0.10   \\
 $^{0}M_{Ks}-5logh$    &            $-21.56 \pm 0.20$ &           $0.07 \pm 0.03$ &                          -1.87 &                                -0.36 &                   -0.15 &                     -0.10  &                    0.02   \\
 $^{0}M_{W1}-5logh$   &            $-21.80 \pm 0.30$ &           $0.03 \pm 0.01$ &                          -2.31 &                                -0.3  &                   -0.04 &                      0.01 &                    0.04 \\
 $^{0}M_{W2}-5logh$   &            $-21.19 \pm 0.30$ &           $0.00    \pm 0.01$ &                          -2.34 &                                -0.32 &                    0.01 &                      0.03 &                    0.02     \\
 $^{0}M_{W3}-5logh$   &            $-22.21 \pm 0.48$ &           $0.03 \pm 0.05$ &                          -1.73 &                                -1.28 &                    0.64 &                      0.05 &                    0.28  \\
 $^{0}M_{W4}-5logh$   &            $-22.39 \pm 0.67$ &           $0.01 \pm 0.04$ &                          -0.75 &                                -1.34 &                    0.49 &                      0.10  &                   -0.01  \\
\hline
\end{tabular}}
\caption{Absolute magnitude estimates for the Milky Way,  updating Table 1 and Table 3 of \citetalias[][]{licquia2015b} with additional absolute magnitude estimates beyond the optical.  GALEX UV, SDSS $ugriz$, 2MASS $J-Ks$ and Wise $W1-W4$ magnitudes are all presented in the AB system. The Johnsons-Cousins $UBVRI$ magnitudes are presented in the Vega system. }
\label{Tab:absmag}
\end{table*}



\section{Testing the Accuracy of Our Gaussian Process Regression Methods}
\label{sec:appendix_test_train}

In order to have confidence in the results we obtain from the Gaussian process regression, it is necessary to test the accuracy of the predicted photometry and error estimates. As in any supervised machine learning problem, we want to ensure that the given inputs map correctly to the given outputs. In order to enable an unbiased evaluation of our model we must split  our data into separate training and test samples, as we do not want to evaluate a model with information that was used to train it. To do this we use the $\tt{scikit-learn}$ function $\tt{sklearn.model\_selection.test\_train\_split}$. We use a split of $75\%$ of our data for training and $25\%$ of the data for testing. We do not construct a separate validation set as we do not have many hyper parameters to tune (the kernel defaults work well for our case, as discussed in \autoref{subsection:kernel}). 

For each photometric property that we study, we train the GPR using only the physical and photometric properties of the galaxies within the training sample. With that trained model we then make predictions for a given photometric property using the testing sample's physical properties. This gives us predicted absolute magnitudes and colours for every galaxy in our testing sample, which we can then compare to their actual observed values.

\begin{figure*}[!htbp]
    \centering
    \includegraphics[width=\linewidth]{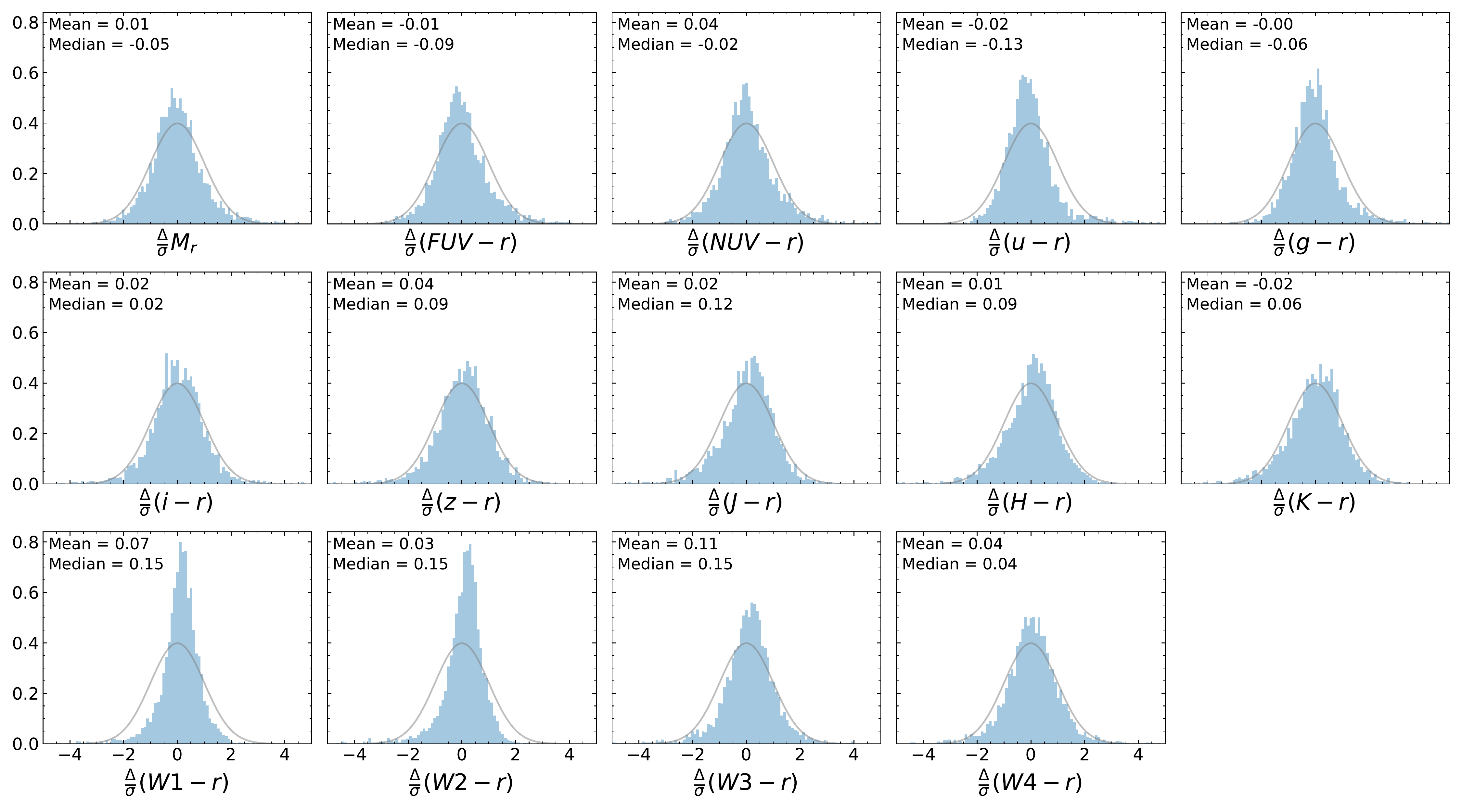}
    \caption{Histograms depicting the difference between the actual colour or absolute magnitude of galaxies in an independent test sample from the value predicted by GPR, normalised by the predicted error resulting from the GPR fit ($\frac{\Delta}{\sigma}$). For reference we have overlaid a Gaussian with a mean of zero and variance of one in each panel; in the ideal case, the normalised residuals would follow this distribution.  For reference we also provide the means and medians of $\frac{\Delta}{\sigma}$ for each panel. The $\frac{\Delta}{\sigma}$ distribution for each property matches the expectation well, with biases of $<0.2 \sigma$ and RMS residuals slightly smaller than would be expected from the predicted errors. Most of the minor skewing is attributed to galaxies with small axis ratios, for which colour changes quickly with inclination. Results for all quantities necessary for deriving the Milky Way SED are shown here. We can conclude that the GPR is able to accurately map from galaxy physical properties to colours and absolute magnitudes.}
    \label{fig:test_train_hist}
\end{figure*}

\autoref{fig:test_train_hist} depicts the results of this training and testing analysis for all of the bands relevant to constructing the SED. We present histograms based on the differences between the intrinsic value of a photometric property and the predicted value from the GPR, $\Delta$, for each galaxy in the testing sample. We then divide these $\Delta$ values by the standard deviation predicted by the GPR for that galaxy, $\sigma$ to obtain the normalised residual, $\frac{\Delta}{\sigma}$. By focusing on this quantity we can assess whether a bias in a given galaxy is present and how large it is compared to the prediction uncertainties, and can additionally evaluate whether the error estimates from the GPR are too small (too optimistic) or too large (too conservative) compared to actual deviations.

In each histogram, we overlay a Gaussian with a mean of zero and a variance of one in grey. One would expect the distribution of $\frac{\Delta}{\sigma}$ to follow this distribution if there is no bias in the predictions and the error values are all accurate. We also list the means and medians of $\frac{\Delta}{\sigma}$ for each band on the corresponding histogram. Some quantities exhibit a small skew, but invariably the mean offsets are $<0.05\sigma$ and the median offsets are $<0.2\sigma$, so any bias is far subdominant to other sources of errors. In general the distribution of $\frac{\Delta}{\sigma}$ is tighter than the expected Gaussian, indicating that the $\sigma$ estimates from GPR are slightly too large (i.e., overly conservative).

In order to assess what may cause the small offsets, we have evaluated the histograms of $\frac{\Delta}{\sigma}$ separately for face-on  galaxies (which we take to be those with axis ratio $b/a > 0.5$) and more inclined objects ($b/a \leq 0.5$). We find that most of the slight skewing that is observable in \autoref{fig:test_train_hist} is driven by the galaxies with higher inclinations. In this regime galaxy colour can change quickly and non-linearly with $b/a$, which can explain why the GPR struggles to accurately predict colours for these galaxies.

Overall, we find that the GPR performs well at mapping from galaxy physical parameter inputs to photometric properties, with biases that are much smaller than the predicted uncertainties, and error estimates that err on the conservative side. We therefore can use the method with confidence to make predictions for the photometric properties of the Milky Way.

\section{$k$-corrections for WISE Bands}
\label{sec:appendix_systematics}

\subsection{Calculating the $k$-correction}
\label{subsection:kcorrections}

When we observe extragalactic objects their spectrum is redshifted into the ``observed'' reference frame. In order to convert from observed flux to rest-frame values, astronomers utilise $k$-corrections, which account for the shifting of restframe light into different passbands as redshift increases. While most of the photometry that we have used in this work (as described in \autoref{subsection:photometry}) has readily available $k$-corrections into restframe bands, measurements in the WISE bands \citep{wise} do not.

Typically, $k$-corrections for a given band are calculated by calculating the effect on observed colours of redshifting a theoretical or empirical spectral energy distribution (SED) template; existing software packages such as \textsc{kcorrect} \citet{blanton2007} have been designed to perform this task. However, this approach is not feasible at wavelengths where SED templates are not available or poorly constrained.

For this reason, in the WISE bands we adopt a new, almost purely empirical approach which builds on the template-based methods presented in \citet{beare2014}. Our $k$-correction techniques will be the subject of an upcoming paper, (Fielder, Newman, \& Andrews, in prep). While we do not go into detail here, we summarise the strategy below.

We rely on the assumption that the $k$-correction needed for a given band is a simple (polynomial) function of a galaxy's restframe colour determined in some pair of bands; i.e., that galaxy SEDs constitute a one-parameter family. This colour can be determined in the optical where more conventional $k$-correction methods are effective; we just need some way to sort SEDs into a one-parameter family. The assumption that one parameter is sufficient is not perfect, as there is some diversity in SEDs at fixed restframe colour, but it is good to first order as variations in dust, specific star formation rate, and metallicity all have coarsely similar effects on the spectrum of a galaxy. Because we are focused on $k$ corrections over a limited redshift range near $z=0$, getting things right to this level is sufficient to make $k$-correction errors far subdominant to other sources of uncertainty in the WISE bands.

Our goal is to produce a function that takes as inputs the apparent magnitude in some band, the redshift, and the restframe colour in some pair of bands (which can be determined by conventional $k$-correction methods, we will use $^0(g-r)$ for this example), and returns the restframe absolute magnitude in the desired band. We determine this function in two steps. First, we determine polynomial relationships between the \textit{observed} colour in a pair of bands (one of which serves as the anchor, and one of which is the desired target band) and redshift, in bins of the restframe colour in another (e.g., optical) pair of bands which is provided as one of the inputs. This relationship is determined via a second order polynomial fit; i.e., we parameterize the observed colour $(X - r)_{\rm observed} = f(z) = a_{0} + a_{1}z + a_{2}z^{2}$. For each bin in restframe colour, we determine the fit coefficients separately. This is similar to the technique applied in \citet{brown2014}; see, e.g., Fig. 2 of that work.

We then fit for the linear dependence of each of the fit coefficients on the mean restframe colour of a bin; i.e., we now treat the coefficients as functions $a_{1}(^0(g-r))$ and $a_{2}(^0(g-r))$. We ignore the value of $a_{0}$ as it corresponds to the predicted mean restframe colour at $z=0$; rather than assuming this mean colour is appropriate for all objects, we will instead only make small adjustments to the observed photometry. To do this, we first fit for $a_{1}(^0(g-r))$ and $a_{2}(^0(g-r))$ via a linear regression algorithm which is robust to outliers, specifically the Huber regression method implemented by $\tt{scikit-learn}$ ($\tt sklearn.linear\_model.HuberRegressor$). We assume a linear dependence between each coefficient and the restframe reference colour ($^{0}(g-r)$ here).

The result of this process is a function that predicts the observed colour in a new pair of bands, as a function of the restframe colour in some other pair of bands and the redshift. At $z=0$, the observed colour and restframe colour must agree, by definition. Hence, the \textit{difference} between the prediction for the observed colour for an object at a given redshift, and the prediction for the observed colour at $z=0$ (which corresponds to the value of $a_0(^0(g-r)$, must be identical to the $k$-correction for the target colour. If one subtracts off the $k$-corrected reference band (which again can be an optical band where $k$-corrections are already well-determined), one is left with the $k$-corrected unknown band on its own. 

We apply this (small) correction to the \textit{observed} photometry 
to map to the restframe equivalent (i.e., we do not simply adopt the colour in the new bands predicted from the restframe colour in known bands -- which could entail large shifts from the observed photometry for some objects -- but rather we only make use of the small \textit{offset} between observed and restframe colours that was fit across the sample). To describe the process symbolically, we use the restframe $^{0}(g-r)$ colour for an object to obtain the $a_{1}$ and $a_{2}$ values for the mapping of colour in some band $X$ relative to $r$, $X - r$. Then we can obtain the restframe colour in the new band $^{0}(X-r) = (X-r)_{\rm observed} - a_{1}z - a_{2}z^{2}$. Finally, we can determine the restframe absolute magnitude in band $X$ via the expression $^{0}M_{X} = ^{0}M_{r} + ^{0}(X-r)$.  

The result of this process is a simple but effective empirical method for obtaining $k$-corrections for bands where SED templates are not well known, which can be applied so long as $k$-corrections have already been determined in another set of better-characterised bands. In this work, we use $^{0}(g-r)$ colour to determine the $k$-correction mappings as a function of restframe colour and redshift, and WISE-$r$ colours as the target in each case. We have selected these bands as $g$ and $r$ both have small photometric uncertainties and well-characterised $k$-corrections.  We will present more details of our procedures and tests of their effectiveness in \citet{fielder2021c}.  

When evaluating the fits for the the $k$-corrections we exclude objects that have large WISE photometric errors in the W1 and W2 channels. We do not place requirements on W3 or W4 errors as they are invariably large. Specifically, in $W1$ we perform the $k$-correction fits restricting to galaxies with errors $\sigma_{m_{W1}}<0.125$, while in $W2$ we require errors $\sigma_{m_{W2}}<0.25$. Objects with larger errors \textit{are} still included when performing our GPR analyses.


\subsection{Photometric Offsets Between SDSS and DESI Legacy Survey Imaging}
\label{subsection:legacy_offset}

As described in \autoref{subsubsection:legacy}, we rely on Legacy Survey catalogues for WISE photometry, as the \textit{Tractor}-based measurements have lower uncertainties than other public catalogues. Because of the use of a matched object model across all passbands, colours in WISE bands relative to optical bands will be determined with greater accuracy than total magnitudes in a single band.  However, because the filters and instruments used for the Legacy Surveys have different transmission and response curves and photometry was performed with differing analysis pipelines (BASS $g$ and $r$ and MOSAIC-3 $z$ filters were used in the northern portion of the Legacy Survey footprint, while in the south DECam was used for $g$, $r$, and $z$), the rest-frame $r$ band absolute magnitudes for a given object should differ between the SDSS and Legacy data.

\begin{figure}
    \centering
    \begin{subfigure}{0.9\linewidth}
        \includegraphics[width=\linewidth]{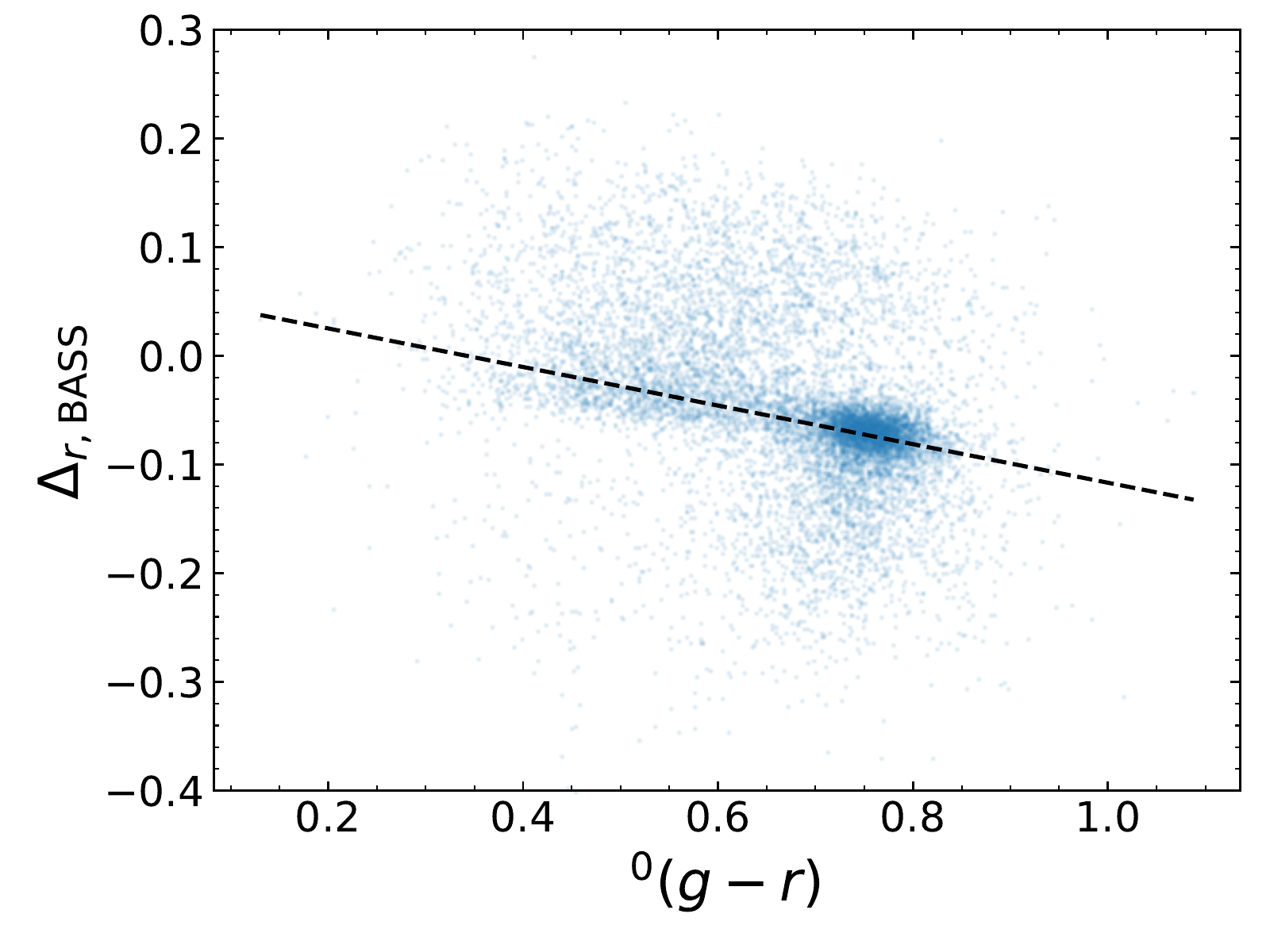}
    \end{subfigure}\\
    \begin{subfigure}{0.9\linewidth}
        \includegraphics[width=\linewidth]{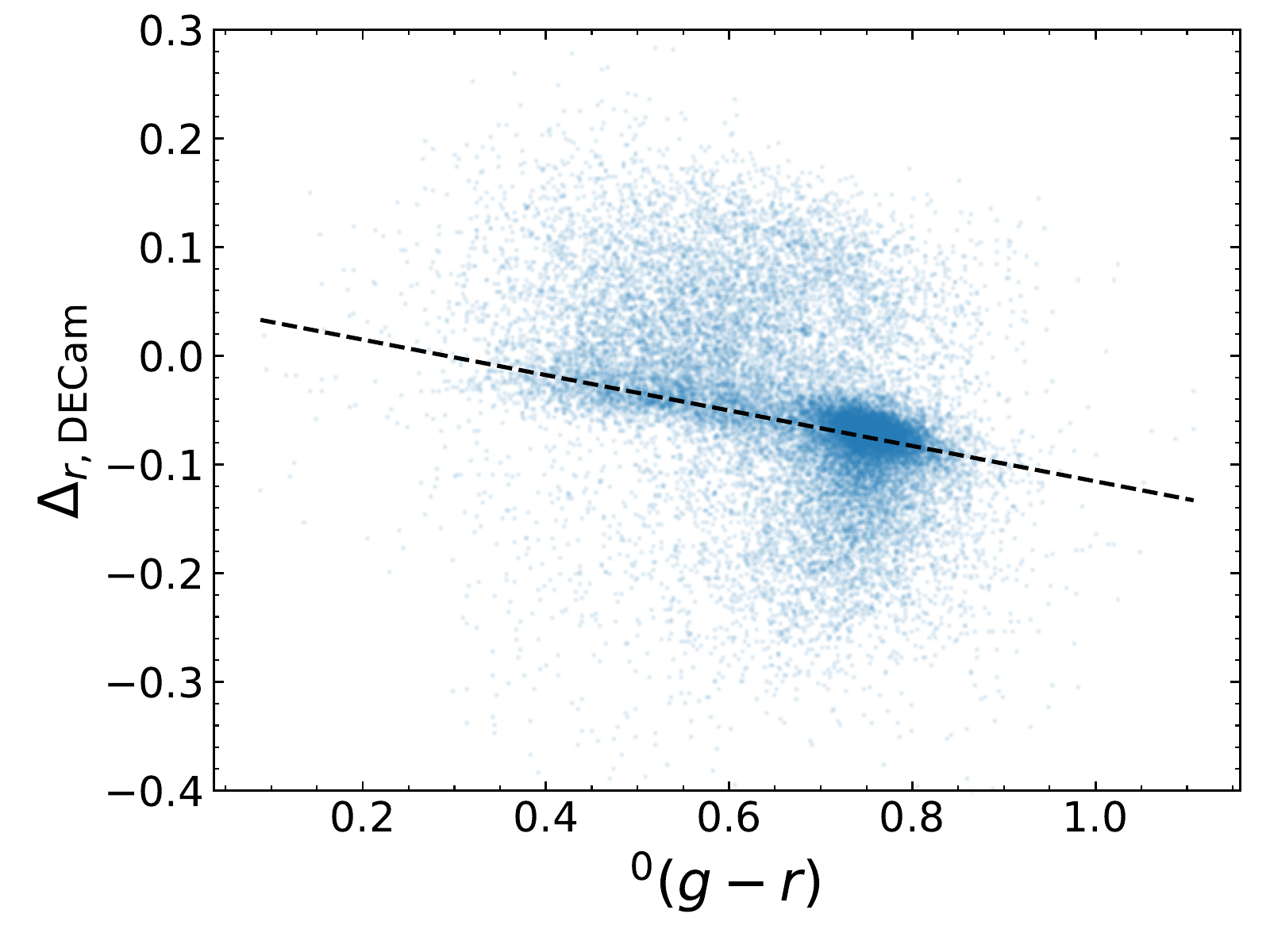}
    \end{subfigure}
\caption{Offsets between SDSS and DESI Legacy Survey restframe $r$-band absolute magnitude, plotted as a function of SDSS restframe colour, $^{0}(g-r)$. The upper panel shows objects in the northern portion of the Legacy Survey, which utilised the BASS filters and instrument in $r$. The lower panel shows objects in the southern Legacy region, where DECam was used. The dashed black line depicts our linear fit for $\Delta_{r}$ as a function of $^{0}(g-r)$. This fit follows the locus of objects well. With this fit we can obtain the offsets between colours measured relative to Legacy Survey and SDSS photometry and subtract those offsets to construct predicted WISE-based colours.}
\label{fig:legacy_offsets}
\end{figure}

To calculate the WISE minus SDSS $r$ band colours used to construct the Milky Way SED, we therefore must take the WISE minus Legacy Survey $r$ band colour (which should be measured self-consistently due to the use of a common model) and apply a small correction to compensate for the differences between Legacy $r$ and SDSS $r$. To do this we proceed as follows:
\begin{enumerate}
    \item \textbf{Calculate absolute magnitudes for DESI Legacy bands --}
        We calculate restframe absolute magnitudes for all objects in the cross-matched catalogue in the BASS and DECam $r$ and SDSS $g$ and $r$ bands, using the \textsc{kcorrect} \citep{blanton2007} software. We base these calculations only on the SDSS $ugriz$ photometry for our galaxy sample.
    \item \textbf{Calculate the filter offsets in $r$ --}
        We now wish to determine the offsets between restframe $r$ absolute magnitude in Legacy Survey and SDSS filters, as a function of restframe colour: $\Delta_{r} = M_{r,\rm{BASS}} - M_{r,\rm{SDSS}}$ for the North, or  $\Delta_{r} = M_{r,\rm{DECam}} - M_{r,\rm{SDSS}}$ for the South. In \autoref{fig:legacy_offsets} we plot $\Delta_{r}$ as a function of SDSS $^{0}(g-r)$ for our sample. Almost all objects fall along a linear relationship between the $r$-band offset and colour.
    \item \textbf{Fit for offsets as a function of $^{0}(g-r)$ --} 
        We perform a robust least-squares fit using the $\tt{scikit-learn}$ \citep{scikit-learn} Huber regression function with $^{0}(g-r)$ as the independent variable and $\Delta_{r}$ as the dependent variable. In \autoref{fig:legacy_offsets} we plot these fits with black dashed lines. 
\end{enumerate}

Using the coefficients resulting from this fit, we can convert a restframe colour referenced to Legacy Survey $r$ (e.g., $^{0}(W1-r_{\rm{DECam}})$ to one referenced to SDSS $r$ ($^{0}(W1-r_{\rm{SDSS}})$, for this example) by evaluating the fit line at the restframe $^{0}(g-r)$ colour for a given object. Our GPR predictions for WISE absolute magnitudes and WISE-$r$ colours require use of this correction; however, for WISE colours such as (W1-W2) the dependence on the offset cancels out. 

We apply this correction in tandem with calculating the $k$-corrected WISE colours, before training the Gaussian process regression. In the previous sub-section we expressed our calculation of absolute magnitudes as $^{0}M_{X} = ^{0}M_{r} + ^{0}(X-r)$. Re-written to explicitly use colour relative to the SDSS $r$ band, this expression becomes
\begin{equation}
    ^{0}M_{X}=\;^{0}M_{r_{\rm{SDSS}}} +\;^{0}(X-r_{\rm{SDSS}}) \\  = ^{0}M_{r_{\rm{SDSS}}} + (^{0}M_X-^{0}M_{r_{\rm{SDSS}}}) .
\end{equation}
However, we have measured the $X - r_{\rm Legacy}$ colour, not $X - r_{\rm SDSS}$.  In order to account for the offset between Legacy and SDSS photometry we therefore require the correction derived above: 
\begin{equation}
    ^{0}M_{X}=\; ^{0}M_{r_{\rm{SDSS}}} + (^{0}M_X-^{0}M_{r_{\rm{Legacy}}}) + (^{0}M_{r_{\rm{Legacy}}}-^{0}M_{r_{\rm{SDSS}}}) , 
\end{equation}
where $(^{0}M_{r_{\rm{Legacy}}}-^{0}M_{r_{\rm{SDSS}}})$ is the offset between restframe SDSS $r$ and DESI Legacy Survey $r$, or $\Delta_{r}$ above. Applying this correction before performing further analysis allows us to train the GPR based upon the combined North and South DESI Legacy catalogues rather than training with each separately.

The magnitude of the offsets applied to the WISE photometric band estimates are small, spanning from $-0.14$ to $0.05$ in magnitude. Compared to the photometric errors of these bands in \autoref{Tab:color_sed} and \autoref{Tab:absmag}, the errors on the photometry far outshine any error attributed to the DESI Legacy - SDSS offset (especially for W3 and W4).

\section{Eddington Bias Corrections}
\label{subsection:eddbias}

Uncertainties in stellar mass, star formation rate, disc scale length, etc. can lead to biases in the inferred photometric properties of the Milky Way. 
When the distribution of objects in parameter space varies non-linearly, scatter from errors will preferentially move objects from more sparsely populated regions to rarer locations compared to the opposite situation. This can lead to an ``Eddington'' bias in inferred properties. For example, because of the rarity of massive galaxies, a galaxy with a large stellar mass measurement is more likely to have an intrinsic stellar mass that is smaller than the measured value than one that is larger, as there is a much larger number of objects that can up-scatter compared to the number that down-scatter. As a result, the stellar mass measurements used when training the Gaussian process regression would then be biased high. Similar effects can occur with star formation rate, disk scale length, etc.


We quantify this bias through an empirical approach based upon the methods presented in \citetalias[][]{licquia2015b}. We perturb the measured values of the galaxy sample used to train the GPR with Gaussian noise sampled from the measured errors. Specifically this means we repeatedly add to each galaxy's measured $\mass$, SFR, $B/T$, etc. a value randomly drawn from a Gaussian distribution centred at zero with a standard deviation of that object's error in the given quantity (e.g., $\sigma_{\log{\mass}}$), and determine the effects that adding errors to each quantity has. We make the simplifying assumption that the bar vote fraction should not significantly contribute to the Eddington bias (as uncertainties in that quantity are difficult to characterise, but also should have limited effects on photometry).

To obtain predicted photometry for a given noise level, we run the GPR in a similar manner to the methods used in \autoref{sec:gpr}, except we now train the regression on the perturbed sample before predicting photometric quantities. We then predict the photometric quantity of interest by evaluating each model assuming the fiducial value for each  Milky Way physical property (i.e., we use a fixed set of values for all evaluations rather than sampling from the MW distributions). In our application, we perform this procedure on samples with Gaussian noise applied to each physical property from one to four times successively.

In the original cross-matched training set, the values for each physical property have been perturbed from their true, intrinsic values due to measurement errors; we define this case as corresponding to having noise applied $n=1$ times. However, we wish to evaluate what the measurement would be if there were no noise; i.e., for the case where $n=0$.  We therefore wish to characterise the difference in the property of interest between the case where noise has been applied $n$ or $n-1$ times, and evaluate for $n=1$ to determine the correction needed to remove Eddington bias.

To determine this value we perform use a least squares quadratic fit to the set of differences between GPR predictions as we add more noise: i.e., we fit a relation of the form $P_n - P_{n-1} = An^{2} + Bn + C$, where $P_n$ is the prediction from GPR when noise has been applied a total of $n$ times and $A$, $B$, and $C$ are coefficients of the fit. We use the resulting fit to extrapolate to $P_1 - P_{0}$, which should correspond to the effect of Eddington bias. The mean Eddington bias at $n=1$ is then quantified as the sum of the coefficients $A+B+C$ from the quadratic fit of the set of differences $P_n - P_{n-1}$. In practice, we perform the entire analysis (adding random noise repeatedly, training the GPR, fitting for $P_n - P_{n-1}$, and adding the coefficients) 25 times. The effect of Eddington bias for a given band is then calculated as the mean of the set of 25 values. We then subtract this value from the GPR-predicted mean for a given photometric property of the Milky Way to obtain a corrected value. The bias removed in each of our predictions is documented in the third column of \autoref{Tab:color_sed}-\autoref{Tab:absmag}.

The uncertainty in each of the 25 Eddington bias estimates (i.e., in the quadratic function evaluated at the point $n=1$) can be calculated via the square root of the sum of the covariance matrix for the coefficients of the least-squares quadratic fit (as when $n=1$ the quadratic result is simply $A+B+C$). The uncertainty in the Eddington bias in each band is therefore equivalent to $\sigma_{\rm{bias}} = \frac{\sqrt{\sum_{i = 1}^{n}{\sigma_i^2}}}{n}$, where $\sigma_i$ is the estimated uncertainty from the i'th sample and here $n=25$. As a check we have also computed errors via the standard error on the mean Eddington bias and find comparable (though somewhat more optimistic) results. The Eddington bias uncertainty is combined in quadrature with the uncertainty estimated from sampling the GPR  to produce the errors on the corrected values in the tables in \autoref{sec:appendix_data}.

\bsp	
\label{lastpage}
\end{document}